\documentclass[%
 reprint,
groupedaddress,
longbibliography,
 amsmath,amssymb,
prx,
]{revtex4-1}

\pdfoutput=1
\usepackage{lipsum}
\usepackage{graphicx}
\usepackage{dcolumn}
\usepackage{bm}
\usepackage{hyperref}

\newcommand{\expct}[1]{\langle #1 \rangle}
\newcommand{\figref}[1]{Fig.~\ref{#1}}
\newcommand{\figsref}[1]{Figs.~\ref{#1}}
\renewcommand{\eqref}[1]{Eq.~(\ref{#1})}
\newcommand{\tbref}[1]{Table~\ref{#1}}
\newcommand{\pref}[1]{(\ref{#1})}

\newcommand{\unit}[1]{~\mathrm{#1}}
\newcommand{\degc}{\unit{{}^\circ{}C}}

\newcommand{\thetan}{\theta_\mathrm{n}}
\newcommand{\thetap}{\theta_\mathrm{p}}

\DeclareMathOperator{\sign}{sign}

\begin{document}

\title{Tilt-induced polar order and topological defects in growing bacterial populations}

\author{Takuro Shimaya}
\email{t.shimaya@noneq.phys.s.u-tokyo.ac.jp}
\affiliation{Department of Physics, The University of Tokyo, 7-3-1 Hongo, Bunkyo-ku, Tokyo, 113-0033, Japan.}%
\author{Kazumasa A. Takeuchi}
\email{kat@kaztake.org}
\affiliation{Department of Physics, The University of Tokyo, 7-3-1 Hongo, Bunkyo-ku, Tokyo, 113-0033, Japan.}%

\date{\today}

\begin{abstract} 
       Rod-shaped bacteria, such as {\sl Escherichia~coli}, commonly live forming mounded colonies.
       They initially grow two-dimensionally on a surface and finally achieve three-dimensional growth.
       While it was recently reported that three-dimensional growth is promoted by topological defects of winding number $+1/2$ in populations of motile bacteria, how cellular alignment plays a role in non-motile cases is largely unknown.
       Here, we investigate the relevance of topological defects in colony formation processes of non-motile {\sl E.~coli} populations, and found that {\sl both} $\pm{}1/2$ topological defects contribute to the three-dimensional growth.
       Analyzing the cell flow in the bottom layer of the colony, we observe that $+1/2$ defects attract cells and $-1/2$ defects repel cells, in agreement with previous studies on motile cells, in the initial stage of the colony growth.
       However, later, cells gradually flow toward $-1/2$ defects as well, exhibiting a sharp contrast to the existing knowledge.
       By investigating three-dimensional cell orientations by confocal microscopy, we find strong vertical tilting of cells near the defects.
       Crucially, this leads to the emergence of a polar order in the otherwise nematic two-dimensional cell orientation.
       We extend the theory of active nematics by incorporating this polar order and the vertical tilting, which successfully explains the influx toward $-1/2$ defects in terms of a polarity-induced force.
       Our work reveals that three-dimensional cell orientations may result in drastic changes in properties of active nematics, especially those of topological defects, which may be generically relevant in active matter systems driven by cellular growth instead of self-propulsion.
\end{abstract}
\maketitle

\begin{figure*}[t]
       \centering
       \includegraphics[width=0.85\hsize]{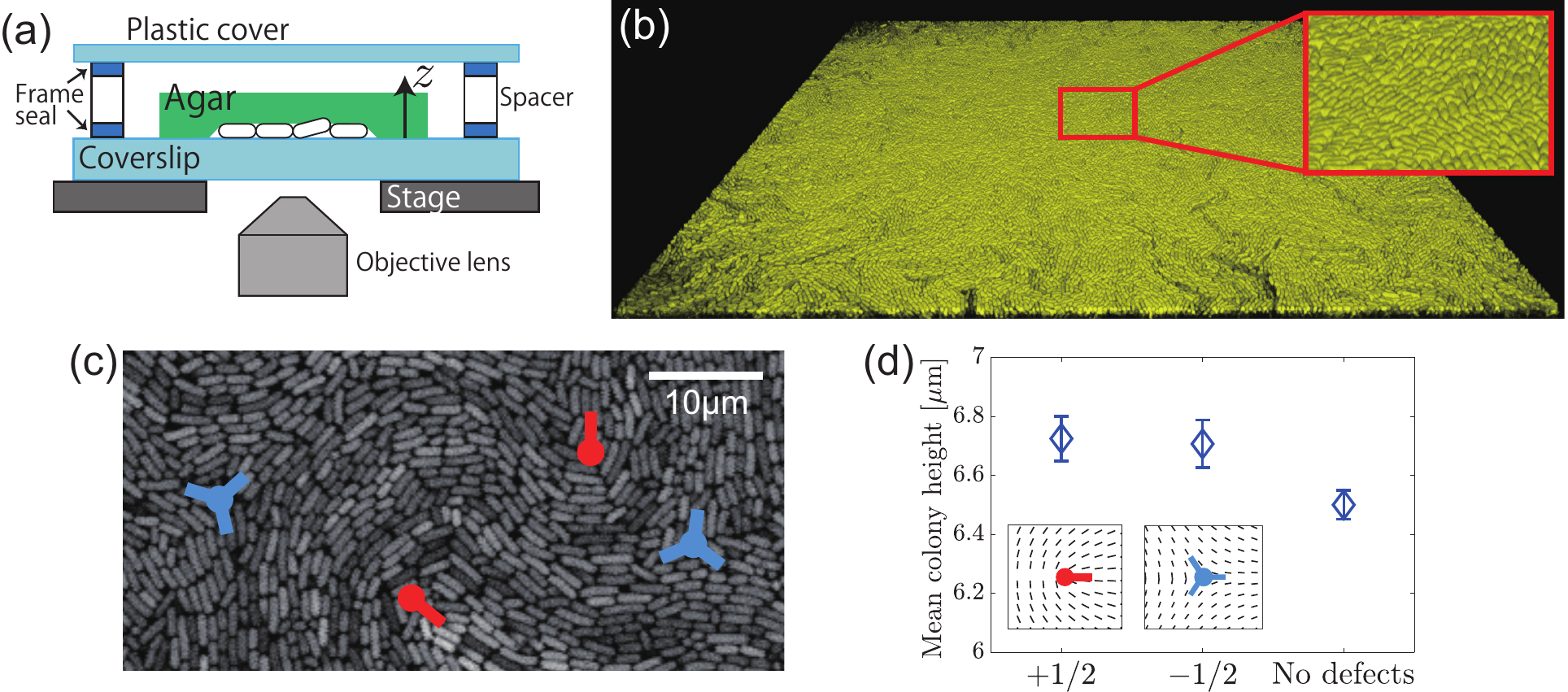}
       \caption{
       \label{fig-1}
       Morphology of three-dimensional colonies formed from numerous cells observed by end-point confocal microscopy.
       The confocal data were taken 14 hours after the cells had filled the bottom plane.
       (a) Experimental setup.
       Bacterial cells were between a coverslip and a nutrient agar pad.
       (b) A three-dimensional image of colonies ($184.52\unit{\mu{}m} \times 184.52 \unit{\mu{}m}$ square), where cells were stacked three-dimensionally.
       The region surrounded by the red rectangle is enlarged and displayed in the upper right side.
       (c) A two-dimensional cross section showing the bottom layer.
       Red comets and blue trefoils indicate $+1/2$ and $-1/2$ defects, respectively. 
       The arms of the symbols reflect the structure of the director field as illustrated in the insets of (b).
       (d) Mean colony height at the location of the defects in the bottom layer and that far from defects. At each position in the $xy$-plane, the height was evaluated by the length of the profile along the $z$-axis whose intensity is higher than 20\% of the maximum.
       The heights at the defect positions were extracted from all of the hundreds of defects that were sufficiently far ($> 9\unit{\mu m}$) from each other.
       For the colony heights far from defects, we randomly picked up 1000 points which were sufficiently far ($> 9\unit{\mu m}$) from any defect (see methods in Appendix\,A).
       The error bars indicate the standard error from the ensemble averaging.
       The optical resolution was about $250\unit{nm}$ in the vertical direction.
       See also Fig.\,S2(a)-(c) for the colony height distributions.
       Here we show the result of a single measurement (uniform colony, end-point confocal \#{}1); see Fig.\,S2(e) for the result of another biological replicate (uniform colony, end-point confocal \#{}2), where we confirmed that the colony height was again higher at the locations of the defects.
       }
\end{figure*}

\section{Introduction}
Numerous species of bacteria live in dense populations, which often take the form of biofilms \cite{Flemming2016}.
Besides being a challenging subject for biologists and physicists, because biofilms cause a variety of problems in medicine, industry, and our daily life \cite{MattilaSandholm1992, Shirtliff2009}, understanding the mechanism of biofilm formation is a crucial mission across diverse disciplines.
In the early stage of biofilm formation processes, two-dimensional colonies are first formed, then a three-dimensional structure is eventually constructed \cite{Flemming2016}.
Because mechanical interactions between cells are important at this stage, many studies have attempted to understand structure formation dynamics from a physical perspective \cite{Allen2018}.

In particular, rod-shaped bacteria, irrespective of whether they are motile or not, are aligned with each other and behave like an active nematic liquid crystal in a dense two-dimensional space \cite{Peng2016,Genkin2017,You2018,Yaman2019,Dell2019,Doostmohammadi2019,Sengupta2020,Meacock2021,Copenhagen2021}.
For motile bacteria, it has recently been reported that $+1/2$ topological defects promote three-dimensional growth of \textit{Myxococcus~xanthus} populations \cite{Copenhagen2021}.
Besides bacteria, it is known that topological defects also play decisive roles in various kinds of cell populations \cite{Saw2018,DOOSTMOHAMMADI2021}, such as epithelial cells \cite{Saw2017}, neural stem cells \cite{Kawaguchi2017}, fibroblasts \cite{Duclos2017, Turiv2020} and actin fibers in \textit{Hydra} \cite{Maroudas-Sacks2021}.
However, for growing but non-motile bacteria, while some studies investigated how non-motile cells initiate three-dimensional growth \cite{Su2012,Farrell2013,Grant2014,Duvernoy2018,Beroz2018,You2019,Neher2019,Hartmann2019,Takatori2020,Dhar2021},
the relevance of local cell alignment to three-dimensional growth, in particular, that of topological defects, remains unknown.

Here, by observing colony formation processes of non-motile {\sl E.~coli} between a coverslip and a nutrient agar pad (\figref{fig-1}(a)),
we find an indication that both $+1/2$ and $-1/2$ topological defects promote three-dimensional growth of colonies.
This finding is put on solid ground by analyses of the two-dimensional velocity field around topological defects, which reveal that cells are transported toward both $+1/2$ and $-1/2$ defects, implying upward growth there.
Remarkably, this influx toward both types of defects is contrary to the existing knowledge that cells escape from $-1/2$ defects \cite{Peng2016,Genkin2017,Saw2017,Kawaguchi2017,Turiv2020,Copenhagen2021}, and cannot be explained by the conventional active nematic theory.
Combining confocal observations and theoretical modeling, we find that the three-dimensional tilting of cells is promoted around topological defects, which can induce additional force around defects.
Crucially, we uncover the formation of a polar order due to three-dimensional asymmetric tilting of cells around defects, which turns out to be the key to theoretically account for the emergence of the influx toward $-1/2$ defects.

\section{Results}
\subsection{Topological defects promote three-dimensional growth of bacterial colonies}  \label{sec:res1}
First we studied the relation between cell orientation and colony structure, using non-motile {\sl E.~coli} placed between a coverslip and a nutrient agar pad (\figref{fig-1}(a); see Appendix\,A for the experimental methods).
We put cell suspension on the coverslip so that cells are initially distributed densely and uniformly.
Then we cultured it for $14\unit{hours}$ after cells had filled the bottom plane and observed the resulting three-dimensional colony, which consisted of multiple layers of tilted cells, by confocal microscopy.
Here we took only a single confocal image at this end point, to take a high quality image without photobleaching.

\begin{figure*}[t]
       \includegraphics[width=\hsize]{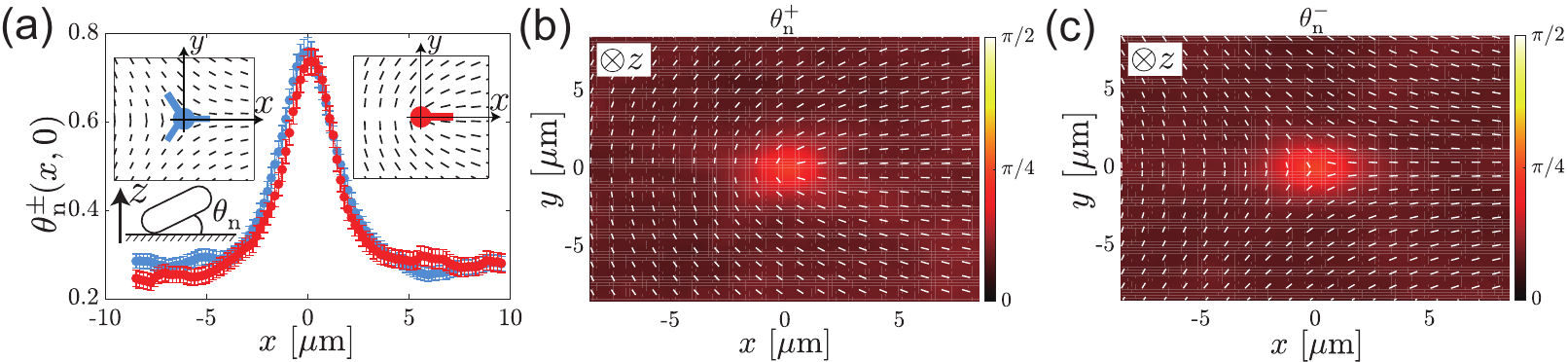}
       \caption{
       \label{fig-3}
       Results on the three-dimensional nematic tilting obtained by the end-point confocal observation.
       To obtain them, we first measured the three-dimensional cell orientations by the structure tensor method and obtained the non-negative tilt angle with respect to the $xy$ plane (see methods in Appendix\,A). We then took the ensemble average over all defects.
       These data were obtained by a single measurement (uniform colony, end-point confocal \#{}1), but the reproducibility was confirmed in Fig.\,S7(a)(b) by using another substrate and agar pad (uniform colony, end-point confocal \#{}2).
       (a) Nematic tilt angle $\thetan^\pm(\bm{r})$ around $\pm 1/2$ defects. The results on the $x$-axis, i.e., $\bm{r}=(x,0)$, are displayed.
       The error bars indicate the standard error from the ensemble averaging.
       The two insets illustrate the definition of the axes for $+1/2$ (right inset) and $-1/2$ (left inset) defects.
       (b,c) Spatial profiles of the nematic tilt angle $\thetan^\pm(\bm{r})$ for the $+1/2$ defects (b) and the $-1/2$ defects (c).
       The white rods represent the nematic director field.
       }
\end{figure*}

To test the relevance of cell alignment to the three-dimensional growth, we investigated whether the presence of topological defects influenced the colony height.
First, we noticed that the orientation of cells in the bottom layer was nearly horizontal (\figref{fig-1}(b)(c)), albeit weakly tilted (typically $\sim 10^\circ$ in this end-point observation; see \figref{fig-3}(a) and descriptions thereof).
Therefore, we can regard this bottom layer as a quasi-two-dimensional active nematic system.
We measured the two-dimensional orientation of cells, $\bm{n}(\bm{R})$ at position $\bm{R}$ in the bottom layer, from the image intensity using the structure tensor method (see Fig.\,S1(a) and methods in Appendix\,A).
We then detected topological defects (\figref{fig-1}(c) and Fig.\,S1), and measured the colony height at the positions of the defects (see methods in Appendix\,A).
For comparison, we also measured the colony height at randomly selected locations that are sufficiently far from topological defects.
We found that the mean colony height is slightly higher at the positions of the defects (\figref{fig-1}(d)), both $+1/2$ and $-1/2$, than in the regions far from the defects.
The statistical significance is confirmed by the Wilcoxon rank-sum test (Fig.\,S2).
For the null hypothesis that the median of the height distribution at the positions of the defects is identical to that far from defects, the p-value was $0.018$ for the $+1/2$ defects and $0.043$ for the $-1/2$ defects.
While these results were obtained from observations of 20 separate regions in a single experiment, the reproducibility was also confirmed by another biological replicate using a different substrate and agar pad (Fig.\,S2(e)).
These results suggest that topological defects promote the vertical growth of colonies.

\begin{figure*}[p]
       \includegraphics[width=.85\hsize]{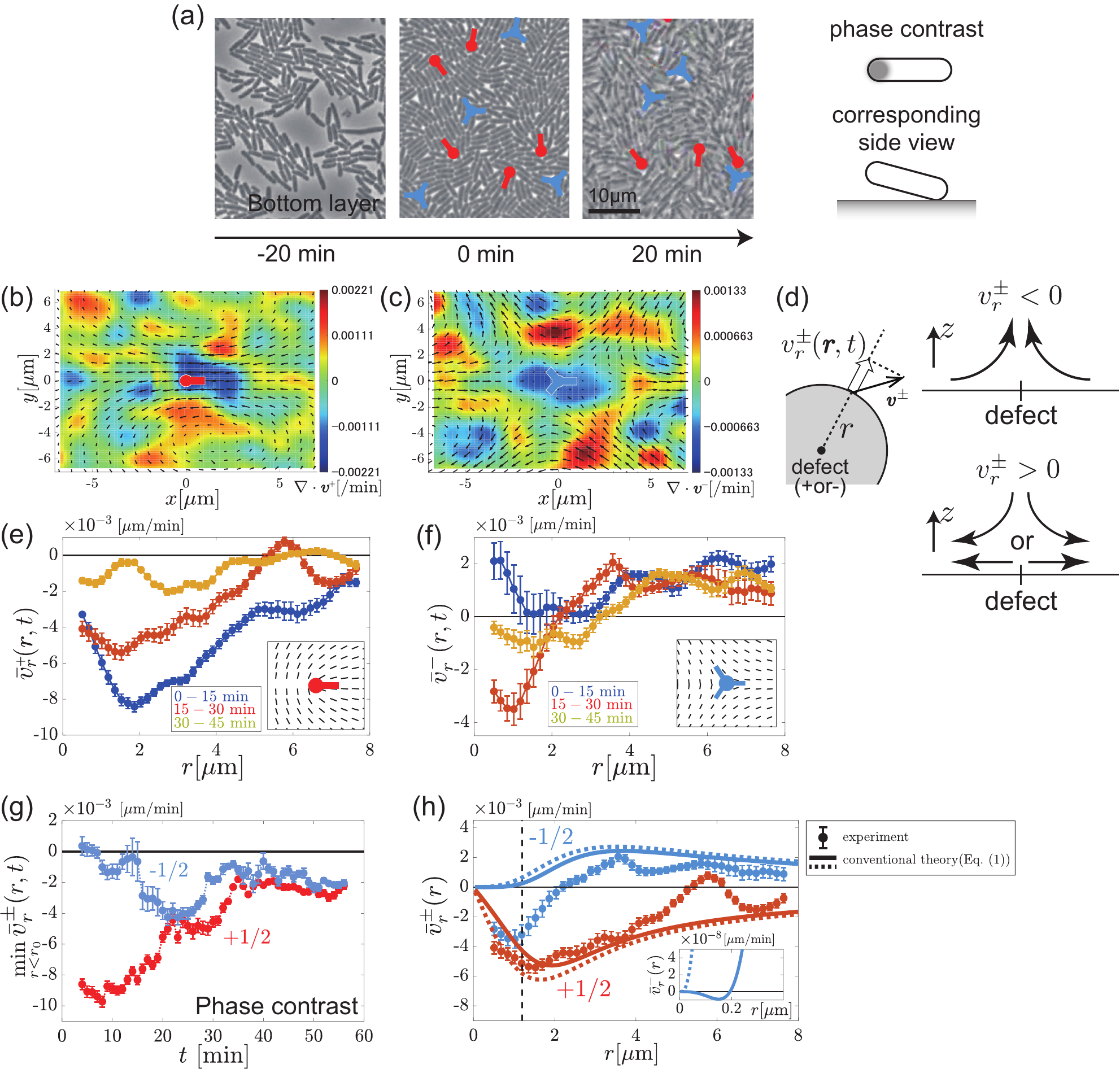}
       \caption{
       \label{fig-2}
       Phase-contrast images and results of two-dimensional velocity analyses for uniform colonies formed from numerous cells observed by phase-contrast microscopy.
       The time interval of the time-lapse observation was $1\unit{min}$.
       The double sign $\pm$ corresponds to the sign of the defects.
       These data were obtained by a single measurement (uniform colony, phase contrast \#{}1), but the reproducibility was confirmed in Fig.\,S5 by using another substrate and agar pad (uniform colony, phase contrast \#{}2).
       (a) Phase-contrast images of cells taken at $t=-20,0,20\unit{min}$, from left to right.
       $t=0$ is the moment when cells filled the bottom plane. 
       See also Videos~1 and 2.
       The sketches on the right side illustrate how three-dimensionally tilted cells appear in the phase-contrast images.
       (b,c) Velocity field $\bm{v}^\pm(\bm{r})$ (black arrows) and its divergence (color map) around $+1/2$ defects (b) and $-1/2$ defects (c).
       (d) Schematic illustration of the definition of the radial velocity $v_r^\pm(\bm{r},t)$.
       With this, the mean radial velocity is defined by $\bar{v}_r^\pm(r,t) \equiv \frac{1}{2\pi}\oint d\phi v_r^\pm(\bm{r},t)$, which corresponds to the average of the radial components of the velocity over a circumference of radius $r$ centered at the defect.
       The sign of $\bar{v}_r^\pm$ indicates the direction of net flow, including the existence of vertical growth if $\bar{v}_r^\pm < 0$.
       (e,f) Time evolution of the mean radial velocity $\bar{v}_r^\pm(r,t)$ around $+1/2$ defects (e) and $-1/2$ defects (f).
       Here we used the velocity field averaged over $0\unit{min} \leq t \leq 15\unit{min}$, $15\unit{min} \leq t \leq 30\unit{min}$, and $30\unit{min} \leq t \leq 45\unit{min}$.
       The error bars indicate the time average of the standard error evaluated from each frame.
       (g) Time evolution of the minimum of $\bar{v}_r^\pm(r,t)$ in the region $r<r_0=2\unit{\mu{}m}$ near the defect.
       The moving average taken from $(t-5\unit{min})$ to $(t+5\unit{min})$ is shown with the corresponding error bar.
       (h) Comparison of the mean radial velocity $\bar{v}_r^\pm(r)$ between the experimental data (symbols) and theoretical curves based on the conventional equation (\eqref{eq:3}) (lines).
       The displayed experimental data (symbols) are identical to those shown in (e,f) for $15\unit{min} \leq t \leq 30\unit{min}$ (for which the influx toward $-1/2$ defects was strongest).
       The vertical dashed lines indicate the defect core radius (see Fig.\,S6(a,b)).
       The dotted lines represent the results for $\epsilon=\epsilon_0$ and $a_\mathrm{n}=a^0_\mathrm{n}$ ($\epsilon=0.25, a_\mathrm{n}/\xi_0=0.055\unit{\mu{}m^2/min}, r_S = 1.2\unit{\mu{}m}$; see Appendix B), which correspond to the conventional case of extensile active nematics.
       The solid lines are the results for \eqref{eq:3} with three-dimensional nematic tilting, i.e., $a_\mathrm{n}=a^0_\mathrm{n}\cos\thetan^\pm(\bm{r})$.
       Specifically, we used $\thetan^\pm(r,\phi)=\thetan^\infty + (\thetan^0-\thetan^\infty)\exp(-r^2/r_\theta^2)$ with $\thetan^\infty=0.3$ and $\thetan^0=0.75$, with the other parameters left unchanged (see Appendix B).
       The inset is a close-up of the results for $-1/2$ defects.
       }
\end{figure*}

\subsection{Two-dimensional velocity fields around topological defects}
To clarify the origin of the promoted three-dimensional growth, we investigate how cells in the bottom layer were displaced near topological defects.
We conducted a time-lapse phase-contrast observation of the bottom layer of cells, cultured from densely and uniformly distributed populations as in the confocal observation (see methods in Appendix\,A).
Cells then filled the two-dimensional plane rather homogeneously, without forming visible microcolonies, and after a short while, cells started to tilt upward, almost simultaneously (\figref{fig-2}(a) and Video~1; the appearance of a dark spot in the cell body indicates the tilting of the cell, as sketched in \figref{fig-2}(a)).
Based on the uniformity of this initial two-dimensional growth, as compared to the growth of a single circular colony discussed later, we shall refer to the present case as ``uniform colony'' in the following.
Using the images after the two-dimensional plane was filled, we detected topological defects from the two-dimensional cell orientation $\bm{n}(\bm{R},t)$ of the bottom layer, where $t \geq 0$ is the time elapsed since the bottom plane was filled.
The density of defects initially increased slightly, then stayed approximately constant from $t \approx 30\unit{min}$ [Fig.\,S3(a)].
As expected from the absence of cell motility, the defects hardly moved in our system, typical displacements being only a few microns over the observation time [Fig.\,S3(b)].

We then measured the velocity field around defects by particle image velocimetry (PIV) (see methods in Appendix\,A).
In \figref{fig-2}(b)(c), the arrows show the velocity field $\bm{v}^\pm(\bm{r},t)$ around $\pm 1/2$ defects, time-averaged over $30\unit{min} \leq t \leq 105\unit{min}$, where $\bm{r}$ indicates the position relative to the defect and the double sign corresponds to the sign of the defect (see also Fig.\,S4(a,b)).
While the structure of $\bm{v}^\pm(\bm{r},t)$ resembles those around defects in typical extensile active nematic systems \cite{Doostmohammadi2018}, their divergence $\nabla \cdot \bm{v}^\pm(\bm{r},t)$ (\figref{fig-2}(b)(c); see also Fig.\,S4(e)) reveals a distinguished character of our system: we found negative divergence around both types of defects, not only around $+1/2$ defects (\figref{fig-2}(b)) as previously reported for systems of motile cell populations \cite{Peng2016,Genkin2017,Saw2017,Kawaguchi2017,Turiv2020,Copenhagen2021},
but even around $-1/2$ defects (\figref{fig-2}(c)), as opposed to those earlier studies.
Since negative divergence indicates influx of cells, this implies that cells are moving toward both types of defects in the bottom layer and pushed out upward.
This is consistent with the result of the confocal observation that the colony height was higher at the positions of the $\pm1/2$ defects.
To inspect the time evolution of this influx, we examined the mean radial velocity at a distance $r$ from $+1/2$ or $-1/2$ defect, $\bar{v}_r^\pm(r,t) \equiv \frac{1}{2\pi}\oint d\phi v_r^\pm(\bm{r},t)$, where $v_r^\pm(\bm{r},t)$ is the radial component of the velocity $\bm{v}^\pm(\bm{r},t)$ at polar coordinates $\bm{r}=(r,\phi)$ centered at the defect (\figref{fig-2}(d)).
For the $+1/2$ defects (\figref{fig-2}(e)), we find that $\bar{v}^+_r(r)$ is essentially negative all the time, but the depth of the minimum decreased with increasing time. 
This may be because of decay of the overall flow speed throughout the colony (Fig.\,S4(f)), possibly due to nutrient starvation, pressure increase and/or quorum sensing.
In contrast, for the $-1/2$ defects (\figref{fig-2}(f)), $\bar{v}^-_r(r)$ was initially positive for all $r$, but decrease near the defect and eventually become negative.
To see the time-dependent influx toward the defects more clearly, we plotted $\min_{r<r_0}\bar{v}_r^\pm(r)$ with $r_0 = 2\unit{\mu{}m}$ in \figref{fig-2}(g).
While the strength of the influx toward the $+1/2$ defect monotonically decreased, that toward the $-1/2$ defect increased until $t\simeq 25\unit{min}$.
These suggest an intrinsic change in the dynamics around the $-1/2$ defect that cannot be explained by the decay of the overall flow speed.
The reproducibility was confirmed by an independent biological replicate (Fig.\,S5).

\begin{figure*}[t]
       \includegraphics[width=\hsize]{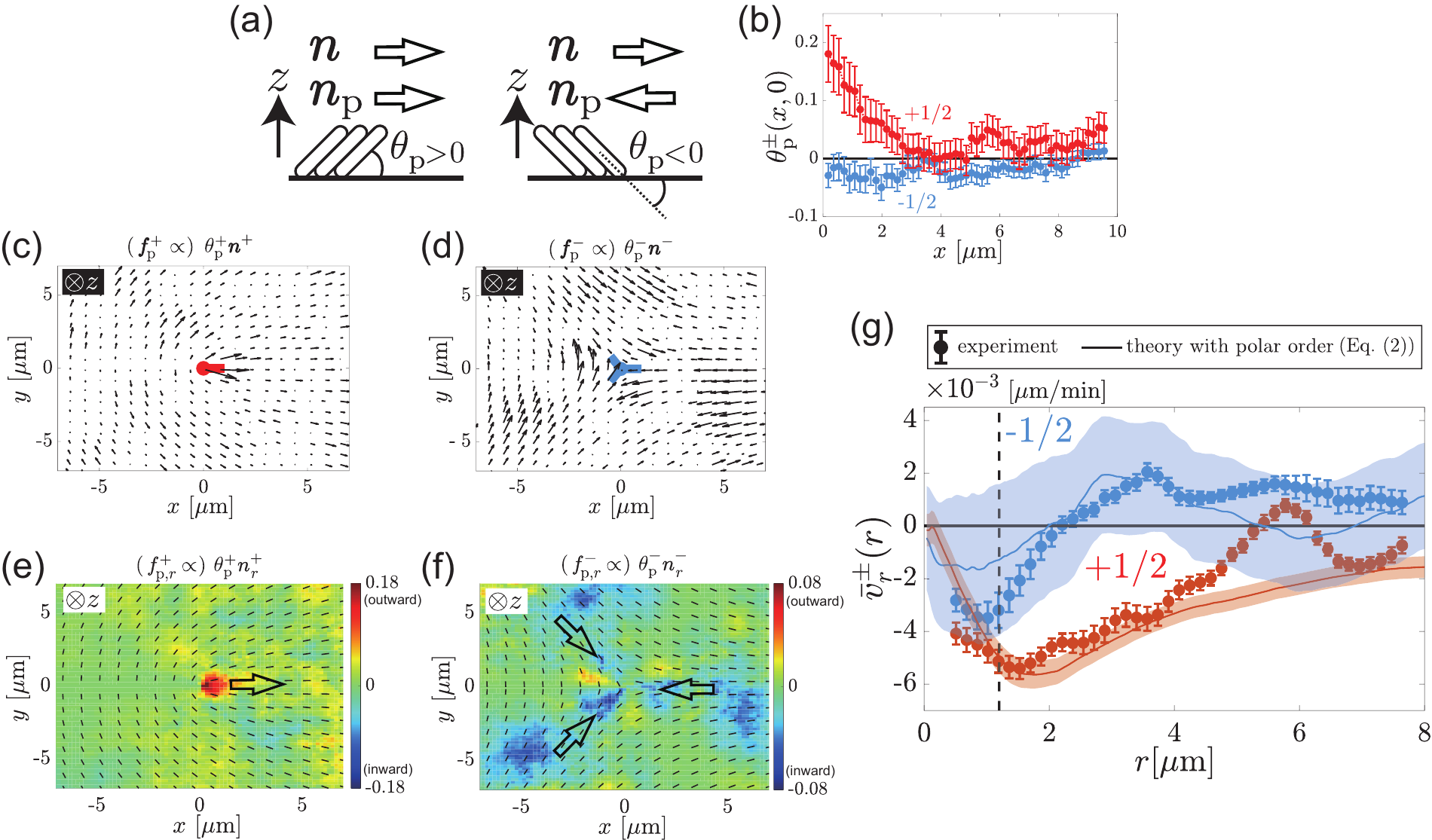}
       \caption{
       \label{fig-4}
       Results on the polar order obtained by the single end-point confocal measurement (uniform colony, end-point confocal \#{}1) (see also Fig.\,S7(c)-(g) for the results of another biological replicate (uniform colony, end-point confocal \#{}2)) and theoretical calculations of the mean radial velocity based on our theory with polar order.
       (a) Illustrations of the polar tilt angle $\thetap$ and the polar director $\bm{n}_\mathrm{p}=\bm{n}\thetap/|\thetap|$.
       By choosing the direction of the head of the nematic director $\bm{n}$, we can uniquely determine the sign of $\thetap$.
       The polar tilt angle $\thetap$ takes a positive (negative) value if the cell end at the head (tail) of the director $\bm{n}$ is lifted above the substrate.
       Note that $\thetap$ changes its sign if $\bm{n}$ is reversed, but $\bm{n}_\mathrm{p}$ remains unchanged, always pointing the direction of the upper end of the cell.
       The polarity-induced force $\bm{f}_\mathrm{p}\propto \thetap\bm{n}$ is oriented toward $\bm{n}_\mathrm{p}$ (see methods in Appendix\,A).
       In the following, we use $\bm{n}^\pm = (\cos(\pm\phi/2), \sin(\pm\phi/2))$ for the director field around $\pm 1/2$ defects, with azimuth $\phi$ of the coordinate $\bm{r}$.
       (b) The polar tilt angle $\thetap^\pm$ measured on the $+x$-axis of $\pm 1/2$ defects (see \figref{fig-3}(a) for the definition of the axis).
       The ensemble average over all defects is shown.
       The error bars indicate the standard error from the ensemble averaging.
       (c,d) $\thetap^\pm \bm{n}^\pm$, representing the strength and direction of the polarity-induced force $\bm{f}_\mathrm{p}^\pm$ around the $+1/2$ defect (c) and the $-1/2$ defect (d).
       (e,f) $\thetap^\pm n_r^\pm$, which is proportional to the radial component of the polarity-induced force, $f_{\mathrm{p},r}^\pm$, around the $+1/2$ defect (e) and the $-1/2$ defect (f).
       The negative radial component indicates that the polarity-induced force is directed toward the defect.
       The black rods represent the nematic director field.
       The outlined arrows illustrate the direction of the radial component of the polarity-induced force near the defect.
       (g) Comparison of the mean radial velocity between the experimental data (symbols) and our theoretical curves with nematic tilting and polar order (lines) (\eqref{eq:4}).
       The shaded bands indicate the range of uncertainty, evaluated from the standard error of the experimental data of $\thetap^\pm(\bm{r},t)$ which were directly used in the theoretical evaluation.
       The experimental data (symbols) are from the phase-contrast observation (those shown in \figref{fig-2}(e)(f)).
       The parameter values were $\epsilon_0=0.25, a^0_\mathrm{n}/\xi_0=0.055\unit{\mu{}m^2/min}$, $a^0_\mathrm{p}/\xi_0=0.8\unit{\mu{}m/min}$, $\thetan^\infty=0.2$, and $\thetan^0=0.25$ (see Appendix B).
       The vertical dashed lines indicate the defect core radius.
       }
\end{figure*}

\subsection{Theoretical analyses and relevance of three-dimensional tilting of cells} \label{sec:res3}
To seek for a possible mechanism of the influx toward $-1/2$ defects, we developed a theory based on two-dimensional extensile active nematics, extended to incorporate characteristics of growing non-motile colonies we observed.
Following earlier studies \cite{Kawaguchi2017, Copenhagen2021, You2018, Dell2019}, we describe the cell alignment by the nematic order tensor $\bm{Q}(\bm{r},t) \equiv S(2\bm{n} \otimes \bm{n} - \bm{1})$, with the scalar nematic order parameter $S(\bm{r},t)$, the director field $\bm{n}(\bm{r},t)$, and the identity matrix $\bm{1}$ (see Appendix B).
As a result of cell growth along the long axis of the cell body, interacting with nearby cells, cells exert the extensile active stress $\bm{\sigma}=-a_\mathrm{n}\bm{Q}$ with the active stress coefficient $a_\mathrm{n}(>0)$ even without the motility \cite{You2018, Dell2019}.
This stress induces the force $\bm{f}=\nabla\cdot\bm{\sigma}$ and drives the velocity field $\bm{v}(\bm{r},t)$.
In the overdamped and low Reynolds number limit, this active force is balanced by the friction originating from cell-substrate interaction, giving the following linearized equation 
\begin{equation}
       \bm{\xi v} = \nabla\cdot(-a_\mathrm{n}\bm{Q}),  \label{eq:3}
\end{equation}
with the friction tensor $\bm{\xi}$.
We assume that the friction is anisotropic with respect to the cell alignment: 
$\bm{\xi}=\xi_0(\bm{1}-\epsilon \bm{Q})$ with the friction anisotropy parameter $\epsilon$.
As suggested in Ref.~\cite{Doumic2020}, we may reasonably assume that it is easier for \textit{E.~coli} cells to slide along their longitudinal axis, hence $\epsilon > 0$.
Setting $\bm{Q}$ with the theoretical director configuration for $\pm 1/2$ defects, $\bm{n}^\pm(\bm{r}) = (\cos(\pm\phi/2), \sin(\pm\phi/2))$ with azimuth $\phi$ of the coordinate $\bm{r}$, and using the experimentally determined core radius (see Appendix B and Fig.\,S6(a)(b)), we calculated the mean radial velocity $\bar{v}_r^\pm(r)$ (\figref{fig-2}(h) dotted lines).
This shows influx only for $+1/2$ defects and outflux only for $-1/2$ defects.
Therefore, to explain the experimentally observed influx toward $-1/2$ defects, we need to extend the existing theoretical framework described so far.
One may consider that this influx might be due to the density heterogeneity, in particular small voids observed at $-1/2$ defects in the early stage of the process (Video~1).
However, this is unlikely to explain the observed influx, because more voids existed at earlier times whereas the influx developed later (compare Video~1 and \figref{fig-2}(g)). 
We also examined the possibility that the cell growth may generate an influx toward $-1/2$ defects, by adding a growth term to the hydrodynamic equation, but this did not yield the influx (see Supplementary Information).

Instead of the growth and the density heterogeneity, here we focus on the three-dimensional orientations of the cells, because the influx toward $-1/2$ defects became strong when cells began to tilt three-dimensionally (\figref{fig-2}(f,g) and Videos~1 and 2) despite the decay of the overall flow speed.
We experimentally measured the tilt angle $\thetan$ of cells from the horizontal plane (see the illustration in \figref{fig-3}(a)), at a late time $t$ from the end-point confocal data, by the structure tensor method for the three-dimensional space, applied to the bottom layer (see methods in Appendix\,A). 
Taking average over the regions around $\pm 1/2$ defects, we obtained a field of the tilt angle, $\thetan^\pm(\bm{r})$ (\figref{fig-3}).
While almost all cells were already tilted (hence $\thetan^\pm(\bm{r}) > 0$ everywhere) at the moment of the end-point observation, we found that three-dimensional tilting was strongest at the core of both defects (\figref{fig-3}; see also Fig.\,S7(a)(b) for the results of another biological replicate).
The peak of $\thetan^\pm(\bm{r})$ is well approximated by a Gaussian function centered at the defect core plus a constant (\figref{fig-3}(a)), and $\thetan^\pm(\bm{r})$ turned out to be essentially isotropic (\figref{fig-3}(b)(c)).

We considered that this tilting may have weakened, in our two-dimensional description, the local active stress and the friction anisotropy around the defects.
More quantitatively, we assume that the local active stress coefficient and the friction anisotropy are given by $a_\mathrm{n}(\bm{r},t)=a^0_\mathrm{n} \cos\thetan(\bm{r},t)$ and $\epsilon(\bm{r},t)=\epsilon_0 \cos\thetan(\bm{r},t)$, respectively, with constants $a^0_\mathrm{n}$ and $\epsilon_0$.
Using this, we solved \eqref{eq:3} and found that the influx toward $-1/2$ defects can emerge (\figref{fig-2}(h) blue solid line and inset; see also Fig.\,S6(c)) within the reasonable range of parameter values.
However, the strength of the influx was too small to account for the experimental result (\figref{fig-2}(h) blue symbols, to be compared with the blue solid line).
This led us to seek for another key factor for the influx toward $-1/2$ defects.

Here, we propose a key mechanism for the strong influx toward $-1/2$ defects.
So far, we assumed that active force is induced only by nematic alignment.
However, when cells are tilted three-dimensionally, the sign of the tilt angle $\thetap$ may break the nematic symmetry and make it possible to develop a polar order (\figref{fig-4}(a)).
If this happens, the violation of the nematic symmetry may result in the generation of an additional force term that is otherwise forbidden, which needs to be included in the force balance equation \pref{eq:3}.
Such a polarity-induced force is expected to be proportional to the strength of the polar order, i.e., $\thetap$, in its lowest order, and act in the direction of the director.
In this context, it is interesting to refer to past experiments on densely packed vibrated granular rods \cite{Blair2003, Volfson2004}, which indeed showed the formation of the polar order due to rod tilting and the resulting horizontal transport of the rods, driven by the polarity-induced force.
This suggests that a similar polarity-induced force may arise in our growing bacterial populations, resulting from the extensile active force of cells, if the polar order is formed.
Inspired by this possibility, we measured $\thetap(\bm{r})$ around both types of defects by end-point confocal microscopy.
Note that the single-cell tilt angles fluctuate largely from cell to cell (see Videos~1 and 2), and this is why $\thetap$ and $\thetan$, i.e., the signed and unsigned averages of the tilt angles, respectively, differ.
The sign of $\thetap$ is determined by choosing the direction of the head of the nematic director $\bm{n}$ (see \figref{fig-4}(a) and Appendix\,A): here we set $\bm{n}^\pm(\bm{r}) = (\cos(\pm\phi/2), \sin(\pm\phi/2))$ for the director field around $\pm 1/2$ defects.
Figure~\ref{fig-4}(b) displays the result on the $+x$-axis.
This shows non-vanishing $\thetap(\bm{r})$ for both $\pm 1/2$ defects, specifically $\thetap(\bm{r})>0$ (upper end oriented outward) for $+1/2$ defects and $\thetap(\bm{r})<0$ (upper end oriented inward) for $-1/2$ defects on the $+x$-axis, demonstrating the emergence of the polar order in our growing bacterial populations.
Consequently, the above-mentioned symmetry argument predicts the polarity-induced force $\bm{f}_\mathrm{p}$ to arise, which satisfies $\bm{f}_\mathrm{p}\propto \thetap\bm{n}$ for small $\thetap$.
In \figref{fig-4}(c)(d), we show $\thetap^\pm(\bm{r}) \bm{n}^\pm(\bm{r})$ around $\pm 1/2$ defects, which represent the strength and the direction of the polarity-induced force $\bm{f}_\mathrm{p}(\bm{r})$.
What contributes to the mean radial velocity is its radial component $f^\pm_{\mathrm{p},r}(\bm{r})$, proportional to $\thetap^\pm(\bm{r}) n_r^\pm(\bm{r})$ which is shown in \figref{fig-4}(e)(f), with $n_r^\pm(\bm{r})$ being the radial component of the director $\bm{n}^\pm(\bm{r})$.
These results show that, while the polarity-induced force around $+1/2$ defects drives the defects toward their comet tail, that around $-1/2$ defects acts inward, leading to the influx toward the defects.
We confirmed the reproducibility of the main structure of the polarity-induced force by taking a biological replicate (Fig.\,S7), while other features such as the apparent chirality in \figref{fig-4}(d) were not.

To quantitatively deal with the effect of the polar order upon the mean radial velocity, we incorporate the polarity-induced force $\bm{f}_p$ into \eqref{eq:3}.
With $a_\mathrm{n}=a^0_\mathrm{n} \cos\thetan$ and $\bm{\xi}=\xi_0(\bm{1}-\epsilon_0 \cos\thetan \bm{Q})$, we obtain the following equation:
\begin{equation}
       \xi_0(\bm{1}-\epsilon_0 \cos\thetan \bm{Q}) \bm{v} = \nabla\cdot(-a^0_\mathrm{n}\cos\thetan\bm{Q}) + a^0_\mathrm{p}\thetap\bm{n}.  \label{eq:4}
\end{equation}
Then we experimentally measured $\thetan(\bm{r},t), \thetap(\bm{r},t)$ and $\bm{Q}(\bm{r},t)$ for $\pm 1/2$ defects by a time-lapse confocal observation, using a time period showing the strongest influx toward $-1/2$ defects (see Appendix B and Fig.\,S8).
We are to determine three unknown parameters, $a^0_\mathrm{n}S_0/\xi_0$, $a^0_\mathrm{p}/\xi_0$ and $\epsilon S_0$, where $S_0$ is the scalar nematic order parameter sufficiently far from defects.
The friction anisotropy $\epsilon$ turned out to hardly affect $\bar{v}_r^\pm(r,t)$, so that we are left with two effective parameters, $a^0_\mathrm{n}S_0/\xi_0$ and $a^0_\mathrm{p}/\xi_0$.
While the nematic contribution solely could not reproduce the experimental result as we described above, we found, remarkably, that the addition of the polar contribution $a^0_\mathrm{p}/\xi_0$ strengthened the influx toward $-1/2$ defects significantly (\figref{fig-4}(g) solid curves).
In particular, we were able to find such values of $a^0_\mathrm{n}S_0/\xi_0$ and $a^0_\mathrm{p}/\xi_0$ that satisfactorily reproduced the experimental data of both $\bar{v}_r^+(r)$ and $\bar{v}_r^-(r)$ simultaneously (see methods in Appendix\,A).
This demonstrates that the three-dimensional tilting and resulting polar order were the keys to understand the unusual influx toward $-1/2$ defects we observed in our growing non-motile bacterial populations.

\section{Relation to circular colonies formed from isolated cells}

Although many earlier studies have already investigated how non-motile bacteria construct three-dimensional structures, most of them have focused on the process where isolated cells grow and form circular colonies \cite{Su2012, Grant2014, Duvernoy2018, You2019, Dhar2021}.
In this situation, it has been reported that the in-plane stress derived from cell growth is maximized at the center of the colony \cite{Volfson2008, Boyer2011, You2019, Echten2020}, which causes a few cells to be verticalized first, locally, near the center \cite{Su2012, Grant2014, You2019}.
This is contrasted to the case of our experiments starting from densely and uniformly distributed cells, in which cells were verticalized almost homogeneously and simultaneously (Video~1).
We checked if cell alignment plays any role in such circular colonies (Video\,3), but detected no significant correlation between the position of the first verticalization and the strength of the local orientational order (Fig.\,S9; see Supplementary Information Sec.\,IV for details).
Instead, we confirmed that shorter cells tend to be verticalized first  (Fig.\,S9(e)), in agreement with the recent theory based on the torque balance \cite{You2019}.
These suggest that, in such isolated circular colonies, the spatially non-uniform stress indeed constitutes a major contribution to the start of the three-dimensional transition, as reported earlier \cite{Su2012,Grant2014,You2019}, regardless of topological defects.
Conversely, by using uniform colonies, we reduced the effect of non-uniform stress and thereby revealed the intriguing role of topological defects in the three-dimensional transition.

Note also that a previous study \cite{Grant2014} on circular colonies reported that collisions between colonies also triggered the cell extrusion in that case.
In the case of uniform colonies from numerous cells we studied, groups of cells merged and filled holes to complete the formation of the two-dimensional bottom layer (Video~1 and Fig.\,S10(a)).
We tested the possible influence of such collisions upon our defect analyses, by examining whether hole filling events affected the defect formation. 
Detecting the locations of the holes at $t = -5\unit{min}$ (Fig.\,S10(a)) and those of the defects at $t = 0$ (Fig.\,S10(b)), we confirmed that hole filling events did not promote the formation of defects (Fig.\,S10(c)(d)). 
Therefore, we conclude that our results on the relevance of topological defects to the cell flow and the three-dimensional growth are not significantly affected by cell collisions that preceded the formation of the complete bottom layer.

\section{Concluding remarks}
In summary, we showed the relevance of topological defects to the three-dimensional growth of growing non-motile \textit{E.~coli} populations, unveiling the emergence of polar order and resulting novel properties endowed with this active nematic system.
When cultured from densely and uniformly distributed populations, cells started to construct the three-dimensional structure a short while after they filled the bottom plane.
Since then, the net influx toward both $+1/2$ and $-1/2$ defects appeared, which may have promoted the vertical growth of colonies.
The influx toward $-1/2$ defects, which grew stronger with time despite the decay of the overall flow speed, was unexpected also from the existing theory of active nematics, but we revealed that this resulted from the three-dimensional tilting of cells around defects and the polar order induced thereby.
We extended the active nematics theory to incorporate these effects and successfully accounted for the experimental observation.

Our results suggest the role of $-1/2$ defects in the formation of three-dimensional structures of non-motile cell populations, which has been overlooked compared to that of $+1/2$ defects supported by many recent studies on motile cells \cite{Peng2016,Genkin2017,Saw2017,Kawaguchi2017,Turiv2020,Copenhagen2021}.
Although the height increase at the defects was not large in our setup, we consider that further vertical growth may have been prevented by the presence of the agar.
Further investigation is needed to see whether the colony height above the defects can grow further, by alternative methods that can stably measure sessile \textit{E. coli} populations for a longer period of time, and whether the orientation and topological defects in intermediate layers may also affect the colony height.
Besides, it is important to contemplate the possibility of physiological significance that topological defects may ultimately have.
In \textit{Bacillus subtilis} colonies, it has been found that the roughness of the colony surface can change the wettability of the biofilm, making it more resistant to droplets that may contain toxic substances \cite{Epstein2011,Trejo2011,Werb2017,HAYTA2021,ZABIEGAJ2021}.
The local vertical growth mediated by topological defects might be involved in the formation of such surface morphology.

Finally, the emerging polar order and the influx toward $-1/2$ defects reported in this work may provide a novel characterization of non-motile but growing active matter, contrasted with the standard active matter for self-propelled particles.
As such, these results may also shed a new light on other cellular systems with three-dimensional structures.
In this context, it is of great importance to elucidate how the polar order is formed when cells start to tilt.
Our observations show that the direction of the polar order (\figref{fig-4}(c,d)) and that of the velocity field (\figref{fig-2}(b,c) arrows, typical of extensile active nematics \cite{Doostmohammadi2018}) tend to be oriented oppositely.
This suggests that the polar order may be driven by the active stress originating from the nematic orientation.
The recently reported instability of the in-plane orientation in extensile active nematics \cite{Nejad.Yeomans-PRL2022} may also be a hint.
It is also important to understand how the absence of motility is involved in this mechanism; qualitatively, we may argue that the lack of cell motility would help maintain the cell tilting.
Further elucidation of the mechanism of the polar order formation and quantitative prediction of the resulting polar angle as well as the polarity-induced force are key tasks left for future studies, which will also clarify the relevance of our findings to other cellular populations.

\appendix
\section{Experimental Methods}
\subsection{Strains, culture media and sample setup}
We used a wild-type {\sl E. coli} strain MG1655 and its mutant MG1655-pZA3R-EYFP that contains a plasmid pZA3R-EYFP expressing enhanced yellow fluorescent proteins.
We used LB broth (tryptone 1 wt\%, sodium chloride 1 wt\% and Yeast extract 0.5 wt\%) and TB+Cm medium (tryptone 1 wt\% , sodium chloride 1 wt\% and chloramphenicol $165\unit{\mu{}g/ml}$).
To prepare nutrient agar pads, we added agar powder to medium, solidified it by a microwave oven, then cut it into squares of size $13\unit{mm} \times 13\unit{mm}$.
For each observation, we inoculated bacterial suspension on a coverslip and put an agar pad on the suspension.
We then attached the following on the coverslip, surrounding the agar pad, to prevent the agar from drying out (\figref{fig-1}(a)): a frame seal (SLF0601, Bio-Rad), a 3D printed PLA spacer ($5\unit{mm}$ height, hollow square, inner dimensions $14\unit{mm} \times 14\unit{mm}$ and outer dimensions $22\unit{mm} \times 22\unit{mm}$), another frame seal, then a plastic cover that enclosed the inner region.
Details on the strain and the culture condition in each experiment are provided below and in \tbref{tb-1}.
The {\sl E. coli} strains we used did not swim at all in our experimental conditions (Videos~1, 2 and 3).

\begin{table*}[t]\centering
       \caption{
       \label{tb-1}
       List of experimental measurements in this study.}
       \begin{tabular}{l|l|c|l}
          Measurement & Strain & Initial cell density & Data \\
          \hline
          uniform colony, end-point confocal \#{}1 & a mutant MG1655-pZA3R-EYFP & high & \figsref{fig-1}, \ref{fig-3}, \ref{fig-4}, S1 and S2 \tabularnewline
          uniform colony, end-point confocal \#{}2 & a mutant MG1655-pZA3R-EYFP & high & Fig.\,S7 \tabularnewline
          uniform colony, phase contrast \#{}1 & a wile-type MG1655 & high & \figsref{fig-2}, S3 and S10 \tabularnewline
          uniform colony, phase contrast \#{}2 & a wile-type MG1655 & high & Fig.\,S5 \tabularnewline
          uniform colony, time-lapse confocal \#{}1 & a mutant MG1655-pZA3R-EYFP & high & Fig.\,S8 \tabularnewline
          circular colony, phase contrast \#{}1 & a wile-type MG1655 & low & Fig.\,S9 \tabularnewline
          circular colony, phase contrast \#{}2 & a wile-type MG1655 & low & Fig.\,S9 \tabularnewline
          \hline
       \end{tabular}
\end{table*}

\subsection{Confocal observations of uniform colonies formed from numerous cells}
We used the mutant strain MG1655-pZA3R-EYFP that expresses enhanced yellow fluorescent proteins.
Before the observations, we inoculated the strain from a glycerol stock into $2\unit{ml}$ TB+Cm medium in a test tube.
After shaking it overnight at $37\degc$, we transferred $20\unit{\mu{}l}$ of the incubated suspension to $2\unit{ml}$ fresh TB+Cm medium and cultured it until OD at $600\unit{nm}$ wavelength reached $0.1$-$0.5$.
The bacterial suspension was finally concentrated to $\mathrm{OD}=5$ by a centrifuge, and $1\unit{\mu{}l}$ of the suspension was inoculated between the coverslip and the agar pad (1.5 wt\% agar).

The sample was placed on the microscope stage, in a stage-top incubator maintained at $37\degc$.
The microscope we used was Leica SP8, equipped with a 63x (N.A. 1.40) oil immersion objective and operated by Leica LasX.
The data shown in \figsref{fig-1}, \ref{fig-3}, \ref{fig-4}, S1 and S2 were obtained by a single end-point observation, in which we cultured the colonies without excitation light until 14 hours after the cells had filled the observation area.
We also show data obtained by another biological replicate with a different substrate and agar pad in Fig.\,S7.
For each set of these data, we captured three-dimensional images of size $184.52\unit{\mu{}m} \times 184.52\unit{\mu{}m} \times 16\unit{\mu{}m}$ from 20 separate regions.
The optical resolution, as evaluated by the formula of the point-spread function, was about $140\unit{nm}$ in the horizontal plane and $250\unit{nm}$ in the vertical direction.
The confocal pinhole size was $0.21$ Airy unit.
For the data shown in Fig.\,S8, we carried out a single time-lapse observation and obtained images of size $184.52\unit{\mu{}m} \times 184.52\unit{\mu{}m} \times 6.4\unit{\mu{}m}$ from 4 separate regions with the time interval $15\unit{min}$.
The image pixel size was $\approx 0.18\unit{\mu{}m}$ in the $xy$ plane and $0.16\unit{\mu{}m}$ along the $z$-axis.

\subsection{Analysis of confocal images}
For each region, we chose the plane corresponding to the bottom layer and measured the two-dimensional cell orientation $\bm{n}(\bm{R})$ by the structure tensor method.
The image pixel size was $\approx 0.18\unit{\mu{}m}$.
After sharpening the images by a high-pass filter, we calculated the structure tensor $J(\bm{R})$ at a given pixel $\bm{R}=(X,Y)$ by
\begin{equation}
J(\bm{R}) = 
       \begin{pmatrix}
         [\Delta_X I,\Delta_X I]_{\bm{R}}, & [\Delta_Y I,\Delta_X I]_{\bm{R}}  \\ \relax
         [\Delta_X I,\Delta_Y I]_{\bm{R}}, & [\Delta_Y I,\Delta_Y I]_{\bm{R}} 
       \end{pmatrix}
     ,  \label{eq:1}
\end{equation}
with the image intensity $I(X,Y)$, $\Delta_X I \equiv I(X+1,Y)-I(X-1,Y)$, $\Delta_Y I \equiv I(X,Y+1)-I(X,Y-1)$, and $[g,h]_{\bm{R}}\equiv\sum_{(X',Y')\in\mathrm{ROI}^\ell_{\bm{R}}}\ g(X',Y')h(X',Y')f_{\bm{R}}^\sigma(X',Y')$. 
Here, the summation is taken over a region of interest ROI${}^\ell_{\bm{R}}$, which is a square of size $\ell \approx 7.2\unit{\mu{}m}$ (40 pixels) centered at $\bm{R}$, and $f_{\bm{R}}^\sigma(X',Y')$ is the Gaussian kernel defined by $f_{\bm{R}}^\sigma(X',Y')\equiv\exp[-\frac{(X'-X)^2+(Y'-Y)^2}{2\sigma^2}]$ with $\sigma \approx 1.8\unit{\mu{}m}$ (10 pixels).
Then the cell orientation $\bm{n}(\bm{R})$ is given by the eigenvector of $J(\bm{R})$ associated with the smallest eigenvalue $\lambda^\mathrm{min}(\bm{R})$.
The orientation $\bm{n}(\bm{R})$ can also be represented by angle $\psi(\bm{R})$ such that $\bm{n}=\pm(\cos\psi, \sin\psi)$ with $-\pi/2\le \psi < \pi/2$.

To detect topological defects, we first calculated the nematic order parameter by
\begin{equation}
       S(\bm{R})=\expct{\sin2\psi}^2_{\mathrm{ROI}^\ell_{\bm{R}}} + \expct{\cos2\psi}^2_{\mathrm{ROI}^\ell_{\bm{R}}},  \label{eq:9}
\end{equation}
where $\expct{\cdot}_{\mathrm{ROI}^\ell_{\bm{R}}}$ denotes the spatial average within ROI${}^\ell_{\bm{R}}$.
Then we located the positions of local minima of $S(\bm{R})$ as candidates of topological defect cores.
For each candidate point, we calculated the topological charge $q=\frac{1}{2\pi}\oint_\mathcal{C} d\psi$, where $\mathcal{C}$ is a square closed path with a side of about $3.6\unit{\mu{}m}$ (20 pixels) centered at the candidate point.
The candidate point is regarded as a topological defect if $q=\pm 1/2$, and dismissed otherwise.
To determine the angle of the arm of each defect (\figref{fig-1}(d) inset), we used the profile of $|\psi-\phi|$ on $\mathcal{C}$, where $\phi$ is the azimuth with respect to the defect core.
A single minimum of $|\psi-\phi|$ exists for each $+1/2$ defect, while there are three local minima for each $-1/2$ defect.
Each minimum point corresponds to an arm of the defect.
Blue trefoils indicating $-1/2$ defects in \figref{fig-1}(c) were drawn by setting one of the arms of the trefoil at the angle of the global minimum, with the other two arms added by rotating the first arm by $120^\circ$.
We thereby obtained the two-dimensional locations of all defects and their signs.

To investigate the dependence of the colony height on topological defects, we picked up hundreds of isolated defects, separated by a distance longer than $9\unit{\mu{}m}$ from the nearest defect.
For comparison, we also randomly selected 1000 points which are separated more than 9$\unit{\mu{}m}$ from defects.
For a given position in the $xy$-plane, we obtained the image intensity profile along the $z$-axis, with the interval of $z$-slices being $0.16\unit{\mu{}m}$.
The height was then determined by the length of the region whose intensity was higher than 20\% of the maximum intensity in this profile.

The three-dimensional tilting of the cells around defects was characterized as follows. 
First, for each defect, we rotated the confocal image horizontally so that the defect arm was orientated in the positive direction of the $x$-axis.
For $-1/2$ defects, we did this rotation for each of their three arms and obtained a set of three images from each defect.
Then, for each rotated confocal image $I(\bm{r})$, where $\bm{r}$ is the coordinate relative to the defect, we obtained the three-dimensional cell orientation $\bm{n}_3(\bm{r})$ by the three-dimensional version of the structure tensor method.
For each pixel $\bm{r}=(x,y,z)$, which was chosen from the plane corresponding to the bottom layer in each region, we calculated the three-dimensional structure tensor:
\begin{equation}
       J(\bm{r}) = \begin{pmatrix}
                [\Delta_x I,\Delta_x I]_{\bm{r}} & [\Delta_y I,\Delta_y I]_{\bm{r}} & [\Delta_z I,\Delta_x I]_{\bm{r}} \\ \relax
                [\Delta_x I,\Delta_y I]_{\bm{r}} & [\Delta_y I,\Delta_y I]_{\bm{r}} & [\Delta_z I,\Delta_y I]_{\bm{r}} \\ \relax
                [\Delta_x I,\Delta_z I]_{\bm{r}} & [\Delta_y I,\Delta_z I]_{\bm{r}} & [\Delta_z I,\Delta_z I]_{\bm{r}} 
              \end{pmatrix},  \label{eq:5}
\end{equation}
where $\Delta_x I \equiv I(x+1,y,z)-I(x-1,y,z)$, $\Delta_y I$ and $\Delta_z I$ are defined likewise, 
$[g,h]_{\bm{r}}\equiv\sum_{\bm{r}'\in\mathrm{ROI}^{\ell_x,\ell_y,\ell_z}_{\bm{r}}}\ g(\bm{r}')h(\bm{r}')f_{\bm{r}}^\sigma(\bm{r}')$.
Here, the summation is taken over a three-dimensional region of interest ROI${}^{\ell_x,\ell_y,\ell_z}_{\bm{r}}$, which is a cuboid of size $\ell_x \times \ell_y \times \ell_z$ centered at $\bm{r}$, with $\ell_x = \ell_y \approx 4.3\unit{\mu{}m}$ (24 pixels) and $\ell_z \approx 3.8\unit{\mu{}m}$ (24 pixels). The Gaussian kernel $f_{\bm{r}}^\sigma(\bm{r}')$ is defined by $f_{\bm{r}}^\sigma(\bm{r}')\equiv\exp[-\frac{(\bm{r}'-\bm{r})^2}{2\sigma^2}]$ with $\sigma = 2.2\unit{\mu{}m}$.
Then the three-dimensional cell orientation $\bm{n}_3(\bm{r})$ is given by the eigenvector of $J(\bm{r})$ associated with the smallest eigenvalue.
The orientation $\bm{n}_3(\bm{r})$ is then represented by angles $\psi(\bm{r})$ and $\theta(\bm{r})$ such that $\bm{n}_3 = (\cos\theta\cos\psi, \cos\theta\sin\psi, \sin\theta)$ with $0 \leq \psi < 2\pi$ and $-\pi/2 \leq \theta < \pi/2$.
As is clear from the definition, the angle $\psi(\bm{r})$ specifies the two-dimensional cell orientation $\bm{n}(\bm{r})$ by $\bm{n} = (\cos\psi, \sin\psi)$ and $\theta(\bm{r})$ indicates the angle between the three-dimensional orientation and the $xy$-plane.
Note that $\bm{n}_3(\bm{r})$ and $-\bm{n}_3(\bm{r})$ are equivalent, so that the sign of $\bm{n}(\bm{r})$ and $\theta(\bm{r})$ can be changed simultaneously.

To investigate statistical properties of the cell tilt angle around $\pm 1/2$ topological defects, we need to define tilt angles whose sign can be determined unambiguously.
The simplest choice is to take the ensemble average of $|\theta(\bm{r})|$, which can be used to detect the presence of the three-dimensional tilting.
We took this average over isolated defects of each sign, separated by a distance longer than $9\unit{\mu{}m}$ from the nearest defect, and this defines our $\thetan^\pm(\bm{r})$.
To characterize the polar order, we need an angle that can take both positive and negative values.
Here we chose such a sign that the tilt angle is positive if the cell end farther from the defect is lifted above the substrate.
More specifically, we use the director field $\bm{n}_\mathrm{ref}^\pm(\bm{r}) \equiv (\cos(\pm\phi/2), \sin(\pm\phi/2))$ around $\pm 1/2$ defects, with the azimuth $\phi$ of the position $\bm{r}$ in the $xy$-plane, and took the average of the field $\theta(\bm{r}) \sign[\bm{n}_\mathrm{ref}^\pm(\bm{r}) \cdot \bm{n}(\bm{r})]$ over isolated defects (with the same criterion on the distance from other defects).
This is our $\thetap^\pm(\bm{r})$ which characterized the polar order.
The polarity-induced force is then 
$\bm{f}_\mathrm{p}^\pm(\bm{r}) \propto \thetap^\pm(\bm{r})\bm{n}_\mathrm{ref}^\pm(\bm{r})$.
This right-hand side is shown in \figref{fig-4}(c)(d), and its radial component in \figref{fig-4}(e)(f).

\subsection{Phase-contrast observation of uniform colonies formed from numerous cells}
We used the wild-type strain MG1655.
Before the time-lapse observation, we inoculated the strain from a glycerol stock into $2\unit{ml}$ LB broth in a test tube.
After shaking it overnight at $37\degc$, we transferred $20\unit{\mu{}l}$ of the incubated suspension to $2\unit{ml}$ fresh LB broth and cultured it until the optical density (OD) at $600\unit{nm}$ wavelength reached $0.1$-$0.3$.
The bacterial suspension was finally concentrated to $\mathrm{OD}=5$ by a centrifuge, and $1\unit{\mu{}l}$ of the suspension was inoculated between the coverslip and the LB agar pad (1.5 wt\% agar).

The sample was placed on the microscope stage, in an incubation box maintained at $37\degc$.
The microscope we used was Leica DMi8, equipped with a 63x (N.A. 1.30) oil immersion objective and a CCD camera (Leica DFC3000G), and operated by Leica LasX.
The image pixel size was $\approx 0.17\unit{\mu{}m}$.
For the data shown in \figsref{fig-2}, S3 and S10, we used a single substrate and carried out a time-lapse observation with the time interval $1\unit{min}$ for 30 separate regions of dimensions $110.03\unit{\mu{}m} \times 81.97\unit{\mu{}m}$.
We also obtained a biological replicate using another substrate for the data shown in Fig.\,S5.
For each region, we determined the frame at $t=0$, i.e., the frame in which cells filled the observation area for the first time.
We then measured the cell orientation $\bm{n}(\bm{R})$ and detected topological defects in all frames, by the method described below.
We used isolated topological defects only, each separated by a distance longer than $9.5\unit{\mu{}m}$ from the nearest defect.
As a result, we obtained hundreds of defects for each time.

\subsection{Phase-contrast observation of circular colonies formed from a few cells}
We used the wild-type strain MG1655.
We cultured bacteria in the same way as for the observation of uniform colonies.
The bacterial suspension was finally diluted to $\mathrm{OD}=0.01$, and $1\unit{\mu{}l}$ of the suspension was inoculated between the coverslip and the LB agar pad (2.0 wt\% agar).

The imaging process and the condition during the observation were the same as those for the observation of uniform colonies.
We carried out time-lapse observations with the time interval $1\unit{min}$ for 30 isolated colonies, which started to form from a few cells.
We repeated the experiments twice using different substrates and acquired data from 60 colonies in total.
From each colony, we chose the frame right before the first extrusion of a cell from the bottom layer took place.
We used 60 such images from the 60 colonies for analysis.
For each colony, we binarized the image, and obtained the area $A$ by the total number of pixels, the center position by the center of mass, and the radius $R_\mathrm{max}$ by $\pi R_\mathrm{max}^2 = A$, using the regionprops function of MATLAB.
The first extruded cell was detected manually, by using a black spot that a tilted cell exhibits in the phase-contrast image (see Video~3).
We manually labeled pixels contained in each extruded cell, and obtained the position as well as the mean and the standard deviation of the coherency over the labeled pixels (see the section ``Analysis of phase-contrast images'' for the method to evaluate the coherency).
To obtain the spatial dependence of the coherency shown by boxplots in Fig.\,S9(b), we divided the space into regions bordered by concentric circles, with the radii that increased by $R/R_\mathrm{max}=0.1$.
The length of cells was evaluated manually from the major axis of each cell by using a painting software.

\subsection{Analysis of phase-contrast images}

Using phase-contrast images from uniform and circular colonies, we measured the two-dimensional cell orientation $\bm{n}(\bm{R})$ and detected topological defects, in the same manner as those for confocal observations.
The image pixel size was $\approx 0.17\unit{\mu{}m}$.
The structure tensor was calculated with the ROI size $\ell \approx 6.8\unit{\mu{}m}$ (40 pixels) and the characteristic length of the Gaussian filter, $\sigma \approx 1.7 \unit{\mu{}m}$ (10 pixels).
The detection of topological defects was carried out with the closed path $\mathcal{C}$ with a side of about $3.4\unit{\mu{}m}$ (20 pixels), as in \figref{fig-2}(a) and Video~2.

In addition to the cell orientation $\bm{n}(\bm{R})$, we also obtained the coherency parameter $C(\bm{R})$ defined by
\begin{equation}
       C(\bm{R}) = \frac{\lambda^\mathrm{max}(\bm{R})-\lambda^\mathrm{min}(\bm{R})}{\lambda^\mathrm{max}(\bm{R})+\lambda^\mathrm{min}(\bm{R})},  \label{eq:2}
\end{equation}
with the largest eigenvalue $\lambda^\mathrm{max}(\bm{R})$.
This quantifies the degree of the local nematic order.

For uniform colonies, we also measured the velocity field of the cells around the detected defects, by particle image velocimetry (PIV).
For this, we used MatPIV \cite{MatPIV} (open source PIV toolbox for MATLAB), with the PIV window set to be a square of size $\approx 2.7\unit{\mu{}m}$ (16 pixels).
To take averages over defects, for each defect we rotated the image so that the defect arm was oriented in the positive direction of the $x$-axis.
For $-1/2$ defects, we did this rotation for each of their three arms, and all of the resulting velocity fields were used for the ensemble average.
We thereby obtained the ensemble-averaged velocity field $\bm{v}(\bm{r},t)$, as a function of the coordinate $\bm{r}=(x,y)$ relative to the defect, and time $t$.

The divergence of $\bm{v}(\bm{r})=(u(\bm{r}),v(\bm{r}))$ was calculated as follows (here we omit $t$ from the argument for simplicity).
We first obtained $D(\bm{r}) = \frac{u(x+1,y)-u(x-1,y)}{2\delta} + \frac{v(x,y+1)-v(x,y-1)}{2\delta}$ with the pixel size $\delta \approx 0.17\unit{\mu{}m}$.
We then calculated the divergence field by
\begin{equation}
       (\nabla \cdot \bm{v})(\bm{r}) = \frac{ \sum_{(x',y')\in\mathrm{ROI}^\ell_{\bm{r}}} D(\bm{r}') f_{\bm{r}}^\sigma(x',y')}{ \sum_{(x',y')\in\mathrm{ROI}^\ell_{\bm{r}'}} f_{\bm{r}}^\sigma(x',y')},  \label{eq:8}
\end{equation}
where $\mathrm{ROI}^\ell_{\bm{r}}$ and the Gaussian kernel $f_{\bm{r}}^\sigma(x',y')$ were defined as above, but with $\ell \approx 2.7\unit{\mu{}m}$ (16 pixels) and $\sigma \approx 0.68\unit{\mu{}m}$ (5 pixels).

\section{Theoretical calculations}

To theoretically account for the experimental result of the mean radial velocity $\bar{v}_r^\pm(r)$, in particular the influx toward $-1/2$ defects shown in \figref{fig-2}(e)(f), we solved the force balance equations \pref{eq:3} and \pref{eq:4}.
While detailed descriptions on the solutions are given in Supplementary Material, here we outline the theoretical assumptions and the methods to obtain the theoretical results shown in \figref{fig-4}(g), which satisfactorily reproduced the experimental data when the influx toward $-1/2$ defects was strongest.

First we assume the director field winding uniformly around a $+1/2$ or $-1/2$ defect, $\bm{n}^\pm(r,\phi) = (\cos(\pm\phi/2), \sin(\pm\phi/2))$, where $(r,\phi)$ is the two-dimensional polar coordinate, centered at the defect core.
The nematic order tensor $\bm{Q}^\pm(r,\phi)$ is then given by
\begin{equation}
       \bm{Q}^\pm(r,\phi) = S(r)
              \begin{pmatrix}
                \cos(\pm\phi) & \sin(\pm\phi)  \\
                \sin(\pm\phi) & -\cos(\pm\phi) 
              \end{pmatrix}, \label{eq:6}
\end{equation}
with the scalar nematic order parameter $S(r)$ left as a free parameter.
Based on the assumption that $\bm{Q}^\pm$ minimizes the nematic free energy, $S(r)$ can be theoretically expressed by the following Pad{\'e} approximant \cite{Copenhagen2021,Prismen2006,Prismen1999}:
\begin{equation}
       S(r) = S_0F(r/r_S), \qquad F(x)\approx x\sqrt{\frac{0.34+0.07x^2}{1+0.41x^2+0.07x^4}} \label{eq:7},
\end{equation}
with the defect core radius $r_S$ and $S_0=S(\infty)$.
To determine the value of $r_S$, we fitted \eqref{eq:7} to the experimental data of the coherency $C(\bm{r})$ (Fig.\,S6(a)(b)) and obtained $r_S = 1.2\unit{\mu{}m}$.
Note that, because the angle field $\psi(\bm{r})$ does not contain information of the defect core, the nematic order parameter evaluated by \eqref{eq:9} is not suitable for estimating $r_S$.
Concerning $S_0$, it always appears as a product with either $\epsilon$ or $a_\mathrm{n}$, so that we fix $S_0=1$ without loss of generality.

The case without three-dimensional cell tilting, described by \eqref{eq:3}, was already dealt with by earlier studies \cite{Kawaguchi2017, Copenhagen2021}. 
Since \eqref{eq:3} is linear, we can readily solve it and obtain, for the mean radial velocity,
\begin{equation}
    \bar{v}^\pm_r(r) = -\epsilon\frac{a_\mathrm{n}}{\xi_0} S(r) \frac{S'(r)\pm S(r)/r}{1-\epsilon^2S(r)^2}.
\end{equation}
Then we can show, with \eqref{eq:7}, that it is negative for $+1/2$ defects and positive for $-1/2$ defects, for all $r>0$ (see Supplementary Material).
In \figref{fig-2}(h), by the dotted lines, we showed $\bar{v}_r^\pm(r)$ for $\epsilon=0.25, a_\mathrm{n}/\xi_0=0.055\unit{\mu{}m^2/min}, r_S = 1.2\unit{\mu{}m}$.

In fact, even in the presence of three-dimensional cell tilting and polar order, i.e., in the case of \eqref{eq:4}, it is linear in $\bm{v}$ and the solution for the case of $\pm 1/2$ defects is given by
\begin{multline}
    \bm{v}^\pm(r,\phi) = \xi_0^{-1}(\bm{1}-\epsilon_0 \cos\thetan^\pm(r,\phi)  \bm{Q}^\pm)^{-1} \\ [\nabla\cdot(-a^0_\mathrm{n}\cos\thetan^\pm(r,\phi)\bm{Q}^\pm) + a^0_\mathrm{p}\thetap^\pm(r,\phi)\bm{n}^\pm].  \label{eq:v}
\end{multline}

Regarding the first term that describes the contribution by non-uniform nematic tilting, we determined $\thetan^\pm(r,\phi)$ by time-lapse and end-point confocal observations.
Because we could not obtain clear spatial profile of $\thetan^\pm(r,\phi)$ from the time-lapse observation due to photobleaching, we used high-quality, end-point confocal images to determine the spatial profile, then calibrated its amplitude by the time-lapse observation to account for the time period of interest.
First, on the spatial profile, our end-point confocal observation (\figref{fig-3}) suggests that $\thetan^\pm(r,\phi)=\thetan^\infty + (\thetan^0-\thetan^\infty)\exp(-r^2/r_\theta^2)$ with constants $\thetan^\infty, \thetan^0, r_\theta$, regardless of $\phi$ and the sign of the defect.
From the spatial profile, we obtained $r_\theta=1\unit{\mu{}m}$.
For the peak height, we used time-lapse observations for $200\unit{min}\leq t \leq 250 \unit{min}$, during which the influx toward $-1/2$ defects was strongest for this strain (Fig.\,S8(a)), and estimated $\thetan^\infty=0.2$ and $\thetan^0=0.25$ (Fig.\,S8(b)).

To see the influence of the nematic tilting, we numerically calculated $\bar{v}_r^\pm(r)$ with $\thetan^\infty=0.3$ and $\thetan^0=0.75$, which were estimated from the end-point confocal observation, without polar order (\figref{fig-2}(h), the solid lines).
The other parameters were $\epsilon_0=0.25, a_\mathrm{n}/\xi_0=0.055\unit{\mu{}m^2/min}$ and $r_S = 1.2\unit{\mu{}m}$.
The strength of the influx toward $-1/2$ defects obtained thereby was smaller than the experimental result, indicating that the nematic tilting is insufficnent to quantitatively explain the influx toward $-1/2$ defects.

For the polar contribution to \eqref{eq:v}, we determined the spatial structure of $\thetap^\pm(r,\phi)$ by the end-point confocal observation (\figref{fig-4}(c)(d)).
Then we calibrated the amplitude by multiplying the ratio of $\expct{\thetap^\pm}_{0<x<10\unit{\mu{}m},y=0}$ from the time-lapse observation for $200\unit{min}\leq t \leq 250\unit{min}$ (Fig.\,S8(c)) to that from the end-point observation (\figref{fig-4}(c)(d)).

We are finally left to determine the following parameters: $\epsilon_0$, $a^0_\mathrm{n}/\xi_0$, and $a^0_\mathrm{p}/\xi_0$.
First, we found that the friction anisotropy $\epsilon$ hardly changed the structure of the velocity field (data not shown), so that we chose $\epsilon_0=0.25$.
Then we tuned $a^0_\mathrm{n}/\xi_0$ and $a^0_\mathrm{p}/\xi_0$ to reproduce the experimental data of $\bar{v}^\pm_r(r)$ and obtained $a^0_\mathrm{n}/\xi_0=0.055\unit{\mu{}m^2/min}$ and $a^0_\mathrm{p}/\xi_0=0.8\unit{\mu{}m/min}$, with the results shown in \figref{fig-4}(g).

\begin{acknowledgments}
We are grateful to S. Ramaswamy for motivating us to investigate polar order in three-dimensional orientations.
We thank Y. T. Maeda and H. Salman for sharing the plasmid DNA pZA3R-EYFP and K. Inoue for producing the strain MG1655-pZA3R-EYFP.
We also acknowledge discussions with K. Kawaguchi, D. Nishiguchi, M. Sano, and Y. Zushi.
This work is supported by KAKENHI from Japan Society for the Promotion of Science (JSPS) (No. 19H05800, 20H00128), by KAKENHI for JSPS Fellows (No. 20J10682), and by JST, PRESTO Grant No. JPMJPR18L6, Japan.
\end{acknowledgments}

\section*{Competing interests}
The authors declare no competing interests.


\section*{Data availability}
The data that support the findings of this study will be available at Github upon acceptance.

\section*{Code availability}
The codes used in this study will be available at Github upon acceptance.

\bibliography{./NNbib}

\providecommand{\noopsort}[1]{}\providecommand{\singleletter}[1]{#1}%
\begin{thebibliography}{46}%
\makeatletter
\providecommand \@ifxundefined [1]{%
 \@ifx{#1\undefined}
}%
\providecommand \@ifnum [1]{%
 \ifnum #1\expandafter \@firstoftwo
 \else \expandafter \@secondoftwo
 \fi
}%
\providecommand \@ifx [1]{%
 \ifx #1\expandafter \@firstoftwo
 \else \expandafter \@secondoftwo
 \fi
}%
\providecommand \natexlab [1]{#1}%
\providecommand \enquote  [1]{``#1''}%
\providecommand \bibnamefont  [1]{#1}%
\providecommand \bibfnamefont [1]{#1}%
\providecommand \citenamefont [1]{#1}%
\providecommand \href@noop [0]{\@secondoftwo}%
\providecommand \href [0]{\begingroup \@sanitize@url \@href}%
\providecommand \@href[1]{\@@startlink{#1}\@@href}%
\providecommand \@@href[1]{\endgroup#1\@@endlink}%
\providecommand \@sanitize@url [0]{\catcode `\\12\catcode `\$12\catcode
  `\&12\catcode `\#12\catcode `\^12\catcode `\_12\catcode `\%12\relax}%
\providecommand \@@startlink[1]{}%
\providecommand \@@endlink[0]{}%
\providecommand \url  [0]{\begingroup\@sanitize@url \@url }%
\providecommand \@url [1]{\endgroup\@href {#1}{\urlprefix }}%
\providecommand \urlprefix  [0]{URL }%
\providecommand \Eprint [0]{\href }%
\providecommand \doibase [0]{http://dx.doi.org/}%
\providecommand \selectlanguage [0]{\@gobble}%
\providecommand \bibinfo  [0]{\@secondoftwo}%
\providecommand \bibfield  [0]{\@secondoftwo}%
\providecommand \translation [1]{[#1]}%
\providecommand \BibitemOpen [0]{}%
\providecommand \bibitemStop [0]{}%
\providecommand \bibitemNoStop [0]{.\EOS\space}%
\providecommand \EOS [0]{\spacefactor3000\relax}%
\providecommand \BibitemShut  [1]{\csname bibitem#1\endcsname}%
\let\auto@bib@innerbib\@empty
\bibitem [{\citenamefont {Flemming}\ \emph {et~al.}(2016)\citenamefont
  {Flemming}, \citenamefont {Wingender}, \citenamefont {Szewzyk}, \citenamefont
  {Steinberg}, \citenamefont {Rice},\ and\ \citenamefont
  {Kjelleberg}}]{Flemming2016}%
  \BibitemOpen
  \bibfield  {author} {\bibinfo {author} {\bibfnamefont {H.}~\bibnamefont
  {Flemming}}, \bibinfo {author} {\bibfnamefont {J.}~\bibnamefont {Wingender}},
  \bibinfo {author} {\bibfnamefont {U.}~\bibnamefont {Szewzyk}}, \bibinfo
  {author} {\bibfnamefont {P.}~\bibnamefont {Steinberg}}, \bibinfo {author}
  {\bibfnamefont {S.~A.}\ \bibnamefont {Rice}}, \ and\ \bibinfo {author}
  {\bibfnamefont {S.}~\bibnamefont {Kjelleberg}},\ }\bibfield  {title}
  {\enquote {\bibinfo {title} {Biofilms: an emergent form of bacterial life},}\
  }\href {\doibase 10.1038/nrmicro.2016.94} {\bibfield  {journal} {\bibinfo
  {journal} {Nat. Rev. Microbiol.}\ }\textbf {\bibinfo {volume} {14}},\
  \bibinfo {pages} {563--575} (\bibinfo {year} {2016})}\BibitemShut {NoStop}%
\bibitem [{\citenamefont {Mattila‐Sandholm}\ and\ \citenamefont
  {Wirtanen}(1992)}]{MattilaSandholm1992}%
  \BibitemOpen
  \bibfield  {author} {\bibinfo {author} {\bibfnamefont {T.}~\bibnamefont
  {Mattila‐Sandholm}}\ and\ \bibinfo {author} {\bibfnamefont
  {G.}~\bibnamefont {Wirtanen}},\ }\bibfield  {title} {\enquote {\bibinfo
  {title} {Biofilm formation in the industry: A review},}\ }\href {\doibase
  10.1080/87559129209540953} {\bibfield  {journal} {\bibinfo  {journal} {Food
  Rev. Int.}\ }\textbf {\bibinfo {volume} {8}},\ \bibinfo {pages} {573--603}
  (\bibinfo {year} {1992})}\BibitemShut {NoStop}%
\bibitem [{\citenamefont {Shirtliff}\ and\ \citenamefont
  {Leid}(2009)}]{Shirtliff2009}%
  \BibitemOpen
  \bibfield  {author} {\bibinfo {author} {\bibfnamefont {M.}~\bibnamefont
  {Shirtliff}}\ and\ \bibinfo {author} {\bibfnamefont {J.~G.}\ \bibnamefont
  {Leid}},\ }\href@noop {} {\emph {\bibinfo {title} {The Role of Biofilms in
  Device-Related Infections}}}\ (\bibinfo  {publisher} {Springer},\ \bibinfo
  {year} {2009})\BibitemShut {NoStop}%
\bibitem [{\citenamefont {Allen}\ and\ \citenamefont
  {Waclaw}(2018)}]{Allen2018}%
  \BibitemOpen
  \bibfield  {author} {\bibinfo {author} {\bibfnamefont {R.~J.}\ \bibnamefont
  {Allen}}\ and\ \bibinfo {author} {\bibfnamefont {B.}~\bibnamefont {Waclaw}},\
  }\bibfield  {title} {\enquote {\bibinfo {title} {Bacterial growth: a
  statistical physicist's guide},}\ }\href {\doibase 10.1088/1361-6633/aae546}
  {\bibfield  {journal} {\bibinfo  {journal} {Rep. Prog. Phys.}\ }\textbf
  {\bibinfo {volume} {82}},\ \bibinfo {pages} {016601} (\bibinfo {year}
  {2018})}\BibitemShut {NoStop}%
\bibitem [{\citenamefont {Peng}\ \emph {et~al.}(2016)\citenamefont {Peng},
  \citenamefont {Turiv}, \citenamefont {Guo}, \citenamefont {Wei},\ and\
  \citenamefont {Lavrentovich}}]{Peng2016}%
  \BibitemOpen
  \bibfield  {author} {\bibinfo {author} {\bibfnamefont {C.}~\bibnamefont
  {Peng}}, \bibinfo {author} {\bibfnamefont {T.}~\bibnamefont {Turiv}},
  \bibinfo {author} {\bibfnamefont {Y.}~\bibnamefont {Guo}}, \bibinfo {author}
  {\bibfnamefont {Q.}~\bibnamefont {Wei}}, \ and\ \bibinfo {author}
  {\bibfnamefont {O.~D.}\ \bibnamefont {Lavrentovich}},\ }\bibfield  {title}
  {\enquote {\bibinfo {title} {Command of active matter by topological defects
  and patterns},}\ }\href {\doibase 10.1126/science.aah6936} {\bibfield
  {journal} {\bibinfo  {journal} {Science}\ }\textbf {\bibinfo {volume}
  {354}},\ \bibinfo {pages} {882--885} (\bibinfo {year} {2016})}\BibitemShut
  {NoStop}%
\bibitem [{\citenamefont {Genkin}\ \emph {et~al.}(2017)\citenamefont {Genkin},
  \citenamefont {Sokolov}, \citenamefont {Lavrentovich},\ and\ \citenamefont
  {Aranson}}]{Genkin2017}%
  \BibitemOpen
  \bibfield  {author} {\bibinfo {author} {\bibfnamefont {M.~M.}\ \bibnamefont
  {Genkin}}, \bibinfo {author} {\bibfnamefont {A.}~\bibnamefont {Sokolov}},
  \bibinfo {author} {\bibfnamefont {O.~D.}\ \bibnamefont {Lavrentovich}}, \
  and\ \bibinfo {author} {\bibfnamefont {I.~S.}\ \bibnamefont {Aranson}},\
  }\bibfield  {title} {\enquote {\bibinfo {title} {Topological defects in a
  living nematic ensnare swimming bacteria},}\ }\href {\doibase
  10.1103/PhysRevX.7.011029} {\bibfield  {journal} {\bibinfo  {journal} {Phys.
  Rev. X}\ }\textbf {\bibinfo {volume} {7}},\ \bibinfo {pages} {011029}
  (\bibinfo {year} {2017})}\BibitemShut {NoStop}%
\bibitem [{\citenamefont {You}\ \emph {et~al.}(2018)\citenamefont {You},
  \citenamefont {Pearce}, \citenamefont {Sengupta},\ and\ \citenamefont
  {Giomi}}]{You2018}%
  \BibitemOpen
  \bibfield  {author} {\bibinfo {author} {\bibfnamefont {Z.}~\bibnamefont
  {You}}, \bibinfo {author} {\bibfnamefont {D.~J.~G.}\ \bibnamefont {Pearce}},
  \bibinfo {author} {\bibfnamefont {A.}~\bibnamefont {Sengupta}}, \ and\
  \bibinfo {author} {\bibfnamefont {L.}~\bibnamefont {Giomi}},\ }\bibfield
  {title} {\enquote {\bibinfo {title} {Geometry and mechanics of microdomains
  in growing bacterial colonies},}\ }\href {\doibase 10.1103/PhysRevX.8.031065}
  {\bibfield  {journal} {\bibinfo  {journal} {Phys. Rev. X}\ }\textbf {\bibinfo
  {volume} {8}},\ \bibinfo {pages} {031065} (\bibinfo {year}
  {2018})}\BibitemShut {NoStop}%
\bibitem [{\citenamefont {Yaman}\ \emph {et~al.}(2019)\citenamefont {Yaman},
  \citenamefont {Demir}, \citenamefont {Vetter},\ and\ \citenamefont
  {Kocabas}}]{Yaman2019}%
  \BibitemOpen
  \bibfield  {author} {\bibinfo {author} {\bibfnamefont {Y.~I.}\ \bibnamefont
  {Yaman}}, \bibinfo {author} {\bibfnamefont {E.}~\bibnamefont {Demir}},
  \bibinfo {author} {\bibfnamefont {R.}~\bibnamefont {Vetter}}, \ and\ \bibinfo
  {author} {\bibfnamefont {A.}~\bibnamefont {Kocabas}},\ }\bibfield  {title}
  {\enquote {\bibinfo {title} {Emergence of active nematics in chaining
  bacterial biofilms},}\ }\href {\doibase 10.1038/s41467-019-10311-z}
  {\bibfield  {journal} {\bibinfo  {journal} {Nat. Commun.}\ }\textbf {\bibinfo
  {volume} {10}},\ \bibinfo {pages} {2285} (\bibinfo {year}
  {2019})}\BibitemShut {NoStop}%
\bibitem [{\citenamefont {Dell’Arciprete}\ \emph {et~al.}(2018)\citenamefont
  {Dell’Arciprete}, \citenamefont {Blow}, \citenamefont {Brown},
  \citenamefont {Farrell}, \citenamefont {Lintuvuori}, \citenamefont {McVey},
  \citenamefont {Marenduzzo},\ and\ \citenamefont {Poon}}]{Dell2019}%
  \BibitemOpen
  \bibfield  {author} {\bibinfo {author} {\bibfnamefont {D.}~\bibnamefont
  {Dell’Arciprete}}, \bibinfo {author} {\bibfnamefont {M.~L.}\ \bibnamefont
  {Blow}}, \bibinfo {author} {\bibfnamefont {A.~T.}\ \bibnamefont {Brown}},
  \bibinfo {author} {\bibfnamefont {F.~D.~C.}\ \bibnamefont {Farrell}},
  \bibinfo {author} {\bibfnamefont {J.~S.}\ \bibnamefont {Lintuvuori}},
  \bibinfo {author} {\bibfnamefont {A.~F.}\ \bibnamefont {McVey}}, \bibinfo
  {author} {\bibfnamefont {D.}~\bibnamefont {Marenduzzo}}, \ and\ \bibinfo
  {author} {\bibfnamefont {W.~C.~K.}\ \bibnamefont {Poon}},\ }\bibfield
  {title} {\enquote {\bibinfo {title} {A growing bacterial colony in two
  dimensions as an active nematic},}\ }\href {\doibase
  10.1038/s41467-018-06370-3} {\bibfield  {journal} {\bibinfo  {journal} {Nat.
  Commun.}\ }\textbf {\bibinfo {volume} {9}},\ \bibinfo {pages} {4190}
  (\bibinfo {year} {2018})}\BibitemShut {NoStop}%
\bibitem [{\citenamefont {Doostmohammadi}\ \emph {et~al.}(2016)\citenamefont
  {Doostmohammadi}, \citenamefont {Thampi},\ and\ \citenamefont
  {Yeomans}}]{Doostmohammadi2019}%
  \BibitemOpen
  \bibfield  {author} {\bibinfo {author} {\bibfnamefont {A.}~\bibnamefont
  {Doostmohammadi}}, \bibinfo {author} {\bibfnamefont {S.~P.}\ \bibnamefont
  {Thampi}}, \ and\ \bibinfo {author} {\bibfnamefont {J.~M.}\ \bibnamefont
  {Yeomans}},\ }\bibfield  {title} {\enquote {\bibinfo {title} {Defect-mediated
  morphologies in growing cell colonies},}\ }\href {\doibase
  10.1103/PhysRevLett.117.048102} {\bibfield  {journal} {\bibinfo  {journal}
  {Phys. Rev. Lett.}\ }\textbf {\bibinfo {volume} {117}},\ \bibinfo {pages}
  {048102} (\bibinfo {year} {2016})}\BibitemShut {NoStop}%
\bibitem [{\citenamefont {Sengupta}(2020)}]{Sengupta2020}%
  \BibitemOpen
  \bibfield  {author} {\bibinfo {author} {\bibfnamefont {A.}~\bibnamefont
  {Sengupta}},\ }\bibfield  {title} {\enquote {\bibinfo {title} {Microbial
  active matter: A topological framework},}\ }\href {\doibase
  10.3389/fphy.2020.00184} {\bibfield  {journal} {\bibinfo  {journal} {Front.
  Phys.}\ }\textbf {\bibinfo {volume} {8}},\ \bibinfo {pages} {184} (\bibinfo
  {year} {2020})}\BibitemShut {NoStop}%
\bibitem [{\citenamefont {Meacock}\ \emph {et~al.}(2021)\citenamefont
  {Meacock}, \citenamefont {Doostmohammadi}, \citenamefont {Foster},
  \citenamefont {M.},\ and\ \citenamefont {Durham}}]{Meacock2021}%
  \BibitemOpen
  \bibfield  {author} {\bibinfo {author} {\bibfnamefont {O.~J.}\ \bibnamefont
  {Meacock}}, \bibinfo {author} {\bibfnamefont {A.}~\bibnamefont
  {Doostmohammadi}}, \bibinfo {author} {\bibfnamefont {K.~R.}\ \bibnamefont
  {Foster}}, \bibinfo {author} {\bibfnamefont {Yeomans~J.}\ \bibnamefont {M.}},
  \ and\ \bibinfo {author} {\bibfnamefont {W.~M.}\ \bibnamefont {Durham}},\
  }\bibfield  {title} {\enquote {\bibinfo {title} {Bacteria solve the problem
  of crowding by moving slowly},}\ }\href {\doibase 10.1038/s41567-020-01070-6}
  {\bibfield  {journal} {\bibinfo  {journal} {Nat. Phys.}\ }\textbf {\bibinfo
  {volume} {17}},\ \bibinfo {pages} {205--210} (\bibinfo {year}
  {2021})}\BibitemShut {NoStop}%
\bibitem [{\citenamefont {Copenhagen}\ \emph {et~al.}(2021)\citenamefont
  {Copenhagen}, \citenamefont {Alert}, \citenamefont {Wingreen},\ and\
  \citenamefont {Shaevitz}}]{Copenhagen2021}%
  \BibitemOpen
  \bibfield  {author} {\bibinfo {author} {\bibfnamefont {K.}~\bibnamefont
  {Copenhagen}}, \bibinfo {author} {\bibfnamefont {R.}~\bibnamefont {Alert}},
  \bibinfo {author} {\bibfnamefont {N.~S.}\ \bibnamefont {Wingreen}}, \ and\
  \bibinfo {author} {\bibfnamefont {J.~W.}\ \bibnamefont {Shaevitz}},\
  }\bibfield  {title} {\enquote {\bibinfo {title} {Topological defects promote
  layer formation in \textit{Myxococcus xanthus} colonies},}\ }\href {\doibase
  10.1038/s41567-020-01056-4} {\bibfield  {journal} {\bibinfo  {journal} {Nat.
  Phys.}\ }\textbf {\bibinfo {volume} {17}},\ \bibinfo {pages} {211--215}
  (\bibinfo {year} {2021})}\BibitemShut {NoStop}%
\bibitem [{\citenamefont {Saw}\ \emph {et~al.}(2018)\citenamefont {Saw},
  \citenamefont {Xi}, \citenamefont {Ladoux},\ and\ \citenamefont
  {Lim}}]{Saw2018}%
  \BibitemOpen
  \bibfield  {author} {\bibinfo {author} {\bibfnamefont {T.~B.}\ \bibnamefont
  {Saw}}, \bibinfo {author} {\bibfnamefont {W.}~\bibnamefont {Xi}}, \bibinfo
  {author} {\bibfnamefont {B.}~\bibnamefont {Ladoux}}, \ and\ \bibinfo {author}
  {\bibfnamefont {C.~T.}\ \bibnamefont {Lim}},\ }\bibfield  {title} {\enquote
  {\bibinfo {title} {Biological tissues as active nematic liquid crystals},}\
  }\href {\doibase https://doi.org/10.1002/adma.201802579} {\bibfield
  {journal} {\bibinfo  {journal} {Adv. Mater.}\ }\textbf {\bibinfo {volume}
  {30}},\ \bibinfo {pages} {1802579} (\bibinfo {year} {2018})}\BibitemShut
  {NoStop}%
\bibitem [{\citenamefont {Doostmohammadi}\ and\ \citenamefont
  {Ladoux}(2021)}]{DOOSTMOHAMMADI2021}%
  \BibitemOpen
  \bibfield  {author} {\bibinfo {author} {\bibfnamefont {A.}~\bibnamefont
  {Doostmohammadi}}\ and\ \bibinfo {author} {\bibfnamefont {B.}~\bibnamefont
  {Ladoux}},\ }\bibfield  {title} {\enquote {\bibinfo {title} {Physics of
  liquid crystals in cell biology},}\ }\href {\doibase
  10.1016/j.tcb.2021.09.012} {\bibfield  {journal} {\bibinfo  {journal} {Trends
  Cell Biol.}\ } (\bibinfo {year} {2021}),\
  10.1016/j.tcb.2021.09.012}\BibitemShut {NoStop}%
\bibitem [{\citenamefont {Saw}\ \emph {et~al.}(2017)\citenamefont {Saw},
  \citenamefont {Doostmohammadi}, \citenamefont {Nier}, \citenamefont
  {Kocgozlu}, \citenamefont {Thampi}, \citenamefont {Toyama}, \citenamefont
  {Marcq}, \citenamefont {Lim}, \citenamefont {Yeomans},\ and\ \citenamefont
  {Ladoux}}]{Saw2017}%
  \BibitemOpen
  \bibfield  {author} {\bibinfo {author} {\bibfnamefont {T.~B.}\ \bibnamefont
  {Saw}}, \bibinfo {author} {\bibfnamefont {A.}~\bibnamefont {Doostmohammadi}},
  \bibinfo {author} {\bibfnamefont {V.}~\bibnamefont {Nier}}, \bibinfo {author}
  {\bibfnamefont {L.}~\bibnamefont {Kocgozlu}}, \bibinfo {author}
  {\bibfnamefont {S.}~\bibnamefont {Thampi}}, \bibinfo {author} {\bibfnamefont
  {Y.}~\bibnamefont {Toyama}}, \bibinfo {author} {\bibfnamefont
  {P.}~\bibnamefont {Marcq}}, \bibinfo {author} {\bibfnamefont {C.~T.}\
  \bibnamefont {Lim}}, \bibinfo {author} {\bibfnamefont {J.~M.}\ \bibnamefont
  {Yeomans}}, \ and\ \bibinfo {author} {\bibfnamefont {B.}~\bibnamefont
  {Ladoux}},\ }\bibfield  {title} {\enquote {\bibinfo {title} {Topological
  defects in epithelia govern cell death and extrusion},}\ }\href {\doibase
  10.1038/nature21718} {\bibfield  {journal} {\bibinfo  {journal} {Nature}\
  }\textbf {\bibinfo {volume} {544}},\ \bibinfo {pages} {212--216} (\bibinfo
  {year} {2017})}\BibitemShut {NoStop}%
\bibitem [{\citenamefont {Kawaguchi}\ \emph {et~al.}(2017)\citenamefont
  {Kawaguchi}, \citenamefont {Kageyama},\ and\ \citenamefont
  {Sano}}]{Kawaguchi2017}%
  \BibitemOpen
  \bibfield  {author} {\bibinfo {author} {\bibfnamefont {K.}~\bibnamefont
  {Kawaguchi}}, \bibinfo {author} {\bibfnamefont {R.}~\bibnamefont {Kageyama}},
  \ and\ \bibinfo {author} {\bibfnamefont {M.}~\bibnamefont {Sano}},\
  }\bibfield  {title} {\enquote {\bibinfo {title} {Topological defects control
  collective dynamics in neural progenitor cell cultures},}\ }\href {\doibase
  10.1038/nature22321} {\bibfield  {journal} {\bibinfo  {journal} {Nature}\
  }\textbf {\bibinfo {volume} {545}},\ \bibinfo {pages} {327--331} (\bibinfo
  {year} {2017})}\BibitemShut {NoStop}%
\bibitem [{\citenamefont {Duclos}\ \emph {et~al.}(2017)\citenamefont {Duclos},
  \citenamefont {Erlenk{\"a}mper}, \citenamefont {Joanny},\ and\ \citenamefont
  {Silberzan}}]{Duclos2017}%
  \BibitemOpen
  \bibfield  {author} {\bibinfo {author} {\bibfnamefont {G.}~\bibnamefont
  {Duclos}}, \bibinfo {author} {\bibfnamefont {C.}~\bibnamefont
  {Erlenk{\"a}mper}}, \bibinfo {author} {\bibfnamefont {J.}~\bibnamefont
  {Joanny}}, \ and\ \bibinfo {author} {\bibfnamefont {P}~\bibnamefont
  {Silberzan}},\ }\bibfield  {title} {\enquote {\bibinfo {title} {{Topological
  defects in confined populations of spindle-shaped cells}},}\ }\href {\doibase
  10.1038/nphys3876} {\bibfield  {journal} {\bibinfo  {journal} {Nat. Phys.}\
  }\textbf {\bibinfo {volume} {13}},\ \bibinfo {pages} {58--62} (\bibinfo
  {year} {2017})}\BibitemShut {NoStop}%
\bibitem [{\citenamefont {Turiv}\ \emph {et~al.}(2020)\citenamefont {Turiv},
  \citenamefont {Krieger}, \citenamefont {Babakhanova}, \citenamefont {Yu},
  \citenamefont {Shiyanovskii}, \citenamefont {Wei}, \citenamefont {Kim},\ and\
  \citenamefont {Lavrentovich}}]{Turiv2020}%
  \BibitemOpen
  \bibfield  {author} {\bibinfo {author} {\bibfnamefont {T.}~\bibnamefont
  {Turiv}}, \bibinfo {author} {\bibfnamefont {J.}~\bibnamefont {Krieger}},
  \bibinfo {author} {\bibfnamefont {G.}~\bibnamefont {Babakhanova}}, \bibinfo
  {author} {\bibfnamefont {H.}~\bibnamefont {Yu}}, \bibinfo {author}
  {\bibfnamefont {S.~V.}\ \bibnamefont {Shiyanovskii}}, \bibinfo {author}
  {\bibfnamefont {Q.}~\bibnamefont {Wei}}, \bibinfo {author} {\bibfnamefont
  {M.}~\bibnamefont {Kim}}, \ and\ \bibinfo {author} {\bibfnamefont {O.~D.}\
  \bibnamefont {Lavrentovich}},\ }\bibfield  {title} {\enquote {\bibinfo
  {title} {Topology control of human fibroblast cells monolayer by liquid
  crystal elastomer},}\ }\href {\doibase 10.1126/sciadv.aaz6485} {\bibfield
  {journal} {\bibinfo  {journal} {Sci. Adv.}\ }\textbf {\bibinfo {volume} {6}}
  (\bibinfo {year} {2020}),\ 10.1126/sciadv.aaz6485}\BibitemShut {NoStop}%
\bibitem [{\citenamefont {Maroudas-Sacks}\ \emph {et~al.}(2021)\citenamefont
  {Maroudas-Sacks}, \citenamefont {Garion}, \citenamefont {Shani-Zerbib},
  \citenamefont {Livshits}, \citenamefont {Braun},\ and\ \citenamefont
  {Keren}}]{Maroudas-Sacks2021}%
  \BibitemOpen
  \bibfield  {author} {\bibinfo {author} {\bibfnamefont {Y.}~\bibnamefont
  {Maroudas-Sacks}}, \bibinfo {author} {\bibfnamefont {L.}~\bibnamefont
  {Garion}}, \bibinfo {author} {\bibfnamefont {L.}~\bibnamefont
  {Shani-Zerbib}}, \bibinfo {author} {\bibfnamefont {A.}~\bibnamefont
  {Livshits}}, \bibinfo {author} {\bibfnamefont {E.}~\bibnamefont {Braun}}, \
  and\ \bibinfo {author} {\bibfnamefont {K.}~\bibnamefont {Keren}},\ }\bibfield
   {title} {\enquote {\bibinfo {title} {Topological defects in the nematic
  order of actin fibres as organization centres of \textit{Hydra}
  morphogenesis},}\ }\href {\doibase 10.1038/s41567-020-01083-1} {\bibfield
  {journal} {\bibinfo  {journal} {Nat. Phys.}\ }\textbf {\bibinfo {volume}
  {17}},\ \bibinfo {pages} {251--259} (\bibinfo {year} {2021})}\BibitemShut
  {NoStop}%
\bibitem [{\citenamefont {Su}\ \emph {et~al.}(2012)\citenamefont {Su},
  \citenamefont {Liao}, \citenamefont {Roan}, \citenamefont {Wang},
  \citenamefont {Chiou},\ and\ \citenamefont {Syu}}]{Su2012}%
  \BibitemOpen
  \bibfield  {author} {\bibinfo {author} {\bibfnamefont {P.}~\bibnamefont
  {Su}}, \bibinfo {author} {\bibfnamefont {C.}~\bibnamefont {Liao}}, \bibinfo
  {author} {\bibfnamefont {J.}~\bibnamefont {Roan}}, \bibinfo {author}
  {\bibfnamefont {S.}~\bibnamefont {Wang}}, \bibinfo {author} {\bibfnamefont
  {A.}~\bibnamefont {Chiou}}, \ and\ \bibinfo {author} {\bibfnamefont
  {W.}~\bibnamefont {Syu}},\ }\bibfield  {title} {\enquote {\bibinfo {title}
  {Bacterial colony from two-dimensional division to three-dimensional
  development},}\ }\href {\doibase 10.1371/journal.pone.0048098} {\bibfield
  {journal} {\bibinfo  {journal} {PLOS ONE}\ }\textbf {\bibinfo {volume} {7}},\
  \bibinfo {pages} {1--10} (\bibinfo {year} {2012})}\BibitemShut {NoStop}%
\bibitem [{\citenamefont {Farrell}\ \emph {et~al.}(2013)\citenamefont
  {Farrell}, \citenamefont {Hallatschek}, \citenamefont {Marenduzzo},\ and\
  \citenamefont {Waclaw}}]{Farrell2013}%
  \BibitemOpen
  \bibfield  {author} {\bibinfo {author} {\bibfnamefont {F.~D.~C.}\
  \bibnamefont {Farrell}}, \bibinfo {author} {\bibfnamefont {O.}~\bibnamefont
  {Hallatschek}}, \bibinfo {author} {\bibfnamefont {D.}~\bibnamefont
  {Marenduzzo}}, \ and\ \bibinfo {author} {\bibfnamefont {B.}~\bibnamefont
  {Waclaw}},\ }\bibfield  {title} {\enquote {\bibinfo {title} {Mechanically
  driven growth of quasi-two-dimensional microbial colonies},}\ }\href
  {\doibase 10.1103/PhysRevLett.111.168101} {\bibfield  {journal} {\bibinfo
  {journal} {Phys. Rev. Lett.}\ }\textbf {\bibinfo {volume} {111}},\ \bibinfo
  {pages} {168101} (\bibinfo {year} {2013})}\BibitemShut {NoStop}%
\bibitem [{\citenamefont {Grant}\ \emph {et~al.}(2014)\citenamefont {Grant},
  \citenamefont {Wac{\l}aw}, \citenamefont {Allen},\ and\ \citenamefont
  {Cicuta}}]{Grant2014}%
  \BibitemOpen
  \bibfield  {author} {\bibinfo {author} {\bibfnamefont {M.~A.~A.}\
  \bibnamefont {Grant}}, \bibinfo {author} {\bibfnamefont {B.}~\bibnamefont
  {Wac{\l}aw}}, \bibinfo {author} {\bibfnamefont {R.~J.}\ \bibnamefont
  {Allen}}, \ and\ \bibinfo {author} {\bibfnamefont {P.}~\bibnamefont
  {Cicuta}},\ }\bibfield  {title} {\enquote {\bibinfo {title} {The role of
  mechanical forces in the planar-to-bulk transition in growing
  \textit{Escherichia coli} microcolonies},}\ }\href {\doibase
  10.1098/rsif.2014.0400} {\bibfield  {journal} {\bibinfo  {journal} {J. R.
  Soc. Interface}\ }\textbf {\bibinfo {volume} {11}},\ \bibinfo {pages}
  {20140400} (\bibinfo {year} {2014})}\BibitemShut {NoStop}%
\bibitem [{\citenamefont {Duvernoy}\ \emph {et~al.}(2018)\citenamefont
  {Duvernoy}, \citenamefont {Mora}, \citenamefont {Ardr{\'e}}, \citenamefont
  {Croquette}, \citenamefont {Bensimon}, \citenamefont {Quilliet},
  \citenamefont {Ghigo}, \citenamefont {Balland}, \citenamefont {Beloin},
  \citenamefont {Lecuyer},\ and\ \citenamefont {Desprat}}]{Duvernoy2018}%
  \BibitemOpen
  \bibfield  {author} {\bibinfo {author} {\bibfnamefont {M.}~\bibnamefont
  {Duvernoy}}, \bibinfo {author} {\bibfnamefont {T.}~\bibnamefont {Mora}},
  \bibinfo {author} {\bibfnamefont {M.}~\bibnamefont {Ardr{\'e}}}, \bibinfo
  {author} {\bibfnamefont {V.}~\bibnamefont {Croquette}}, \bibinfo {author}
  {\bibfnamefont {D.}~\bibnamefont {Bensimon}}, \bibinfo {author}
  {\bibfnamefont {C.}~\bibnamefont {Quilliet}}, \bibinfo {author}
  {\bibfnamefont {J.}~\bibnamefont {Ghigo}}, \bibinfo {author} {\bibfnamefont
  {M.}~\bibnamefont {Balland}}, \bibinfo {author} {\bibfnamefont
  {C.}~\bibnamefont {Beloin}}, \bibinfo {author} {\bibfnamefont
  {S.}~\bibnamefont {Lecuyer}}, \ and\ \bibinfo {author} {\bibfnamefont
  {N.}~\bibnamefont {Desprat}},\ }\bibfield  {title} {\enquote {\bibinfo
  {title} {Asymmetric adhesion of rod-shaped bacteria controls microcolony
  morphogenesis},}\ }\href {\doibase 10.1038/s41467-018-03446-y} {\bibfield
  {journal} {\bibinfo  {journal} {Nat. Commun.}\ }\textbf {\bibinfo {volume}
  {9}},\ \bibinfo {pages} {1120} (\bibinfo {year} {2018})}\BibitemShut
  {NoStop}%
\bibitem [{\citenamefont {Beroz}\ \emph {et~al.}(2018)\citenamefont {Beroz},
  \citenamefont {Yan}, \citenamefont {Meir}, \citenamefont {Sabass},
  \citenamefont {Stone}, \citenamefont {Bassler},\ and\ \citenamefont
  {Wingreen}}]{Beroz2018}%
  \BibitemOpen
  \bibfield  {author} {\bibinfo {author} {\bibfnamefont {F.}~\bibnamefont
  {Beroz}}, \bibinfo {author} {\bibfnamefont {J.}~\bibnamefont {Yan}}, \bibinfo
  {author} {\bibfnamefont {Y.}~\bibnamefont {Meir}}, \bibinfo {author}
  {\bibfnamefont {B.}~\bibnamefont {Sabass}}, \bibinfo {author} {\bibfnamefont
  {H.~A.}\ \bibnamefont {Stone}}, \bibinfo {author} {\bibfnamefont {B.~L.}\
  \bibnamefont {Bassler}}, \ and\ \bibinfo {author} {\bibfnamefont {N.~S.}\
  \bibnamefont {Wingreen}},\ }\bibfield  {title} {\enquote {\bibinfo {title}
  {Verticalization of bacterial biofilms},}\ }\href {\doibase
  10.1038/s41567-018-0170-4} {\bibfield  {journal} {\bibinfo  {journal} {Nat.
  Phys.}\ }\textbf {\bibinfo {volume} {14}},\ \bibinfo {pages} {954--960}
  (\bibinfo {year} {2018})}\BibitemShut {NoStop}%
\bibitem [{\citenamefont {You}\ \emph {et~al.}(2019)\citenamefont {You},
  \citenamefont {Pearce}, \citenamefont {Sengupta},\ and\ \citenamefont
  {Giomi}}]{You2019}%
  \BibitemOpen
  \bibfield  {author} {\bibinfo {author} {\bibfnamefont {Z.}~\bibnamefont
  {You}}, \bibinfo {author} {\bibfnamefont {D.~J.~G.}\ \bibnamefont {Pearce}},
  \bibinfo {author} {\bibfnamefont {A.}~\bibnamefont {Sengupta}}, \ and\
  \bibinfo {author} {\bibfnamefont {L.}~\bibnamefont {Giomi}},\ }\bibfield
  {title} {\enquote {\bibinfo {title} {Mono- to multilayer transition in
  growing bacterial colonies},}\ }\href {\doibase
  10.1103/PhysRevLett.123.178001} {\bibfield  {journal} {\bibinfo  {journal}
  {Phys. Rev. Lett.}\ }\textbf {\bibinfo {volume} {123}},\ \bibinfo {pages}
  {178001} (\bibinfo {year} {2019})}\BibitemShut {NoStop}%
\bibitem [{\citenamefont {Warren}\ \emph {et~al.}(2019)\citenamefont {Warren},
  \citenamefont {Sun}, \citenamefont {Yan}, \citenamefont {Cremer},
  \citenamefont {Li},\ and\ \citenamefont {Hwa}}]{Neher2019}%
  \BibitemOpen
  \bibfield  {author} {\bibinfo {author} {\bibfnamefont {M.~R.}\ \bibnamefont
  {Warren}}, \bibinfo {author} {\bibfnamefont {H.}~\bibnamefont {Sun}},
  \bibinfo {author} {\bibfnamefont {Y.}~\bibnamefont {Yan}}, \bibinfo {author}
  {\bibfnamefont {J.}~\bibnamefont {Cremer}}, \bibinfo {author} {\bibfnamefont
  {B.}~\bibnamefont {Li}}, \ and\ \bibinfo {author} {\bibfnamefont
  {T.}~\bibnamefont {Hwa}},\ }\bibfield  {title} {\enquote {\bibinfo {title}
  {Spatiotemporal establishment of dense bacterial colonies growing on hard
  agar},}\ }\href {\doibase 10.7554/eLife.41093} {\bibfield  {journal}
  {\bibinfo  {journal} {eLife}\ }\textbf {\bibinfo {volume} {8}},\ \bibinfo
  {pages} {e41093} (\bibinfo {year} {2019})}\BibitemShut {NoStop}%
\bibitem [{\citenamefont {Hartmann}\ \emph {et~al.}(2019)\citenamefont
  {Hartmann}, \citenamefont {K.~Singh}, \citenamefont {Pearce}, \citenamefont
  {Mok}, \citenamefont {Song}, \citenamefont {D\'{i}az-Pascual}, \citenamefont
  {Dunkel},\ and\ \citenamefont {Drescher}}]{Hartmann2019}%
  \BibitemOpen
  \bibfield  {author} {\bibinfo {author} {\bibfnamefont {R.}~\bibnamefont
  {Hartmann}}, \bibinfo {author} {\bibfnamefont {P.}~\bibnamefont {K.~Singh}},
  \bibinfo {author} {\bibfnamefont {P.}~\bibnamefont {Pearce}}, \bibinfo
  {author} {\bibfnamefont {R.}~\bibnamefont {Mok}}, \bibinfo {author}
  {\bibfnamefont {B.}~\bibnamefont {Song}}, \bibinfo {author} {\bibfnamefont
  {F.}~\bibnamefont {D\'{i}az-Pascual}}, \bibinfo {author} {\bibfnamefont
  {J.}~\bibnamefont {Dunkel}}, \ and\ \bibinfo {author} {\bibfnamefont
  {K.}~\bibnamefont {Drescher}},\ }\bibfield  {title} {\enquote {\bibinfo
  {title} {{Emergence of three-dimensional order and structure in growing
  biofilms}},}\ }\href {\doibase 10.1038/s41567-018-0356-9} {\bibfield
  {journal} {\bibinfo  {journal} {Nat. Phys.}\ }\textbf {\bibinfo {volume}
  {15}},\ \bibinfo {pages} {251--256} (\bibinfo {year} {2019})}\BibitemShut
  {NoStop}%
\bibitem [{\citenamefont {Takatori}\ and\ \citenamefont
  {Mandadapu}()}]{Takatori2020}%
  \BibitemOpen
  \bibfield  {author} {\bibinfo {author} {\bibfnamefont {S.~C.}\ \bibnamefont
  {Takatori}}\ and\ \bibinfo {author} {\bibfnamefont {K.~K.}\ \bibnamefont
  {Mandadapu}},\ }\bibfield  {title} {\enquote {\bibinfo {title}
  {Motility-induced buckling and glassy dynamics regulate three-dimensional
  transitions of bacterial monolayers},}\ }\href
  {https://arxiv.org/abs/2003.05618} {\bibinfo  {journal} {arXiv:2003.05618}\
  }\BibitemShut {NoStop}%
\bibitem [{\citenamefont {Dhar}\ \emph {et~al.}()\citenamefont {Dhar},
  \citenamefont {Thai}, \citenamefont {Ghoshal}, \citenamefont {Giomi},\ and\
  \citenamefont {Sengupta}}]{Dhar2021}%
  \BibitemOpen
\bibfield  {journal} {  }\bibfield  {author} {\bibinfo {author} {\bibfnamefont
  {J.}~\bibnamefont {Dhar}}, \bibinfo {author} {\bibfnamefont {A.~L.~P.}\
  \bibnamefont {Thai}}, \bibinfo {author} {\bibfnamefont {A.}~\bibnamefont
  {Ghoshal}}, \bibinfo {author} {\bibfnamefont {L.}~\bibnamefont {Giomi}}, \
  and\ \bibinfo {author} {\bibfnamefont {A.}~\bibnamefont {Sengupta}},\
  }\bibfield  {title} {\enquote {\bibinfo {title} {Trade-offs in phenotypic
  noise synchronize emergent topology to actively enhance transport in
  microbial environments},}\ }\href {https://arxiv.org/abs/2105.00465}
  {\bibinfo  {journal} {arXiv:2105.00465}\ }\BibitemShut {NoStop}%
\bibitem [{\citenamefont {Doostmohammadi}\ \emph {et~al.}(2018)\citenamefont
  {Doostmohammadi}, \citenamefont {Ign{\'e}s-Mullol}, \citenamefont {Yeomans},\
  and\ \citenamefont {Sagu{\'e}s}}]{Doostmohammadi2018}%
  \BibitemOpen
\bibfield  {journal} {  }\bibfield  {author} {\bibinfo {author} {\bibfnamefont
  {A.}~\bibnamefont {Doostmohammadi}}, \bibinfo {author} {\bibfnamefont
  {J.}~\bibnamefont {Ign{\'e}s-Mullol}}, \bibinfo {author} {\bibfnamefont
  {J.~M.}\ \bibnamefont {Yeomans}}, \ and\ \bibinfo {author} {\bibfnamefont
  {F.}~\bibnamefont {Sagu{\'e}s}},\ }\bibfield  {title} {\enquote {\bibinfo
  {title} {Active nematics},}\ }\href {\doibase 10.1038/s41467-018-05666-8}
  {\bibfield  {journal} {\bibinfo  {journal} {Nat. Commun.}\ }\textbf {\bibinfo
  {volume} {9}},\ \bibinfo {pages} {3246} (\bibinfo {year} {2018})}\BibitemShut
  {NoStop}%
\bibitem [{\citenamefont {Doumic}\ \emph {et~al.}(2020)\citenamefont {Doumic},
  \citenamefont {Hecht},\ and\ \citenamefont {Peurichard}}]{Doumic2020}%
  \BibitemOpen
  \bibfield  {author} {\bibinfo {author} {\bibfnamefont {M.}~\bibnamefont
  {Doumic}}, \bibinfo {author} {\bibfnamefont {S.}~\bibnamefont {Hecht}}, \
  and\ \bibinfo {author} {\bibfnamefont {D.}~\bibnamefont {Peurichard}},\
  }\bibfield  {title} {\enquote {\bibinfo {title} {A purely mechanical model
  with asymmetric features for early morphogenesis of rod-shaped bacteria
  micro-colony},}\ }\href {\doibase 10.3934/mbe.2020356} {\bibfield  {journal}
  {\bibinfo  {journal} {Math. Biosci. Eng.}\ }\textbf {\bibinfo {volume}
  {17}},\ \bibinfo {pages} {6873--6908} (\bibinfo {year} {2020})}\BibitemShut
  {NoStop}%
\bibitem [{\citenamefont {Blair}\ \emph {et~al.}(2003)\citenamefont {Blair},
  \citenamefont {Neicu},\ and\ \citenamefont {Kudrolli}}]{Blair2003}%
  \BibitemOpen
  \bibfield  {author} {\bibinfo {author} {\bibfnamefont {D.~L.}\ \bibnamefont
  {Blair}}, \bibinfo {author} {\bibfnamefont {T.}~\bibnamefont {Neicu}}, \ and\
  \bibinfo {author} {\bibfnamefont {A.}~\bibnamefont {Kudrolli}},\ }\bibfield
  {title} {\enquote {\bibinfo {title} {Vortices in vibrated granular rods},}\
  }\href {\doibase 10.1103/PhysRevE.67.031303} {\bibfield  {journal} {\bibinfo
  {journal} {Phys. Rev. E}\ }\textbf {\bibinfo {volume} {67}},\ \bibinfo
  {pages} {031303} (\bibinfo {year} {2003})}\BibitemShut {NoStop}%
\bibitem [{\citenamefont {Volfson}\ \emph {et~al.}(2004)\citenamefont
  {Volfson}, \citenamefont {Kudrolli},\ and\ \citenamefont
  {Tsimring}}]{Volfson2004}%
  \BibitemOpen
  \bibfield  {author} {\bibinfo {author} {\bibfnamefont {D.}~\bibnamefont
  {Volfson}}, \bibinfo {author} {\bibfnamefont {A.}~\bibnamefont {Kudrolli}}, \
  and\ \bibinfo {author} {\bibfnamefont {L.~S.}\ \bibnamefont {Tsimring}},\
  }\bibfield  {title} {\enquote {\bibinfo {title} {Anisotropy-driven dynamics
  in vibrated granular rods},}\ }\href {\doibase 10.1103/PhysRevE.70.051312}
  {\bibfield  {journal} {\bibinfo  {journal} {Phys. Rev. E}\ }\textbf {\bibinfo
  {volume} {70}},\ \bibinfo {pages} {051312} (\bibinfo {year}
  {2004})}\BibitemShut {NoStop}%
\bibitem [{\citenamefont {Volfson}\ \emph {et~al.}(2008)\citenamefont
  {Volfson}, \citenamefont {Cookson}, \citenamefont {Hasty},\ and\
  \citenamefont {Tsimring}}]{Volfson2008}%
  \BibitemOpen
  \bibfield  {author} {\bibinfo {author} {\bibfnamefont {D.}~\bibnamefont
  {Volfson}}, \bibinfo {author} {\bibfnamefont {S.}~\bibnamefont {Cookson}},
  \bibinfo {author} {\bibfnamefont {J.}~\bibnamefont {Hasty}}, \ and\ \bibinfo
  {author} {\bibfnamefont {L.~S.}\ \bibnamefont {Tsimring}},\ }\bibfield
  {title} {\enquote {\bibinfo {title} {Biomechanical ordering of dense cell
  populations},}\ }\href {\doibase 10.1073/pnas.0706805105} {\bibfield
  {journal} {\bibinfo  {journal} {Proc. Natl. Acad. Sci. USA}\ }\textbf
  {\bibinfo {volume} {105}},\ \bibinfo {pages} {15346--15351} (\bibinfo {year}
  {2008})}\BibitemShut {NoStop}%
\bibitem [{\citenamefont {Boyer}\ \emph {et~al.}(2011)\citenamefont {Boyer},
  \citenamefont {Mather}, \citenamefont {Mondrag\'on-Palomino}, \citenamefont
  {Orozco-Fuentes}, \citenamefont {Danino}, \citenamefont {Hasty},\ and\
  \citenamefont {Tsimring}}]{Boyer2011}%
  \BibitemOpen
  \bibfield  {author} {\bibinfo {author} {\bibfnamefont {D.}~\bibnamefont
  {Boyer}}, \bibinfo {author} {\bibfnamefont {W.}~\bibnamefont {Mather}},
  \bibinfo {author} {\bibfnamefont {O.}~\bibnamefont {Mondrag\'on-Palomino}},
  \bibinfo {author} {\bibfnamefont {S.}~\bibnamefont {Orozco-Fuentes}},
  \bibinfo {author} {\bibfnamefont {T.}~\bibnamefont {Danino}}, \bibinfo
  {author} {\bibfnamefont {J.}~\bibnamefont {Hasty}}, \ and\ \bibinfo {author}
  {\bibfnamefont {L.~S.}\ \bibnamefont {Tsimring}},\ }\bibfield  {title}
  {\enquote {\bibinfo {title} {Buckling instability in ordered bacterial
  colonies},}\ }\href {\doibase 10.1088/1478-3975/8/2/026008} {\bibfield
  {journal} {\bibinfo  {journal} {Phys. Biol.}\ }\textbf {\bibinfo {volume}
  {8}},\ \bibinfo {pages} {026008} (\bibinfo {year} {2011})}\BibitemShut
  {NoStop}%
\bibitem [{\citenamefont {van Holthe~tot Echten}\ \emph {et~al.}()\citenamefont
  {van Holthe~tot Echten}, \citenamefont {Nordemann}, \citenamefont {Wehrens},
  \citenamefont {Tans},\ and\ \citenamefont {Idema}}]{Echten2020}%
  \BibitemOpen
  \bibfield  {author} {\bibinfo {author} {\bibfnamefont {D.}~\bibnamefont {van
  Holthe~tot Echten}}, \bibinfo {author} {\bibfnamefont {G.}~\bibnamefont
  {Nordemann}}, \bibinfo {author} {\bibfnamefont {M.}~\bibnamefont {Wehrens}},
  \bibinfo {author} {\bibfnamefont {S.}~\bibnamefont {Tans}}, \ and\ \bibinfo
  {author} {\bibfnamefont {T.}~\bibnamefont {Idema}},\ }\bibfield  {title}
  {\enquote {\bibinfo {title} {Defect dynamics in growing bacterial
  colonies},}\ }\href {https://arxiv.org/abs/2003.10509} {\bibinfo  {journal}
  {arXiv:2003.10509}\ }\BibitemShut {NoStop}%
\bibitem [{\citenamefont {Epstein}\ \emph {et~al.}(2011)\citenamefont
  {Epstein}, \citenamefont {Pokroy}, \citenamefont {Seminara},\ and\
  \citenamefont {Aizenberg}}]{Epstein2011}%
  \BibitemOpen
\bibfield  {journal} {  }\bibfield  {author} {\bibinfo {author} {\bibfnamefont
  {A.~K.}\ \bibnamefont {Epstein}}, \bibinfo {author} {\bibfnamefont
  {B.}~\bibnamefont {Pokroy}}, \bibinfo {author} {\bibfnamefont
  {A.}~\bibnamefont {Seminara}}, \ and\ \bibinfo {author} {\bibfnamefont
  {J.}~\bibnamefont {Aizenberg}},\ }\bibfield  {title} {\enquote {\bibinfo
  {title} {Bacterial biofilm shows persistent resistance to liquid wetting and
  gas penetration},}\ }\href {\doibase 10.1073/pnas.1011033108} {\bibfield
  {journal} {\bibinfo  {journal} {Proc. Natl. Acad. Sci. USA}\ }\textbf
  {\bibinfo {volume} {108}},\ \bibinfo {pages} {995--1000} (\bibinfo {year}
  {2011})}\BibitemShut {NoStop}%
\bibitem [{\citenamefont {Trejo}\ \emph {et~al.}(2013)\citenamefont {Trejo},
  \citenamefont {Douarche}, \citenamefont {Bailleux}, \citenamefont {Poulard},
  \citenamefont {Mariot}, \citenamefont {Regeard},\ and\ \citenamefont
  {Raspaud}}]{Trejo2011}%
  \BibitemOpen
  \bibfield  {author} {\bibinfo {author} {\bibfnamefont {M.}~\bibnamefont
  {Trejo}}, \bibinfo {author} {\bibfnamefont {C.}~\bibnamefont {Douarche}},
  \bibinfo {author} {\bibfnamefont {V.}~\bibnamefont {Bailleux}}, \bibinfo
  {author} {\bibfnamefont {C.}~\bibnamefont {Poulard}}, \bibinfo {author}
  {\bibfnamefont {S.}~\bibnamefont {Mariot}}, \bibinfo {author} {\bibfnamefont
  {C.}~\bibnamefont {Regeard}}, \ and\ \bibinfo {author} {\bibfnamefont
  {E.}~\bibnamefont {Raspaud}},\ }\bibfield  {title} {\enquote {\bibinfo
  {title} {Elasticity and wrinkled morphology of bacillus subtilis
  pellicles},}\ }\href {\doibase 10.1073/pnas.1217178110} {\bibfield  {journal}
  {\bibinfo  {journal} {Proc. Natl. Acad. Sci. USA}\ }\textbf {\bibinfo
  {volume} {110}},\ \bibinfo {pages} {2011--2016} (\bibinfo {year}
  {2013})}\BibitemShut {NoStop}%
\bibitem [{\citenamefont {Werb}\ \emph {et~al.}(2017)\citenamefont {Werb},
  \citenamefont {Garc\'{i}a}, \citenamefont {Bach}, \citenamefont {Grumbein},
  \citenamefont {Sieber}, \citenamefont {Opitz},\ and\ \citenamefont
  {Lieleg}}]{Werb2017}%
  \BibitemOpen
  \bibfield  {author} {\bibinfo {author} {\bibfnamefont {M.}~\bibnamefont
  {Werb}}, \bibinfo {author} {\bibfnamefont {C.~F.}\ \bibnamefont
  {Garc\'{i}a}}, \bibinfo {author} {\bibfnamefont {N.~C.}\ \bibnamefont
  {Bach}}, \bibinfo {author} {\bibfnamefont {S.}~\bibnamefont {Grumbein}},
  \bibinfo {author} {\bibfnamefont {S.~A.}\ \bibnamefont {Sieber}}, \bibinfo
  {author} {\bibfnamefont {M.}~\bibnamefont {Opitz}}, \ and\ \bibinfo {author}
  {\bibfnamefont {O.}~\bibnamefont {Lieleg}},\ }\bibfield  {title} {\enquote
  {\bibinfo {title} {Surface topology affects wetting behavior of
  \textit{Bacillus subtilis} biofilms},}\ }\href {\doibase
  10.1038/s41522-017-0018-1} {\bibfield  {journal} {\bibinfo  {journal} {NPJ
  Biofilms Microbiomes}\ }\textbf {\bibinfo {volume} {3}},\ \bibinfo {pages}
  {11} (\bibinfo {year} {2017})}\BibitemShut {NoStop}%
\bibitem [{\citenamefont {Hayta}\ \emph {et~al.}(2021)\citenamefont {Hayta},
  \citenamefont {Rickert},\ and\ \citenamefont {Lieleg}}]{HAYTA2021}%
  \BibitemOpen
  \bibfield  {author} {\bibinfo {author} {\bibfnamefont {E.~N.}\ \bibnamefont
  {Hayta}}, \bibinfo {author} {\bibfnamefont {C.~A.}\ \bibnamefont {Rickert}},
  \ and\ \bibinfo {author} {\bibfnamefont {O.}~\bibnamefont {Lieleg}},\
  }\bibfield  {title} {\enquote {\bibinfo {title} {Topography quantifications
  allow for identifying the contribution of parental strains to physical
  properties of co-cultured biofilms},}\ }\href {\doibase
  10.1016/j.bioflm.2021.100044} {\bibfield  {journal} {\bibinfo  {journal}
  {Biofilm}\ }\textbf {\bibinfo {volume} {3}},\ \bibinfo {pages} {100044}
  (\bibinfo {year} {2021})}\BibitemShut {NoStop}%
\bibitem [{\citenamefont {Zabiegaj}\ \emph {et~al.}(2021)\citenamefont
  {Zabiegaj}, \citenamefont {Hajirasouliha}, \citenamefont {Duilio},
  \citenamefont {Guido}, \citenamefont {Caserta}, \citenamefont {Kostoglou},
  \citenamefont {Petala}, \citenamefont {Karapantsios},\ and\ \citenamefont
  {Trybala}}]{ZABIEGAJ2021}%
  \BibitemOpen
  \bibfield  {author} {\bibinfo {author} {\bibfnamefont {D.}~\bibnamefont
  {Zabiegaj}}, \bibinfo {author} {\bibfnamefont {F.}~\bibnamefont
  {Hajirasouliha}}, \bibinfo {author} {\bibfnamefont {A.}~\bibnamefont
  {Duilio}}, \bibinfo {author} {\bibfnamefont {S.}~\bibnamefont {Guido}},
  \bibinfo {author} {\bibfnamefont {S.}~\bibnamefont {Caserta}}, \bibinfo
  {author} {\bibfnamefont {M.}~\bibnamefont {Kostoglou}}, \bibinfo {author}
  {\bibfnamefont {M.}~\bibnamefont {Petala}}, \bibinfo {author} {\bibfnamefont
  {T.}~\bibnamefont {Karapantsios}}, \ and\ \bibinfo {author} {\bibfnamefont
  {A.}~\bibnamefont {Trybala}},\ }\bibfield  {title} {\enquote {\bibinfo
  {title} {Wetting/spreading on porous media and on deformable, soluble
  structured substrates as a model system for studying the effect of morphology
  on biofilms wetting and for assessing anti-biofilm methods},}\ }\href
  {\doibase 10.1016/j.cocis.2021.101426} {\bibfield  {journal} {\bibinfo
  {journal} {Curr. Opin. Colloid Interface Sci.}\ }\textbf {\bibinfo {volume}
  {53}},\ \bibinfo {pages} {101426} (\bibinfo {year} {2021})}\BibitemShut
  {NoStop}%
\bibitem [{\citenamefont {Nejad}\ and\ \citenamefont
  {Yeomans}(2022)}]{Nejad.Yeomans-PRL2022}%
  \BibitemOpen
  \bibfield  {author} {\bibinfo {author} {\bibfnamefont {Mehrana~R.}\
  \bibnamefont {Nejad}}\ and\ \bibinfo {author} {\bibfnamefont {Julia~M.}\
  \bibnamefont {Yeomans}},\ }\bibfield  {title} {\enquote {\bibinfo {title}
  {Active extensile stress promotes 3d director orientations and flows},}\
  }\href {\doibase 10.1103/PhysRevLett.128.048001} {\bibfield  {journal}
  {\bibinfo  {journal} {Phys. Rev. Lett.}\ }\textbf {\bibinfo {volume} {128}},\
  \bibinfo {pages} {048001} (\bibinfo {year} {2022})}\BibitemShut {NoStop}%
\bibitem [{\citenamefont {Sveen}(2004)}]{MatPIV}%
  \BibitemOpen
  \bibfield  {author} {\bibinfo {author} {\bibfnamefont {J.~K.}\ \bibnamefont
  {Sveen}},\ }\href {https://www.duo.uio.no/handle/10852/10196} {\enquote
  {\bibinfo {title} {An introduction to {MatPIV} 1.6.1},}\ } (\bibinfo {year}
  {2004}),\ \bibinfo {note} {eprint no. 2, ISSN 0809-4403, Dept. of
  Mathematics, University of Oslo.}\BibitemShut {Stop}%
\bibitem [{\citenamefont {Pismen}(2006)}]{Prismen2006}%
  \BibitemOpen
  \bibfield  {author} {\bibinfo {author} {\bibfnamefont {L.~M.}\ \bibnamefont
  {Pismen}},\ }\href@noop {} {\emph {\bibinfo {title} {Patterns and interfaces
  in dissipative dynamics}}}\ (\bibinfo  {publisher} {Springer},\ \bibinfo
  {year} {2006})\BibitemShut {NoStop}%
\bibitem [{\citenamefont {Pismen}(1999)}]{Prismen1999}%
  \BibitemOpen
  \bibfield  {author} {\bibinfo {author} {\bibfnamefont {L.~M.}\ \bibnamefont
  {Pismen}},\ }\href@noop {} {\emph {\bibinfo {title} {Vortices in Nonlinear
  Fields. From Liquid Crystals to Superfluids. From Non-Equilibrium Patterns to
  Cosmic Strings}}}\ (\bibinfo  {publisher} {Oxford University Press},\
  \bibinfo {year} {1999})\BibitemShut {NoStop}%
\end{thebibliography}%


\providecommand{\noopsort}[1]{}\providecommand{\singleletter}[1]{#1}%
\begin{thebibliography}{12}%
\makeatletter
\providecommand \@ifxundefined [1]{%
 \@ifx{#1\undefined}
}%
\providecommand \@ifnum [1]{%
 \ifnum #1\expandafter \@firstoftwo
 \else \expandafter \@secondoftwo
 \fi
}%
\providecommand \@ifx [1]{%
 \ifx #1\expandafter \@firstoftwo
 \else \expandafter \@secondoftwo
 \fi
}%
\providecommand \natexlab [1]{#1}%
\providecommand \enquote  [1]{``#1''}%
\providecommand \bibnamefont  [1]{#1}%
\providecommand \bibfnamefont [1]{#1}%
\providecommand \citenamefont [1]{#1}%
\providecommand \href@noop [0]{\@secondoftwo}%
\providecommand \href [0]{\begingroup \@sanitize@url \@href}%
\providecommand \@href[1]{\@@startlink{#1}\@@href}%
\providecommand \@@href[1]{\endgroup#1\@@endlink}%
\providecommand \@sanitize@url [0]{\catcode `\\12\catcode `\$12\catcode
  `\&12\catcode `\#12\catcode `\^12\catcode `\_12\catcode `\%12\relax}%
\providecommand \@@startlink[1]{}%
\providecommand \@@endlink[0]{}%
\providecommand \url  [0]{\begingroup\@sanitize@url \@url }%
\providecommand \@url [1]{\endgroup\@href {#1}{\urlprefix }}%
\providecommand \urlprefix  [0]{URL }%
\providecommand \Eprint [0]{\href }%
\providecommand \doibase [0]{http://dx.doi.org/}%
\providecommand \selectlanguage [0]{\@gobble}%
\providecommand \bibinfo  [0]{\@secondoftwo}%
\providecommand \bibfield  [0]{\@secondoftwo}%
\providecommand \translation [1]{[#1]}%
\providecommand \BibitemOpen [0]{}%
\providecommand \bibitemStop [0]{}%
\providecommand \bibitemNoStop [0]{.\EOS\space}%
\providecommand \EOS [0]{\spacefactor3000\relax}%
\providecommand \BibitemShut  [1]{\csname bibitem#1\endcsname}%
\let\auto@bib@innerbib\@empty
\bibitem [{\citenamefont {Kawaguchi}\ \emph {et~al.}(2017)\citenamefont
  {Kawaguchi}, \citenamefont {Kageyama},\ and\ \citenamefont
  {Sano}}]{Kawaguchi2017}%
  \BibitemOpen
  \bibfield  {author} {\bibinfo {author} {\bibfnamefont {K.}~\bibnamefont
  {Kawaguchi}}, \bibinfo {author} {\bibfnamefont {R.}~\bibnamefont {Kageyama}},
  \ and\ \bibinfo {author} {\bibfnamefont {M.}~\bibnamefont {Sano}},\ }\href
  {\doibase 10.1038/nature22321} {\bibfield  {journal} {\bibinfo  {journal}
  {Nature}\ }\textbf {\bibinfo {volume} {545}},\ \bibinfo {pages} {327}
  (\bibinfo {year} {2017})}\BibitemShut {NoStop}%
\bibitem [{\citenamefont {Copenhagen}\ \emph {et~al.}(2021)\citenamefont
  {Copenhagen}, \citenamefont {Alert}, \citenamefont {Wingreen},\ and\
  \citenamefont {Shaevitz}}]{Copenhagen2021}%
  \BibitemOpen
  \bibfield  {author} {\bibinfo {author} {\bibfnamefont {K.}~\bibnamefont
  {Copenhagen}}, \bibinfo {author} {\bibfnamefont {R.}~\bibnamefont {Alert}},
  \bibinfo {author} {\bibfnamefont {N.~S.}\ \bibnamefont {Wingreen}}, \ and\
  \bibinfo {author} {\bibfnamefont {J.~W.}\ \bibnamefont {Shaevitz}},\ }\href
  {\doibase 10.1038/s41567-020-01056-4} {\bibfield  {journal} {\bibinfo
  {journal} {Nat. Phys.}\ }\textbf {\bibinfo {volume} {17}},\ \bibinfo {pages}
  {211} (\bibinfo {year} {2021})}\BibitemShut {NoStop}%
\bibitem [{\citenamefont {Pismen}(2006)}]{Prismen2006}%
  \BibitemOpen
  \bibfield  {author} {\bibinfo {author} {\bibfnamefont {L.~M.}\ \bibnamefont
  {Pismen}},\ }\href@noop {} {\emph {\bibinfo {title} {Patterns and interfaces
  in dissipative dynamics}}}\ (\bibinfo  {publisher} {Springer},\ \bibinfo
  {year} {2006})\BibitemShut {NoStop}%
\bibitem [{\citenamefont {Pismen}(1999)}]{Prismen1999}%
  \BibitemOpen
  \bibfield  {author} {\bibinfo {author} {\bibfnamefont {L.~M.}\ \bibnamefont
  {Pismen}},\ }\href@noop {} {\emph {\bibinfo {title} {Vortices in Nonlinear
  Fields. From Liquid Crystals to Superfluids. From Non-Equilibrium Patterns to
  Cosmic Strings}}}\ (\bibinfo  {publisher} {Oxford University Press},\
  \bibinfo {year} {1999})\BibitemShut {NoStop}%
\bibitem [{\citenamefont {You}\ \emph {et~al.}(2018)\citenamefont {You},
  \citenamefont {Pearce}, \citenamefont {Sengupta},\ and\ \citenamefont
  {Giomi}}]{You2018}%
  \BibitemOpen
  \bibfield  {author} {\bibinfo {author} {\bibfnamefont {Z.}~\bibnamefont
  {You}}, \bibinfo {author} {\bibfnamefont {D.~J.~G.}\ \bibnamefont {Pearce}},
  \bibinfo {author} {\bibfnamefont {A.}~\bibnamefont {Sengupta}}, \ and\
  \bibinfo {author} {\bibfnamefont {L.}~\bibnamefont {Giomi}},\ }\href
  {\doibase 10.1103/PhysRevX.8.031065} {\bibfield  {journal} {\bibinfo
  {journal} {Phys. Rev. X}\ }\textbf {\bibinfo {volume} {8}},\ \bibinfo {pages}
  {031065} (\bibinfo {year} {2018})}\BibitemShut {NoStop}%
\bibitem [{\citenamefont {Grant}\ \emph {et~al.}(2014)\citenamefont {Grant},
  \citenamefont {Wac{\l}aw}, \citenamefont {Allen},\ and\ \citenamefont
  {Cicuta}}]{Grant2014}%
  \BibitemOpen
  \bibfield  {author} {\bibinfo {author} {\bibfnamefont {M.~A.~A.}\
  \bibnamefont {Grant}}, \bibinfo {author} {\bibfnamefont {B.}~\bibnamefont
  {Wac{\l}aw}}, \bibinfo {author} {\bibfnamefont {R.~J.}\ \bibnamefont
  {Allen}}, \ and\ \bibinfo {author} {\bibfnamefont {P.}~\bibnamefont
  {Cicuta}},\ }\href {\doibase 10.1098/rsif.2014.0400} {\bibfield  {journal}
  {\bibinfo  {journal} {J. R. Soc. Interface}\ }\textbf {\bibinfo {volume}
  {11}},\ \bibinfo {pages} {20140400} (\bibinfo {year} {2014})}\BibitemShut
  {NoStop}%
\bibitem [{\citenamefont {You}\ \emph {et~al.}(2019)\citenamefont {You},
  \citenamefont {Pearce}, \citenamefont {Sengupta},\ and\ \citenamefont
  {Giomi}}]{You2019}%
  \BibitemOpen
  \bibfield  {author} {\bibinfo {author} {\bibfnamefont {Z.}~\bibnamefont
  {You}}, \bibinfo {author} {\bibfnamefont {D.~J.~G.}\ \bibnamefont {Pearce}},
  \bibinfo {author} {\bibfnamefont {A.}~\bibnamefont {Sengupta}}, \ and\
  \bibinfo {author} {\bibfnamefont {L.}~\bibnamefont {Giomi}},\ }\href
  {\doibase 10.1103/PhysRevLett.123.178001} {\bibfield  {journal} {\bibinfo
  {journal} {Phys. Rev. Lett.}\ }\textbf {\bibinfo {volume} {123}},\ \bibinfo
  {pages} {178001} (\bibinfo {year} {2019})}\BibitemShut {NoStop}%
\bibitem [{\citenamefont {Su}\ \emph {et~al.}(2012)\citenamefont {Su},
  \citenamefont {Liao}, \citenamefont {Roan}, \citenamefont {Wang},
  \citenamefont {Chiou},\ and\ \citenamefont {Syu}}]{Su2012}%
  \BibitemOpen
  \bibfield  {author} {\bibinfo {author} {\bibfnamefont {P.}~\bibnamefont
  {Su}}, \bibinfo {author} {\bibfnamefont {C.}~\bibnamefont {Liao}}, \bibinfo
  {author} {\bibfnamefont {J.}~\bibnamefont {Roan}}, \bibinfo {author}
  {\bibfnamefont {S.}~\bibnamefont {Wang}}, \bibinfo {author} {\bibfnamefont
  {A.}~\bibnamefont {Chiou}}, \ and\ \bibinfo {author} {\bibfnamefont
  {W.}~\bibnamefont {Syu}},\ }\href {\doibase 10.1371/journal.pone.0048098}
  {\bibfield  {journal} {\bibinfo  {journal} {PLOS ONE}\ }\textbf {\bibinfo
  {volume} {7}},\ \bibinfo {pages} {1} (\bibinfo {year} {2012})}\BibitemShut
  {NoStop}%
\bibitem [{\citenamefont {Dell’Arciprete}\ \emph {et~al.}(2018)\citenamefont
  {Dell’Arciprete}, \citenamefont {Blow}, \citenamefont {Brown},
  \citenamefont {Farrell}, \citenamefont {Lintuvuori}, \citenamefont {McVey},
  \citenamefont {Marenduzzo},\ and\ \citenamefont {Poon}}]{Dell2019}%
  \BibitemOpen
  \bibfield  {author} {\bibinfo {author} {\bibfnamefont {D.}~\bibnamefont
  {Dell’Arciprete}}, \bibinfo {author} {\bibfnamefont {M.~L.}\ \bibnamefont
  {Blow}}, \bibinfo {author} {\bibfnamefont {A.~T.}\ \bibnamefont {Brown}},
  \bibinfo {author} {\bibfnamefont {F.~D.~C.}\ \bibnamefont {Farrell}},
  \bibinfo {author} {\bibfnamefont {J.~S.}\ \bibnamefont {Lintuvuori}},
  \bibinfo {author} {\bibfnamefont {A.~F.}\ \bibnamefont {McVey}}, \bibinfo
  {author} {\bibfnamefont {D.}~\bibnamefont {Marenduzzo}}, \ and\ \bibinfo
  {author} {\bibfnamefont {W.~C.~K.}\ \bibnamefont {Poon}},\ }\href {\doibase
  10.1038/s41467-018-06370-3} {\bibfield  {journal} {\bibinfo  {journal} {Nat.
  Commun.}\ }\textbf {\bibinfo {volume} {9}},\ \bibinfo {pages} {4190}
  (\bibinfo {year} {2018})}\BibitemShut {NoStop}%
\bibitem [{\citenamefont {Volfson}\ \emph {et~al.}(2008)\citenamefont
  {Volfson}, \citenamefont {Cookson}, \citenamefont {Hasty},\ and\
  \citenamefont {Tsimring}}]{Volfson2008}%
  \BibitemOpen
  \bibfield  {author} {\bibinfo {author} {\bibfnamefont {D.}~\bibnamefont
  {Volfson}}, \bibinfo {author} {\bibfnamefont {S.}~\bibnamefont {Cookson}},
  \bibinfo {author} {\bibfnamefont {J.}~\bibnamefont {Hasty}}, \ and\ \bibinfo
  {author} {\bibfnamefont {L.~S.}\ \bibnamefont {Tsimring}},\ }\href {\doibase
  10.1073/pnas.0706805105} {\bibfield  {journal} {\bibinfo  {journal} {Proc.
  Natl. Acad. Sci. USA}\ }\textbf {\bibinfo {volume} {105}},\ \bibinfo {pages}
  {15346} (\bibinfo {year} {2008})}\BibitemShut {NoStop}%
\bibitem [{\citenamefont {Boyer}\ \emph {et~al.}(2011)\citenamefont {Boyer},
  \citenamefont {Mather}, \citenamefont {Mondrag\'on-Palomino}, \citenamefont
  {Orozco-Fuentes}, \citenamefont {Danino}, \citenamefont {Hasty},\ and\
  \citenamefont {Tsimring}}]{Boyer2011}%
  \BibitemOpen
  \bibfield  {author} {\bibinfo {author} {\bibfnamefont {D.}~\bibnamefont
  {Boyer}}, \bibinfo {author} {\bibfnamefont {W.}~\bibnamefont {Mather}},
  \bibinfo {author} {\bibfnamefont {O.}~\bibnamefont {Mondrag\'on-Palomino}},
  \bibinfo {author} {\bibfnamefont {S.}~\bibnamefont {Orozco-Fuentes}},
  \bibinfo {author} {\bibfnamefont {T.}~\bibnamefont {Danino}}, \bibinfo
  {author} {\bibfnamefont {J.}~\bibnamefont {Hasty}}, \ and\ \bibinfo {author}
  {\bibfnamefont {L.~S.}\ \bibnamefont {Tsimring}},\ }\href {\doibase
  10.1088/1478-3975/8/2/026008} {\bibfield  {journal} {\bibinfo  {journal}
  {Phys. Biol.}\ }\textbf {\bibinfo {volume} {8}},\ \bibinfo {pages} {026008}
  (\bibinfo {year} {2011})}\BibitemShut {NoStop}%
\bibitem [{\citenamefont {van Holthe~tot Echten}\ \emph {et~al.}()\citenamefont
  {van Holthe~tot Echten}, \citenamefont {Nordemann}, \citenamefont {Wehrens},
  \citenamefont {Tans},\ and\ \citenamefont {Idema}}]{Echten2020}%
  \BibitemOpen
  \bibfield  {author} {\bibinfo {author} {\bibfnamefont {D.}~\bibnamefont {van
  Holthe~tot Echten}}, \bibinfo {author} {\bibfnamefont {G.}~\bibnamefont
  {Nordemann}}, \bibinfo {author} {\bibfnamefont {M.}~\bibnamefont {Wehrens}},
  \bibinfo {author} {\bibfnamefont {S.}~\bibnamefont {Tans}}, \ and\ \bibinfo
  {author} {\bibfnamefont {T.}~\bibnamefont {Idema}},\ }\href
  {https://arxiv.org/abs/2003.10509} {\bibinfo  {journal} {arXiv:2003.10509}\
  }\BibitemShut {NoStop}%
\end{thebibliography}%

\end{document}


\title{Supplementary Information for\\``Tilt-induced polar order and topological defects in growing bacterial populations''}

\author{Takuro Shimaya}
\email{t.shimaya@noneq.phys.s.u-tokyo.ac.jp}
\affiliation{Department of Physics, The University of Tokyo, 7-3-1 Hongo, Bunkyo-ku, Tokyo, 113-0033, Japan.}%
\author{Kazumasa A. Takeuchi}
\email{kat@kaztake.org}
\affiliation{Department of Physics, The University of Tokyo, 7-3-1 Hongo, Bunkyo-ku, Tokyo, 113-0033, Japan.}%

\maketitle

\section{mean radial velocity without cell tilting and proliferation}
\label{sec:original}
Here we derive the analytical solution of the mean radial velocity $\bar{v}^\pm_r(r)$ for the case without three-dimensional tilting of the cells and cell proliferation.
This case is described by Eq.~(1), which reads
\begin{equation}
    \bm{\xi v} = \nabla\cdot(-a_\mathrm{n}\bm{Q}),
    \label{eq:Sf}
\end{equation}
where the friction tensor is $\bm{\xi} = \xi_0(\bm{1} - \epsilon \bm{Q})$ with the friction anisotropy $\epsilon$ ($0<\epsilon<1$).
Around a $\pm 1/2$ defect, we assume the simplest director field
\begin{equation}
    \bm{n}^\pm(r,\phi) = \begin{pmatrix} \cos(\pm\phi/2) \\ \sin(\pm\phi/2) \end{pmatrix},
\end{equation}
where $(r,\phi)$ is the two-dimensional polar coordinate centered at the defect core.
Consequently, the nematic order tensor $\bm{Q}^\pm(\bm{r})$ is given by
\begin{equation}
       \bm{Q^\pm}(r,\phi) = S(r)
              \begin{pmatrix}
                \cos(\pm\phi), & \sin(\pm\phi)  \\
                \sin(\pm\phi), & -\cos(\pm\phi) 
              \end{pmatrix},  \label{eq:Sg}
\end{equation}
 with a function $S(r)$, which is the scalar nematic order parameter.
Under the assumption that the active stress coefficient $a_\mathrm{n}$ is constant, following the previous works \cite{Kawaguchi2017, Copenhagen2021}, we obtain the velocity $\bm{v}^\pm(r,\phi)$ as follows:
\begin{align}
    \bm{v}^+(r,\phi) & = -\frac{a_\mathrm{n}}{\xi_0}G^+(r) \[\begin{pmatrix} 1 \\ 0 \end{pmatrix} + \epsilon S(r)\begin{pmatrix} \cos\phi \\ \sin\phi \end{pmatrix} \], \label{eq:Sa} \\
    \bm{v}^-(r,\phi) & = -\frac{a_\mathrm{n}}{\xi_0}G^-(r) \[\begin{pmatrix} \cos2\phi \\ -\sin2\phi \end{pmatrix} + \epsilon S(r)\begin{pmatrix} \cos\phi \\ \sin\phi \end{pmatrix} \], \label{eq:Sb}
\end{align}
with 
\begin{equation}
    G^\pm(r) = \frac{S'(r)\pm S(r)/r}{1-\epsilon^2S(r)^2},
    \label{eq:Sc}
\end{equation}
where $S'(r) = \diff{S}{r}$.
Then the mean radial velocity $\bar{v}^\pm_r(r)$ can be calculated as
\begin{equation}
    \bar{v}^\pm_r(r) = \frac{1}{2\pi} \int^{2\pi}_0 d\phi\ \begin{pmatrix} \cos\phi \\ \sin\phi \end{pmatrix}\cdot\bm{v}^\pm(r,\phi) = -\epsilon\frac{a_\mathrm{n}}{\xi_0} S(r) \frac{S'(r)\pm S(r)/r}{1-\epsilon^2S(r)^2}. \label{eq:Se}
\end{equation}
Since the parameters satisfy $a_\mathrm{n}>0$, $\xi_0>0$, $0<\epsilon<1$, and $0 \leq S \leq 1$, the direction of the mean radial velocity around the defect is determined by $\sign[-(S'(r)\pm S(r)/r)]$.

To proceed, we need to determine $S(r)$.
Here we use the following Pad\'e approximant due to earlier studies \cite{Copenhagen2021,Prismen2006,Prismen1999}, obtained by assuming that $\bm{Q}^\pm(r,\phi)$ minimizes the nematic free energy:
\begin{equation}
       S(r) = S_0F(r/r_S), \qquad F(x)\simeq x\sqrt{\frac{0.34+0.07x^2}{1+0.41x^2+0.07x^4}} \label{eq:S16},
\end{equation}
with the defect core radius $r_S$ and $S_0=S(\infty)$.
Then, $\sign[\bar{v}_r^\pm(r)] = -\sign[S'(r)\pm S(r)/r] = -\sign[F'(x)\pm F(x)/x]$ and we can show that it is negative for $+1/2$ defects and positive for $-1/2$ defects, for all $r>0$, as previously known.
In Fig.\,2(h), by the dotted lines, we showed $\bar{v}_r^\pm(r)$ for $\epsilon S_0=0.25, a_\mathrm{n}S_0/\xi_0=0.055\unit{\mu{}m^2/min}, r_S = 1.2\unit{\mu{}m}$.

\section{Theory with cell proliferation}

Here we show that cell proliferation does not change the direction of the mean radial velocity around defects.
Following ref.\cite{You2018}, we start from the following hydrodynamic equations:
\begin{align}
    \frac{D\rho}{Dt} & = \lambda \rho + \mathcal{D}\nabla^2 \rho, \label{eq:S1} \\
    \frac{D(\rho\bm{v})}{Dt} & = \nabla\cdot \bm{\sigma} - \bm{\xi} \bm{v}, \label{eq:S2}
\end{align}
with the material derivative $D/Dt = \partial_t + \bm{v}\cdot\nabla+(\nabla\cdot\bm{v})$, the cell density $\rho$, the proliferation (growth) rate $\lambda$ and the diffusion constant $\mathcal{D}$.
The material derivative of the momentum density $\rho\bm{v}$ can be expressed as
\begin{align}
    \frac{D(\rho\bm{v})}{Dt} & = \frac{\partial(\rho\bm{v})}{\partial t} + (\bm{v}\cdot\nabla)(\rho\bm{v}) + (\nabla\cdot\bm{v})\rho\bm{v} \notag \\
    & = \rho\frac{\partial\bm{v}}{\partial t} + \bm{v}\frac{\partial \rho}{\partial t} + \rho(\bm{v}\cdot \nabla)\bm{v} + \bm{v}(\bm{v}\cdot \nabla)\rho + (\nabla\cdot\bm{v})\rho\bm{v} \notag \\
    & = \rho\frac{\partial\bm{v}}{\partial t} + \rho(\bm{v}\cdot \nabla)\bm{v} + \bm{v}\frac{D\rho}{Dt}. \label{eq:S3}
\end{align}
In the overdamped and low Reynolds number limit, the first and second terms in the right-hand side can be neglected.
Combining with Eqs.~(\ref{eq:S1}) and (\ref{eq:S2}) , we obtain
\begin{equation}
    \bm{v}(\lambda\rho + \mathcal{D}\nabla^2\rho) = \nabla\cdot\bm{\sigma}-\bm{\xi}\bm{v}.
    \label{eq:S4}
\end{equation}
Assuming that the diffusion of the density is weak compared to the cell growth, \eqref{eq:S4} can be rewritten as
\begin{equation}
    \lambda\rho\bm{v} = \nabla\cdot\bm{\sigma}-\bm{\xi}\bm{v}.
    \label{eq:S5}
\end{equation}
With the active stress field $\bm{\sigma}=-a_\mathrm{n}\bm{Q}$ and the friction $\bm{\xi} = \xi_0(\bm{1} - \epsilon \bm{Q})$, we obtain the following equation for the velocity field $\bm{v}^\pm(r,\phi)$ around a $\pm 1/2$ topological defect:
\begin{equation}
    \bm{v} = [(\xi_0+\lambda\rho)\bm{1} - \epsilon \xi_0 \bm{Q}]^{-1}\nabla\cdot(-a_\mathrm{n}\bm{Q}).
    \label{eq:S6}
\end{equation}
Assuming that $a_\mathrm{n}$ is constant, and using \eqref{eq:Sg} for the nematic order tensor, we obtain the following expression for the velocity field $\bm{v}^\pm(r,\phi)$ around a $\pm 1/2$ topological defect:
\begin{align}
    \bm{v}^+(r,\phi) & = -\frac{a_n}{\xi_0} H^+(r) \[ (1+\lambda\rho/\xi_0) \begin{pmatrix} 1 \\ 0 \end{pmatrix} + \epsilon S(r) \begin{pmatrix} \cos\phi \\ \sin\phi \end{pmatrix} \] \label{eq:S7} \\
    \bm{v}^-(r,\phi) & = -\frac{a_n}{\xi_0} H^-(r) \[ (1+\lambda\rho/\xi_0) \begin{pmatrix} \cos2\phi \\ -\sin2\phi \end{pmatrix} + \epsilon S(r) \begin{pmatrix} \cos\phi \\ \sin\phi \end{pmatrix} \] \label{eq:S8}
\end{align}
with
\begin{equation}
    H^\pm(r) = \frac{S'(r)\pm S(r)/r}{(1+\lambda\rho/\xi_0)^2 - \epsilon^2S(r)^2}.
    \label{eq:S9}
\end{equation}
Then we obtain the mean radial velocity $\bar{v}^\pm_r$ as follows:
\begin{equation}
    \bar{v}^\pm_r(r) = \frac{1}{2\pi} \int^{2\pi}_0 d\phi\ \begin{pmatrix} \cos\phi \\ \sin\phi \end{pmatrix}\cdot\bm{v}^\pm(r,\phi) = -\epsilon\frac{a_\mathrm{n}}{\xi_0} S(r) \frac{S'(r)\pm S(r)/r}{(1+\lambda\rho/\xi_0)^2 - \epsilon^2S(r)^2}.
    \label{eq:S10}
\end{equation}
Comparing with \eqref{eq:Se}, \eqref{eq:S10} changed the denominator only, which is still positive because $\lambda, \rho > 0$. Therefore, the direction of the mean radial velocity is still determined by $\sign[-(S'(r)\pm S(r)/r)]$ and the cell proliferation cannot change the direction of the mean radial velocity.

\section{Theory with non-uniform nematic tilting}

\label{sec:tilting}
Here we show that non-uniform nematic tilting, i.e., non-uniform $\thetan(r,\phi)$, can realize the influx toward $-1/2$ defects without the 3D-induced polar order.
Assuming $a_\mathrm{n}(r,\phi) = a^0_\mathrm{n}\cos\thetan(r,\phi)$, we have the following force balance equation:
\begin{equation}
    \xi_0(\bm{1} - \epsilon_0 \cos\thetan \bm{Q}) \bm{v} = \nabla\cdot(-a^0_\mathrm{n}\cos\thetan\bm{Q}).
    \label{eq:S11}
\end{equation}
Concerning the functional form of $\thetan(r,\phi)$, our experimental data around $\pm 1/2$ defects (Fig.\,3) suggest that it hardly depends on $\phi$ and the sign of the defect, i.e., $\thetan^\pm(r,\phi) = \thetan(r)$, and
\begin{equation}
    \thetan(r)=\thetan^\infty + (\thetan^0-\thetan^\infty)\exp(-r^2/r_\theta^2),
    \label{eq:S12}
\end{equation}
with constants $\thetan^\infty, \thetan^0 (>\thetan^\infty), r_\theta$.
From this, one can intuitively see that the active stress coefficient $a_\mathrm{n}(r,\phi) = a^0_\mathrm{n}\cos\thetan(r,\phi)$ becomes smaller near the defect core, which may generate a force toward the defect. 

In the following, we analytically show the emergence of the influx toward $-1/2$ defects, due to the non-uniform $\thetan(r)$ given by \eqref{eq:S12}.
First, we can rewrite \eqref{eq:S11} as
\begin{equation}
    \xi_0(\bm{1}-\epsilon_0\bm{\tilde{Q}}^\pm)\bm{v^\pm} = \nabla\cdot(-a^0_\mathrm{n}\bm{\tilde{Q}^\pm}),
    \label{eq:S13}
\end{equation}
with
\begin{equation}
    \bm{\tilde{Q}}^\pm(r,\phi) = \tilde{S}(r)
        \begin{pmatrix}
          \cos(\pm\phi) & \sin(\pm\phi)  \\
          \sin(\pm\phi) & -\cos(\pm\phi) 
        \end{pmatrix},  \label{eq:S14}
\end{equation}
and $\tilde{S}(r)=S(r)\cos\thetan(r)$.
In other words, introducing the $r$-dependence to the active stress coefficient is equivalent to changing the scalar nematic order $S(r)$.
Therefore, we obtain the mean radial velocity $\bar{v}^\pm_r(r)$ in the same form as \eqref{eq:Se}:
\begin{equation}
    \bar{v}^\pm_r(r) = -\epsilon_0 \frac{a_\mathrm{n}}{\xi_0} \tilde{S}(r) \frac{\tilde{S}'(r)\pm \tilde{S}(r)/r}{1-{\epsilon_0}^2 \tilde{S}(r)^2}. \label{eq:S15}
\end{equation}
As we described in Sec.~\ref{sec:original}, the sign of the mean radial velocity is determined by $\sign[-(\tilde{S}'(r)\pm \tilde{S}(r)/r)]$, meaning that this direction is sensitive to the functional form of $\tilde{S}(r)$.

Now we evaluate the sign of the mean radial velocity $\bar{v}_r^-(r)$ for $-1/2$ defects. It is determined by $-(\tilde{S}'(r) - \tilde{S}(r)/r)$, where $\tilde{S}(r)=S(r)\cos\thetan(r)$ with $S(r)$ given by \eqref{eq:S16} and $\thetan(r)$ by \eqref{eq:S12}.
We first deal with the case without nematic tilting, i.e., $\thetan(r)=0$ and $\tilde{S}(r)=S(r)$.
Using $x \equiv r/r_S$, we have $-[S'(r)\pm S(r)/r] = -(S_0/r_S)[F'(x) \pm F(x)/x]$, with $F(x)$ given by \eqref{eq:S16} and
\begin{equation}
    F'(x) \simeq \frac{0.34 + 0.14 x^2 + 0.0049 x^4}{(1 + 0.41 x^2 + 0.07 x^4)^2} \(\frac{0.34+0.07x^2}{1+0.41x^2+0.07x^4}\)^{-1/2}.
\end{equation}
Therefore, for the $-1/2$ defects, we obtain
\begin{align}
   -\(S'(r) - \frac{S(r)}{r}\) &= -\frac{S_0}{r_S}\(F'(x) - \frac{F(x)}{x}\) \notag \\
   &\simeq \frac{S_0}{r_S} \frac{x^2(0.069 + 0.048 x^2 + 0.0049 x^4)}{(1 + 0.41 x^2 + 0.07 x^4)^2} \(\frac{0.34+0.07x^2}{1+0.41x^2+0.07x^4}\)^{-1/2}.
\end{align}
In particular, near the defect core, $x \ll 1$, we have
\begin{equation}
    -\(S'(r)- \frac{S(r)}{r}\) \simeq 0.12 \frac{S_0}{r_S} x^2 + \mathcal{O}(x^4) > 0. \qquad \(x = \frac{r}{r_S} \ll 1\) \label{eq:S18}
\end{equation}
Therefore, there arises outflux from $-1/2$ defects, as already known, in the case without nematic tilting.

In the presence of non-uniform nematic tilting, we replace $S(r)$ with $\tilde{S}(r)=S(r)\cos\thetan(r)$.
Since
\begin{equation}
    -\(\tilde{S}'(r)- \frac{\tilde{S}(r)}{r}\) = -\(S'(r)- \frac{S(r)}{r}\)\cos\thetan(r) + S(r)\thetan'(r)\sin\thetan(r) \label{eq:S19}
\end{equation}
and
\begin{equation}
    \thetan'(r) = -(\thetan^0-\thetan^\infty)\frac{2r}{r_\theta^2} \exp(-r^2/r_\theta^2), \label{eq:S20}
\end{equation}
we have, for $r \ll r_S$ and $r \ll r_\theta$,
\begin{equation}
    -\(S'(r)- \frac{S(r)}{r}\) \simeq 1.2\frac{S_0}{r_S}\[ 0.1 \cos\thetan^0 \(\frac{r}{r_S}\)^2 - (\thetan^0-\thetan^\infty) \sin\thetan^0 \(\frac{r}{r_\theta}\)^2 \].
\end{equation}
Therefore, its sign is determined by
\begin{equation}
    \sign[\bar{v}^-_r(r)] = \sign\[0.1\frac{\cos\thetan^0}{r_S^2} - (\thetan^0-\thetan^\infty)\frac{\sin\thetan^0}{r_\theta^2}\]. \qquad (r \ll r_S, r_\theta) \label{eq:S22}.
\end{equation}
This result shows that the influx toward $-1/2$ defects appears if the peak of the tilt angle is sufficiently high or sharp, i.e., if $\thetan^0-\thetan^\infty$ is sufficiently large or $r_\theta$ is sufficiently small.
This is demonstrated by our numerical evaluation with $r_S=1.2\unit{\mu{}m}$, $r_\theta=1\unit{\mu{}m}$, $\thetan^\infty=0.3$ and for various $\thetan^0$ ($0.3\leq \thetan^0 \leq 1.5$), shown in Fig.\,S8(c), where $\bar{v}^-_r(r)$ starts to show a negative minimum when $\thetan^0$ becomes sufficiently large.

Note, however, that the maximum strength of the influx toward $-1/2$ defects that we could achieve by the theory presented in this section, within a reasonable range of parameter values, was significantly weaker than the experimental observation. 
For example, the result with $\thetan^\infty=0.3$ and $\thetan^0=0.75$, which are comparable to the experimentally observed values in Fig.\,3(a), exhibits very little influx (Fig.\,2(h), a blue solid line).
This gap is cleared by taking into account the 3D-induced polar order (see the main text and Appendix B).

\section{Experiments on circular colonies formed from a few cells}
\label{sec:circular}

Here, we prepared cell suspensions at sufficiently low concentration to start with isolated cells, and observed the growth of circular colonies in the two-dimensional plane (Fig.\,S9(a)).
After a while, cells near the center of the colony were extruded from the bottom layer and pushed upward (Video~3), similarly to observations in previous studies \cite{Grant2014,You2019}.
We then collected and analyzed the frames just before the first extrusion events in 60 independent colonies.
To see whether cell extrusion is correlated with local cellular alignment, we measured the orientation of cells using the structure tensor method (Fig.\,S9(a)).
We thereby obtained, at each position $\bm{R}$, the local cell orientation $\bm{n}(\bm{R})$ and the coherency parameter $C(\bm{R})$ which quantifies the degree of the local nematic order ($C(\bm{R})=1$ for perfect alignment and $C(\bm{R})=0$ for random orientations).
We first obtained the distribution of $C(\bm{R})$ in the region at a given normalized distance $R/R_\mathrm{max}$ from the colony center, where $R$ is the radial distance and $R_\mathrm{max}$ is the approximate radius of each colony (Fig.\,S9(b) blue boxplot).
The result shows that $C(\bm{R})$ is constant near the center ($R/R_\mathrm{max}<0.65$) and relatively high at the edge of the colony, which may be because cells at the periphery tend to align tangentially \cite{Su2012,Dell2019}.
We also collected 60 sets of the position and the coherency at the location where the first extrusion occurred in each colony (Fig.\,S9(b), red symbols), and found higher extrusion frequency near the center as suggested by previous studies \cite{Grant2014,You2019}.
Since the coherency $C(\bm{R})$ is essentially uniform near the center ($R/R_\mathrm{max}<0.65$), we simply compare the distribution of $C(\bm{R})$ from all pixels in the region $R/R_\mathrm{max}<0.65$ and that from 60 extrusion events (Fig.\,S9(c) top and (d)).
The result does not show significant difference between the two distributions, indicating that local cell alignment is unlikely to affect the first extrusion.

In contrast, based on a theoretical argument on torque balance around cells, it was proposed that the first extrusion is more likely to happen to short cells \cite{You2019}.
We therefore compare the distributions of cell lengths (Fig.\,S9(c) bottom and (e)) and find that the extruded cells seem to be significantly shorter.
These results suggest that, compared to the cell length or the radial position, the local cellular alignment does not seem to play a major role in the first extrusion in circular colonies.
We also investigated the relation between the coherency and the cell length for each of the first extruded cells, and did not find significant correlation (Fig.\,S9(f)).
These suggests that, in such isolated expanding colonies, the in-plane stress which becomes strongest at the center due to cell growth \cite{Volfson2008, Boyer2011, You2019, Echten2020} indeed constitutes a major factor in the start of the three-dimensional growth, as reported earlier \cite{Grant2014,You2019}, regardless of topological defects. 
Conversely, in uniform colonies where cells fill the two-dimensional space homogeneously, it is reasonable to consider that the stress non-uniformity is reduced and this probably allowed us to see the intriguing role of topological defects in this case.

\clearpage
\section{Supplementary figures}

\begin{figure}[h]
    \centering
    \includegraphics[width=\textwidth]{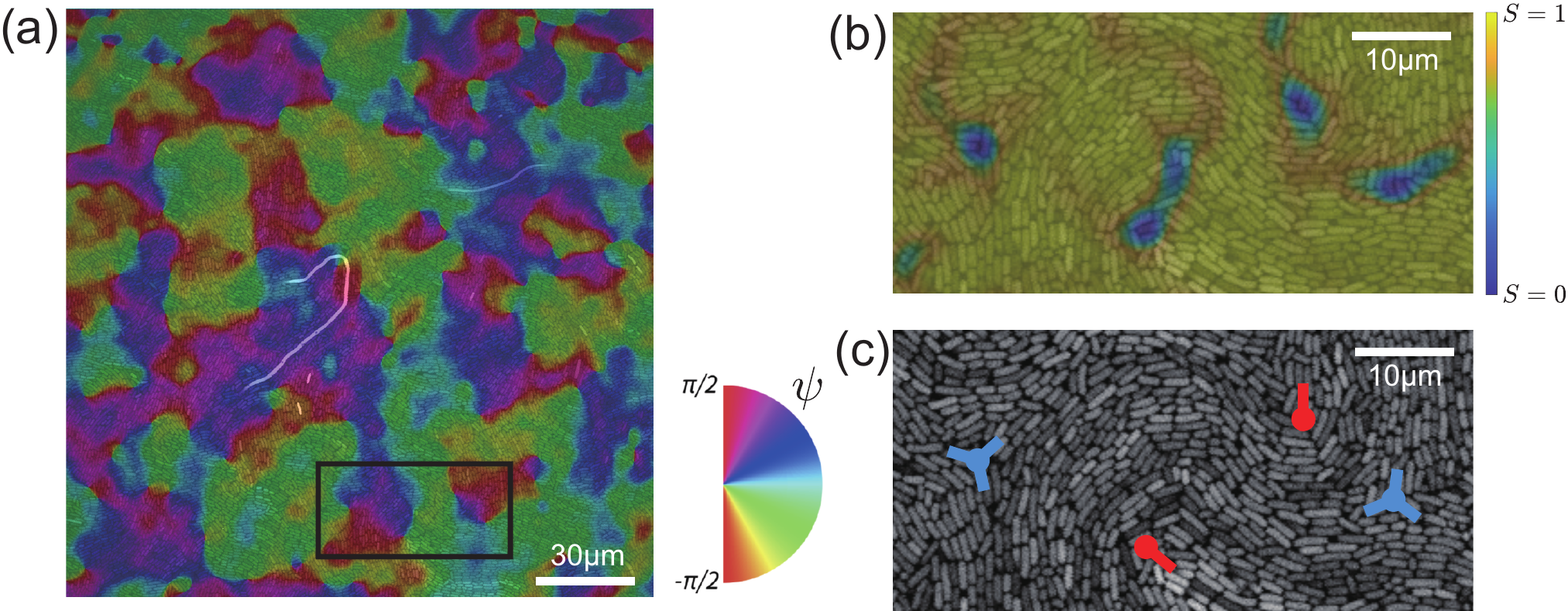}
    \caption{Analysis of the cell alignment from experimentally obtained images.
    The images shown here was obtained by end-point confocal observation.
    (a) The cell orientation angle $\psi(\bm{R})$ in the cross section at the bottom layer.
    (b) The local nematic order parameter $S(\bm{R})$, for the region depicted by the black square in (a).
    (c) Topological defects, detected automatically from (b) at the local minima of $S(\bm{R})$.
    Red comets and blue trefoils indicate $+1/2$ and $-1/2$ defects, respectively.
    The arms of the symbols reflect the structure of the director field as illustrated in the insets of Fig.\,1(d).
    }
\end{figure}

\begin{figure}[h]
    \centering
    \includegraphics[width=\textwidth]{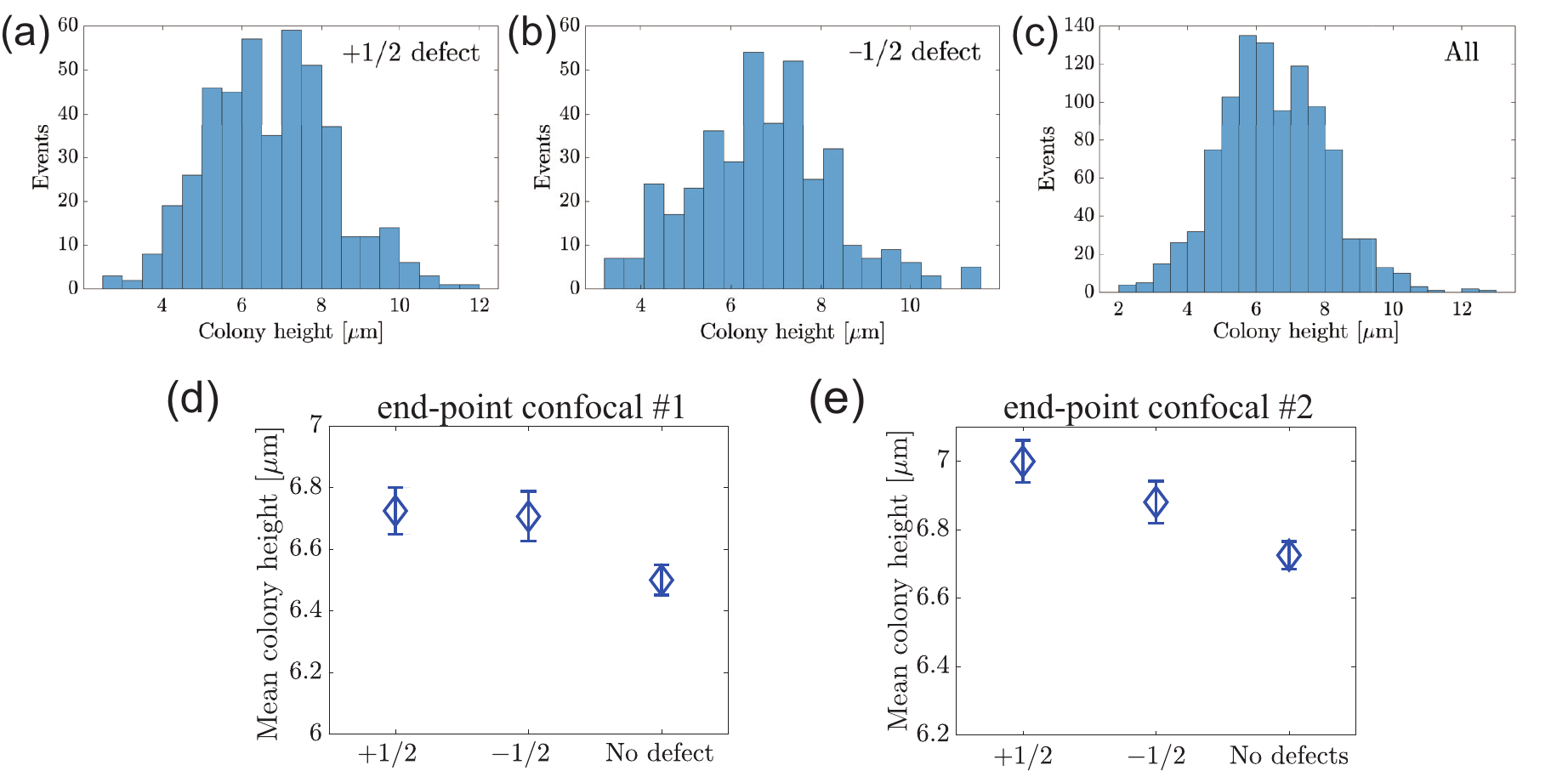}
    \caption{
    Statistical test on the relation between the local colony height and the presence of topological defects in the bottom layer.
    (a-c)
    Histograms of the local colony height above the $+1/2$ defects (a, sample size $N=437$), above the $-1/2$ defects (b, sample size $N=384$), and at positions randomly chosen from regions separated more than 9$\unit{\mu{}m}$ from defects (c, sample size $N=1000$).
    These data were obtained by a single measurement (end-point confocal \#{}1).
    (d,e) Mean colony height at the location of the defects in the bottom layer and that far from defects, from two different biological replicates (end-point confocal \#{}1 (d, same as Fig.\,1(d)) and \#{}2 (e)). At each position in the $xy$-plane, the height was evaluated by the length of the profile along the $z$-axis whose intensity is higher than 20\% of the maximum.
    The heights at the defect positions were extracted from all of the hundreds of defects that were sufficiently far ($> 9\unit{\mu m}$) from each other (sample size $N=437$ (d) and $N=458$ (e) for $+1/2$ defects, and $N=384$ (d) and $N=436$ (e) for $-1/2$ defects).
    For the colony heights far from defects, we randomly picked up 1000 points which were sufficiently far ($> 9\unit{\mu m}$) from any defect (see Appendix\,A in the main paper).
    The error bars indicate the standard error from the ensemble averaging.
    The optical resolution was about $250\unit{nm}$ in the vertical direction.
    The p-value of the hypothesis that the median of the distribution at the positions of the defects is identical to that far from defects was $0.018$ (d) and $4.6\times10^{-5}$ (e) for the $+1/2$ defects, and $0.043$ (d) and $0.0033$ (e) for the $-1/2$ defects.
}
\end{figure}

\begin{figure}[h]
    \centering
    \includegraphics[width=0.8\textwidth]{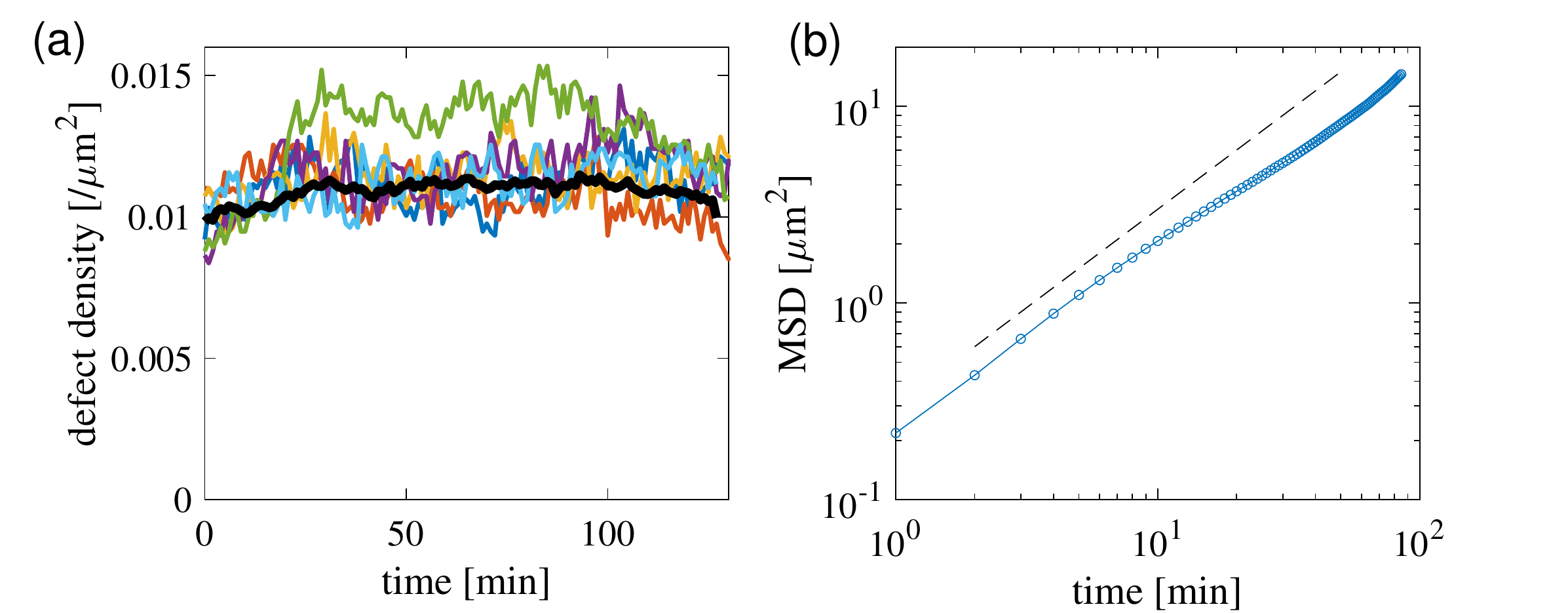}
    \caption{
    Time evolution of topological defects in uniform colonies formed from numerous cells, observed by a single phase-contrast measurement (uniform phase-contrast \#{}1).
    (a) Time series of the number density of topological defects.
    Each colored line represents data from a single observation region, while the black bold line shows the ensemble average over 30 regions.
    The defect density increases until nearly $t=30\unit{min}$.
    (b) The mean-square displacement (MSD) of topological defects. The data are shown in the time region where more than $1\%$ of the detected trajectories can be used to evaluate the MSD, to ensure the reliability of the results. The dashed line is a line of unit slope.
}
\end{figure}

\begin{figure}[h]
    \centering
    \includegraphics[width=0.8\textwidth]{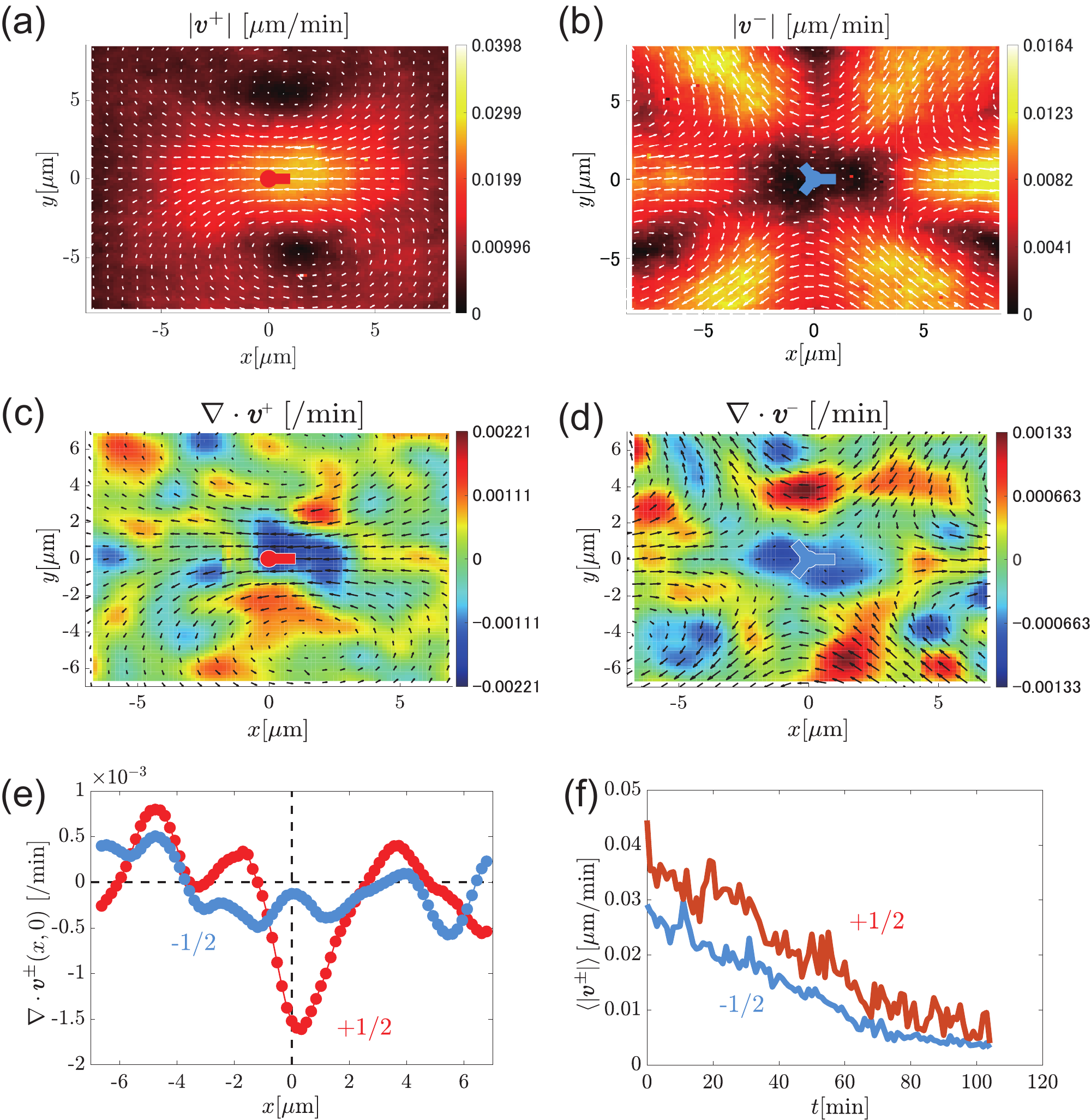}
    \caption{
    Supplemental results of two-dimensional velocity analyses on uniform colonies formed from numerous cells, observed by phase contrast microscopy.
    (a,b) Velocity field $\bm{v}^\pm(\bm{r})$ (white arrows) and its norm (color map) around $+1/2$ defects (a) and $-1/2$ defects (b).
    We took the time average over $30\unit{min} \leq t \leq 105\unit{min}$, during which the defect density was constant (Fig.\,S3(a)).
    (c,d) Velocity field $\bm{v}^\pm(\bm{r})$ (black arrows) and its divergence (color map) around $+1/2$ defects (c) and $-1/2$ defects (d).
    These are identical to Fig.\,2(b)(c) but reprinted here for comparison.
    (e) Divergence on the x-axis.
    (f) Time evolution of the spatially averaged norm $\expct{\overline{|\bm{v}^\pm(\bm{r})|}}$ around defects.
    The spatial average was taken in the region shown in (a)-(d).
}
\end{figure}

\begin{figure}[h]
    \centering
    \includegraphics[width=0.75\textwidth]{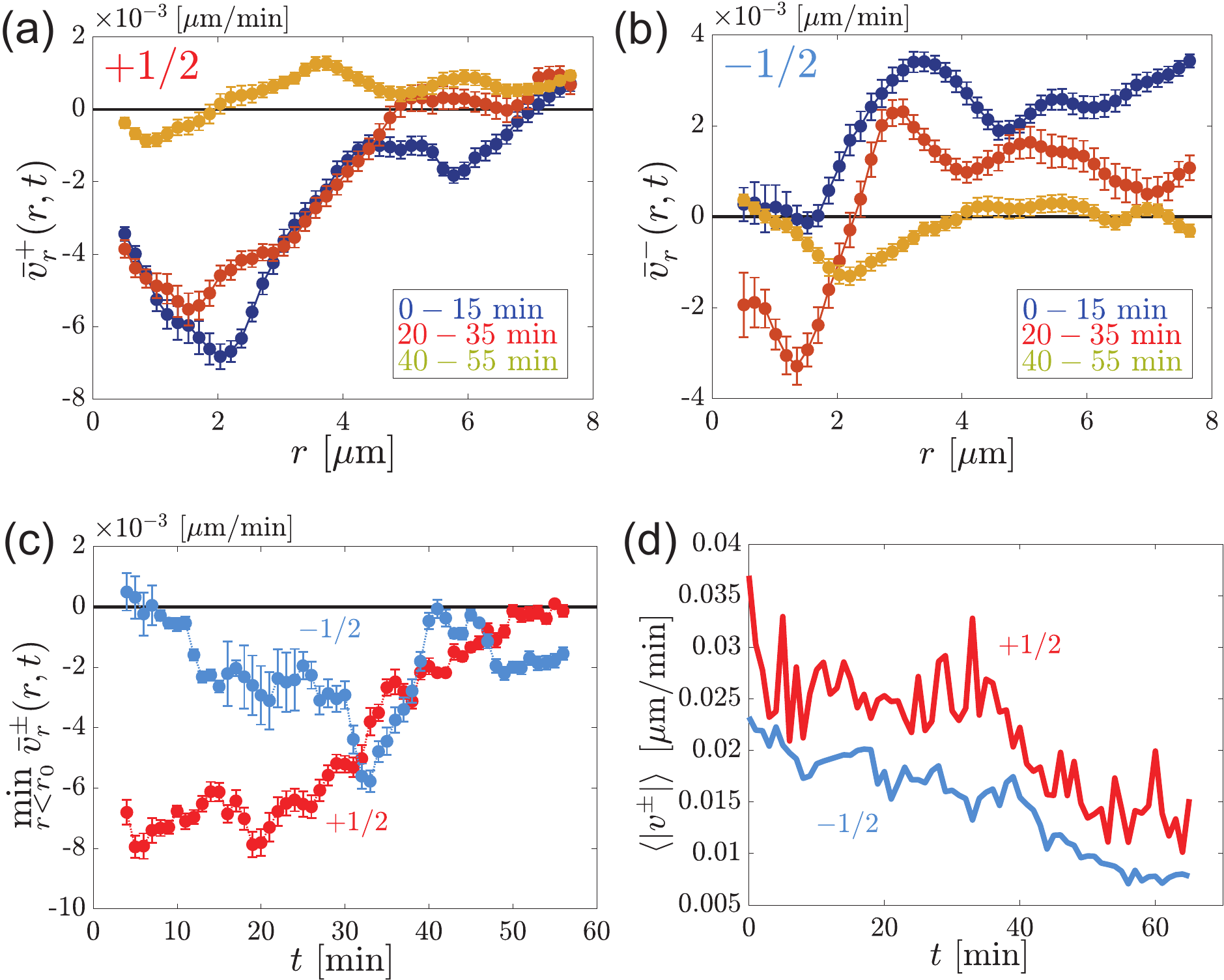}
    \caption{
    Two-dimensional velocity analyses on uniform colonies obtained by a biological replicate (uniform phase-contrast \#{}2).
    See Fig.\,2(e,f,g) and Fig.\,S4(f) for the results obtained by the other replicate (uniform phase-contrast \#{}1).
    (a,b) Time evolution of the mean radial velocity $\bar{v}_r^\pm(r,t)$ around $+1/2$ defects (a) and $-1/2$ defects (b).
    Here we show the velocity field averaged over $0\unit{min} \leq t \leq 15\unit{min}$, $20\unit{min} \leq t \leq 35\unit{min}$, and $40\unit{min} \leq t \leq 55\unit{min}$.
    The error bars indicate the time average of the standard error evaluated from each frame.
    (c) Time evolution of the minimum of $\bar{v}_r^\pm(r,t)$ in the region $r<r_0=2\unit{\mu{}m}$ near the defect.
    The moving average taken from $(t-5\unit{min})$ to $(t+5\unit{min})$ is shown with the corresponding error bar.
    (d) Time evolution of the spatially averaged norm $\expct{\overline{|\bm{v}^\pm(\bm{r})|}}$ around defects.
    The spatial average was taken in the region shown in Fig.\,S4(a)(b).
    }
\end{figure}

\begin{figure}[h]
    \centering
    \includegraphics[width=\textwidth]{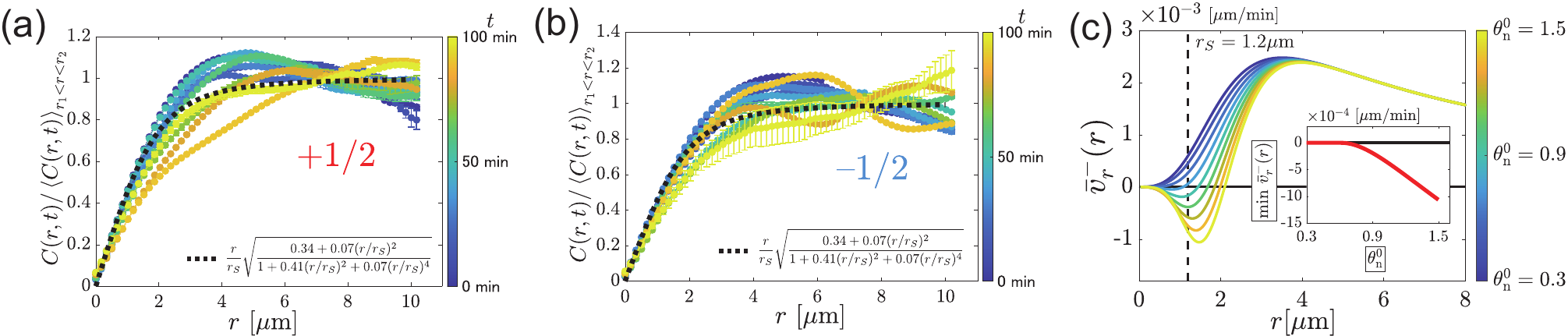}
    \caption{
    Supplementary data related to the theoretical calculations.
    (a,b) The coherency $C(\bm{r},t)$ around topological defects, obtained by phase-contrast observations by a single measurement (uniform colony, phase contrast \#{}1).
    Different colors correspond to different times $t$.
    To obtain $C(\bm{r},t)$, we measured the spatial profile of the coherency around each defect, and averaged it over defects at each time.
    The ordinates indicate $C(\bm{r},t)$ normalized by the spatial average taken in the region $r_1 < r < r_2$ with $r_1 = 5\unit{\mu{}m}$ and $r_2 = 10\unit{\mu{}m}$.
    The error bars indicate the standard error from the ensemble averaging.
    The dotted lines represent \eqref{eq:S16}, from which we estimated the defect core size at $r_S=1.2\unit{\mu{}m}$.
    (c) Theoretical results of the mean radial velocity $\bar{v}_r^-(r)$ around $-1/2$ defects, predicted from \eqref{eq:S11} for the case with nematic tilting but without polar order. 
    The curves were obtained for $\thetan^\infty=0.3$ and $\thetan^0$ varying in the range $0.3\leq \thetan^0 \leq 1.5$ from top to bottom, with $\epsilon_0=0.25, a^0_\mathrm{n}/\xi_0=0.055\unit{\mu{}m^2/min}, r_S=1.2\unit{\mu{}m}$ and $r_\theta=1\unit{\mu{}m}$.
    The vertical dashed line indicates the defect core radius obtained in (a)(b).
    (Inset) The dependence of the minimum of $\bar{v}^-_r(r)$ on $\thetan^0$.
    It becomes negative for sufficiently large $\thetan^0$, as predicted by \eqref{eq:S22}.
    }
\end{figure}

\begin{figure}[h]
    \includegraphics[width=\textwidth]{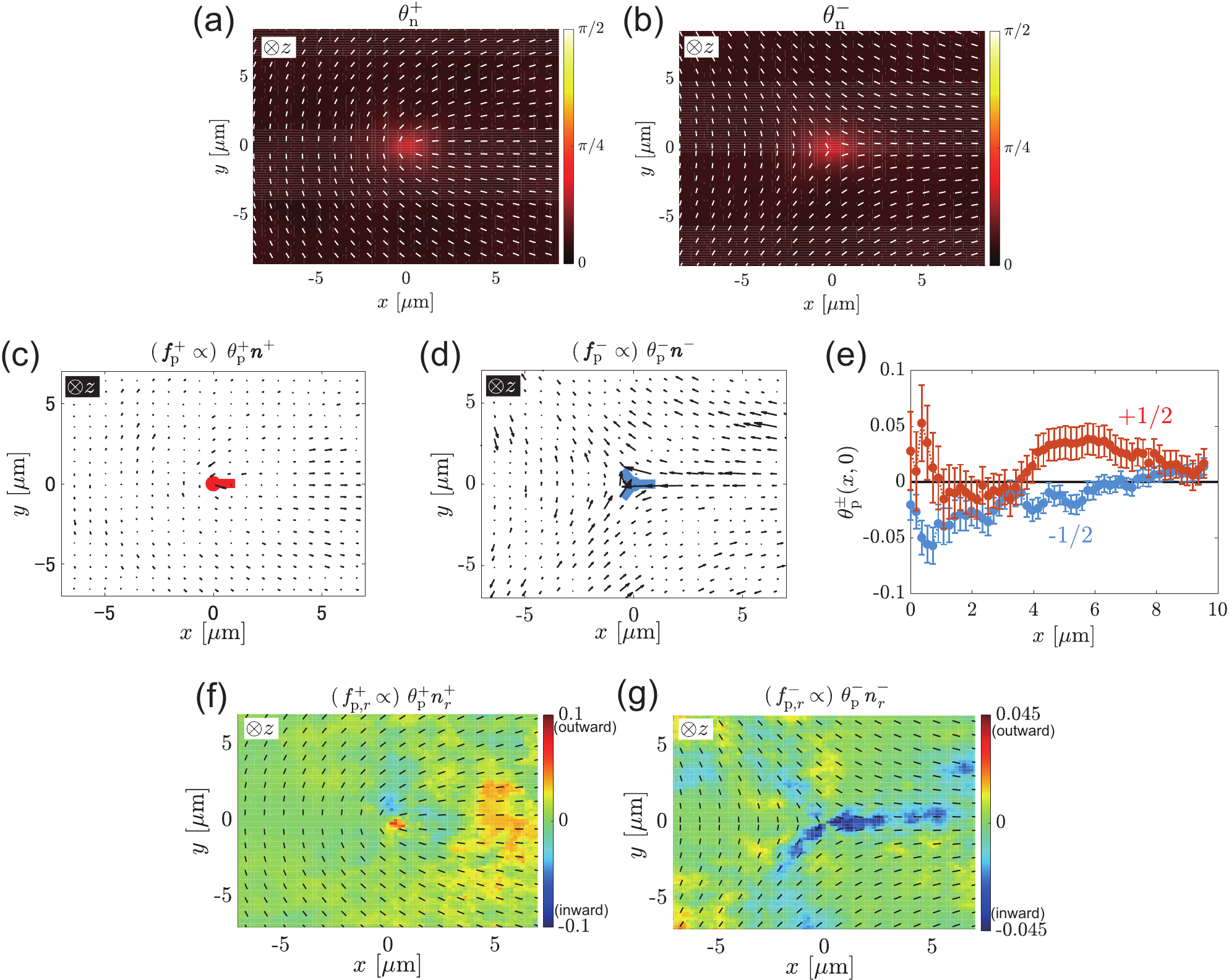}
    \caption{
    Results on the confocal observations of three-dimensional colonies obtained by a biological replicate (end-point confocal \#{}2).
    (a,b) Spatial profiles of the nematic tilt angle $\thetan^\pm(\bm{r})$ for the $+1/2$ defects (a) and the $-1/2$ defects (b).
    The white rods represent the nematic director field.
    (c,d) $\thetap^\pm \bm{n}^\pm$, representing the strength and direction of the polarity-induced force $\bm{f}_\mathrm{p}^\pm$ around the $+1/2$ defect (c) and the $-1/2$ defect (d).
    (e) The polar tilt angle $\thetap^\pm$ measured on the $+x$-axis of $\pm 1/2$ defects.
    The ensemble average over all defects is shown.
    The error bars indicate the standard error from the ensemble averaging.
    (f,g) $\thetap^\pm n_r^\pm$, which is proportional to the radial component of the polarity-induced force, $f_{\mathrm{p},r}^\pm$, around the $+1/2$ defect (f) and the $-1/2$ defect (g).
    The negative radial component indicates that the polarity-induced force is directed toward the defect.
    The black rods represent the nematic director field.
    }
\end{figure}

\begin{figure*}[h]
       \includegraphics[width=\hsize]{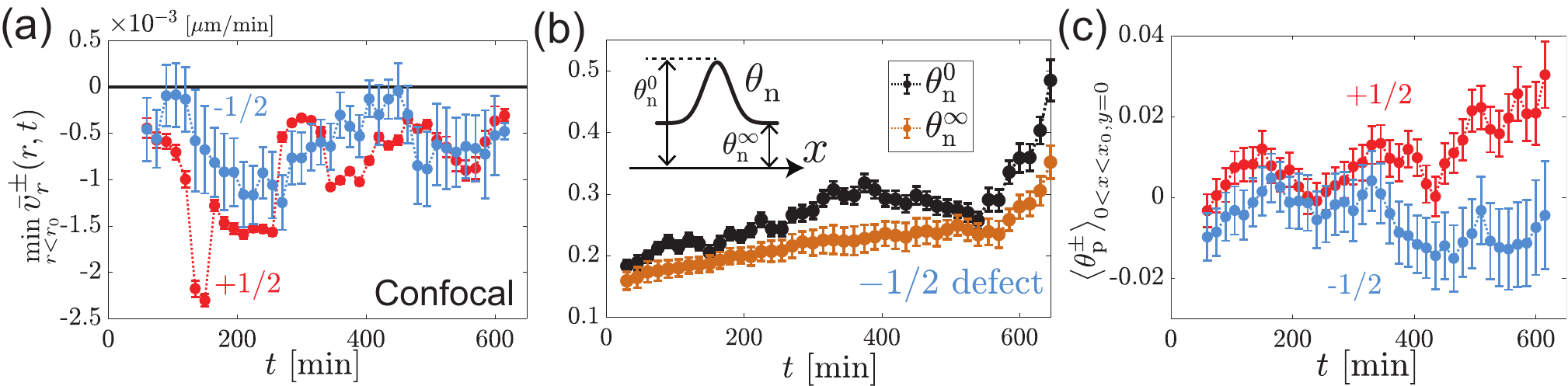}
       \caption{
       Results on uniform colonies formed from numerous cells, obtained by a single time-lapse confocal measurement (uniform colony, time-lapse confocal \#{}1).
       (a) Time evolution of the minimum of the mean radial velocity $\bar{v}_r^\pm(r)$, in the region $r < r_0$ with $r_0 = 3.6\unit{\mu{}m}$.
       The time average from $(t-75\unit{min})$ to $(t+75\unit{min})$ was taken. 
       We consider that the influx toward $+1/2$ defects at early times was underestimated because fast flow could not be detected accurately by PIV because of the relatively long time interval ($15\unit{min}$) chosen here.
       (b) Time evolution of the peak value $\thetan^0$ and the plateau value $\thetan^\infty$ of the nematic tilt angle $\thetan^-(\bm{r},t)$ for $-1/2$ defects.
       We evaluated $\thetan^0$ by the maximum of $\thetan^-(\bm{r},t)$ and $\thetan^\infty$ by the spatial average of $\thetan^-(\bm{r},t)$ in the plateau region far from the peak ($|\bm{r}|>5\unit{\mu{}m})$.
       The field $\thetan^-(\bm{r},t)$ and its standard error were obtained by ensemble averaging over all isolated defects (separated by a distance longer than $9\unit{\mu{}m}$ from other defects) at each time, followed by time averaging from $(t-30\unit{min})$ to $(t+30\unit{min})$.
       The error bars of $\thetan^0$ indicate the time-averaged standard error.
       The error bars of $\thetan^\infty$ indicate the standard deviation of $\thetan^-(\bm{r},t)$ obtained in the plateau region $|\bm{r}|>5\unit{\mu{}m}$.
       (c) Time evolution of the spatially averaged $\thetap^\pm$ measured on the $+x$-axis in the range $0<x<x_0$ with $x_0=10\unit{\mu{}m}$. 
       To obtain this, we first measured $\thetap^\pm(\bm{r})$ and its standard error for each time, as we did for Fig.\,4(b).
       We then took the spatial average on the $+x$-axis in the range $0<x<x_0$, followed by time averaging from $(t-60\unit{min})$ to $(t+60\unit{min})$.
       The error bars indicate the spatial and time-averaged standard error.
       }
\end{figure*}

\begin{figure*}[t]
       \centering
       \includegraphics[width=\textwidth]{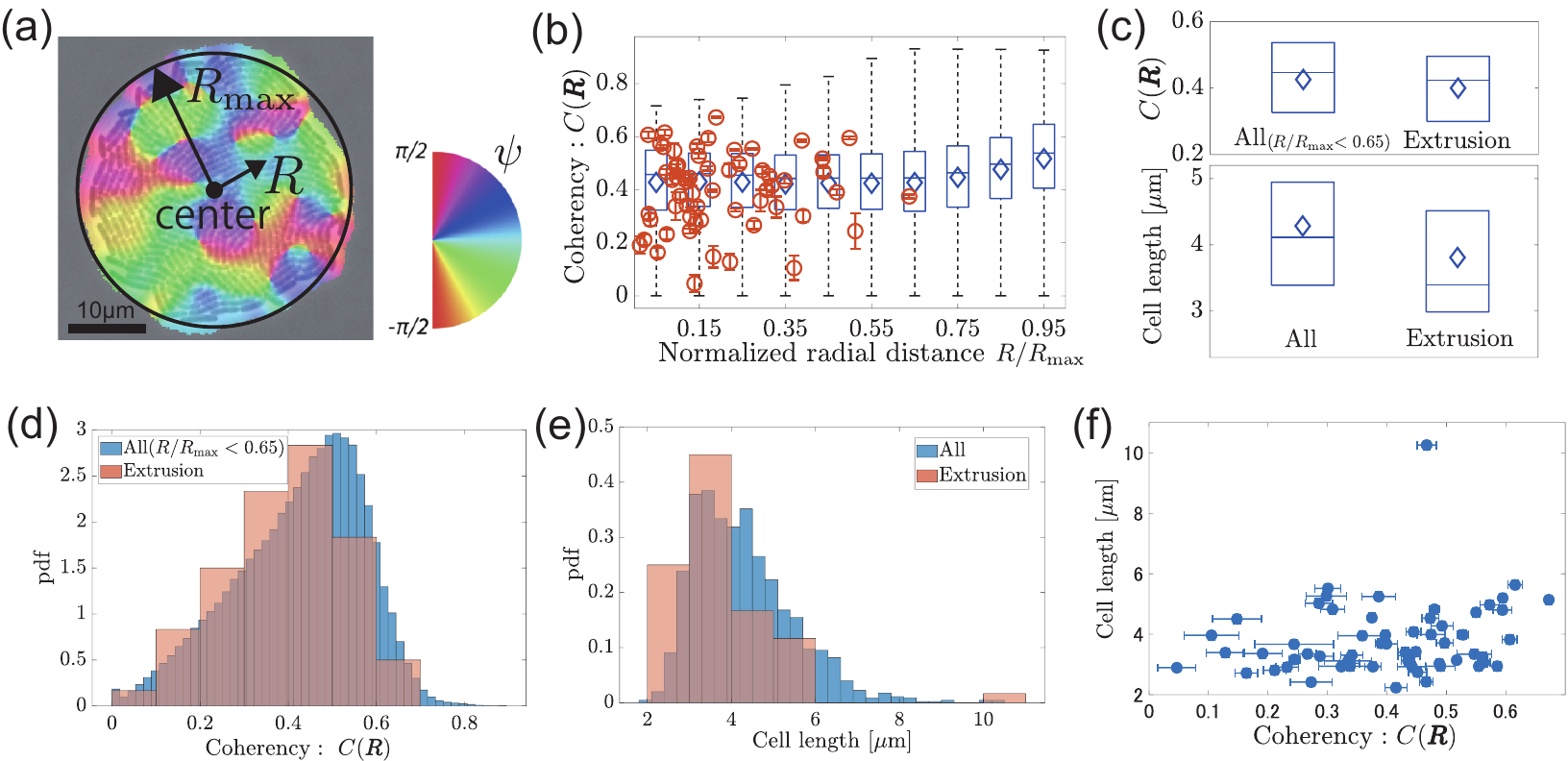}
       \caption{
       Results on circular colonies formed from a few cells. We investigated the first extruded cell (labeled as extrusion) in each of 60 colonies, which we collected from two experiments (circular colony, phase contrast \#{}1 and \#{}2) with 30 colonies each.
       (a) A circular colony formed from a single cell (see also Video~3).
       The color map indicates the cell orientation angle $\psi(\bm{R})$ obtained by the structure tensor method.
       $R_\mathrm{max}$ is defined as the radius of a circle that has the same area as the colony (see Appendix A).
       (b) Spatial distribution of the coherency $C(\bm{R})$ in the region at a given $R/R_\mathrm{max}$ (boxplots) and for the first extruded cells (red circles).
       The boxplots in this paper indicate the median by the horizontal lines, the lower and upper quartiles by the lower and upper box edges, respectively, the minimum and maximum by the error bars, and the mean value by the diamonds.
       The error bars of the red circles indicate the standard deviation of $C(\bm{R})$ over the pixels contained in the extruded cells (see Appendix A).
       The uncertainty of the position is smaller than the symbol size.,
       (c) Comparison of $C(\bm{R})$ (top) and the cell length (bottom) between all cells and the first extruded cells.
       For the coherency distribution, data for all cells were collected in the region $R/R_\mathrm{max}<0.65$.
       (d) Histogram of $C(\bm{R})$ for all cells in the region $R/R_\mathrm{max}<0.65$ (blue, sample size $N=1045164$) and that for the first extruded cells (red, sample size $N=60$).
       (e) Histogram of the cell length for all cells (blue, sample size $N=1507$, collected from 6 independent colonies) and that for the first extruded cells (red, sample size $N=60$).
       (f) Scattered plot of the coherency $C(\bm{R})$ and the length of the first extruded cells.
       The horizontal error bars indicate the standard deviations of $C(\bm{R})$ in the pixels contained in the extruded cells.
       The vertical error bars, which indicate segmentation uncertainty in the image analysis ($\pm 0.15\unit{\mu{}m}$), are smaller than the symbol size.
       }
\end{figure*}

\begin{figure}[h]
    \centering
    \includegraphics[width=0.7\textwidth]{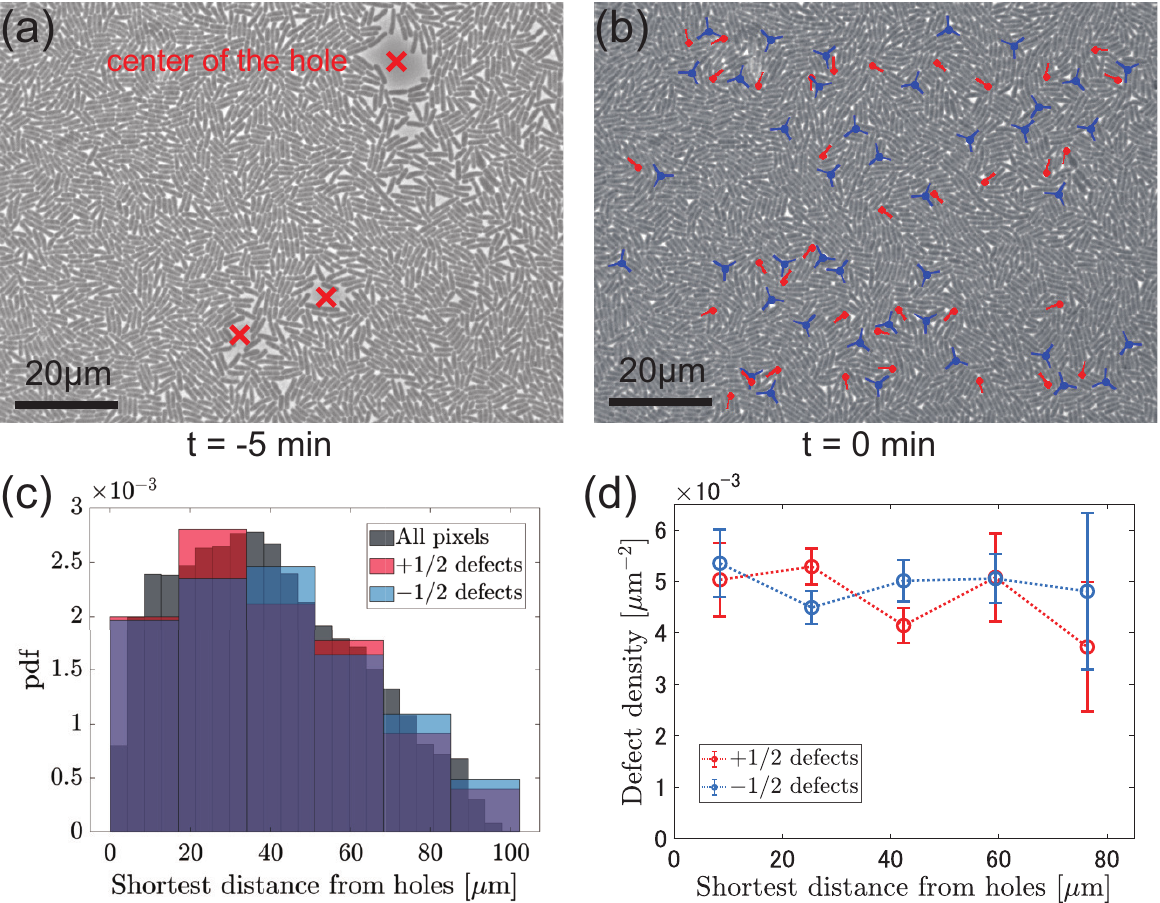}
    \caption{
    Absence of correlation between the defect density and the distance from holes before two-dimensional space filling.
    We investigated the defect density from 6 independent regions of dimensions $110.03\unit{\mu{}m} \times 81.97\unit{\mu{}m}$.
    These data were obtained by a single measurement (uniform phase-contrast \#{}1).
    (a,b) Snapshots of uniform colonies at $t=-5\unit{min}$ (a) and $t=0$ (b), the latter being the moment at which cells filled the two-dimensional bottom plane.
    (c) Probability density functions (pdf) of the shortest distance from holes, for the $+1/2$ defects, the $-1/2$ defects, and all pixels.
    To obtain the pdf's, we first manually detected the center position of the holes (1-5 holes for each independent region) from snapshots at $t=-5\unit{min}$.
    For each position of the defects or pixels, we measured the distance from the center of the closest hole.
    The sampling number was 175 for $+1/2$ defects, 183 for $-1/2$ defects, and 1243620 for the case of all pixels.
    (d) Defect density against the shortest distance from holes.
    The defect density was calculated from the raw data of the histograms of the shortest distance used in (c), as follows.
    First we obtained the defect density for each independent region, by dividing the number of counts of the histogram for $\pm 1/2$ defects by that of the histogram for all pixels.
    Then we took average over all regions and this is the result shown in (d).
    The error bars indicate the standard error.
    }
\end{figure}

\clearpage
\section{Video descriptions}
\noindent
{\bf Video~1:}\\
A phase-contrast movie of a uniform colony formed from numerous cells ({\sl E. coli}, MG1655) between a coverslip and an LB agar pad.
The movie is played 600 times faster than the real speed.\\
{\bf Video~2:}\\
A phase-contrast movie of a uniform colony formed from numerous cells ({\sl E. coli}, MG1655) between a coverslip and an LB agar pad, after cells filled a two-dimensional bottom plane.
Topological defects are also shown.
The movie is played 600 times faster than the real speed.\\
{\bf Video~3:}\\
A phase-contrast movie of a circular colony formed from a few cells ({\sl E. coli}, MG1655) between a coverslip and an LB agar pad.
The movie is played 600 times faster than the real speed.

\bibliography{NNbib}